\documentclass[useAMS,usenatbib]{mn2e}

\usepackage{amsmath}
\usepackage{graphicx}
\usepackage[german,british]{babel}
\usepackage[varg]{txfonts}
\usepackage{biblio}
\usepackage{bibentry}
\usepackage{natbib}
\usepackage{color}
\usepackage{booktabs}
\usepackage{microtype}

\usepackage[bookmarks=false]{hyperref}
\providecommand{\eprint}[1]{\href{http://arxiv.org/abs/#1}{#1}}

\newlength{\colwidth}
\newcommand{\g}{{\rm g}}
\newcommand{\cm}{{\rm cm}}
\newcommand{\psc}{{\rm cm^{-2}}}
\newcommand{\pcc}{{\rm cm^{-3}}}

\newcommand{\kms}{{\, \rm km}\,{\rm s}^{-1}}
\newcommand{\K}{{\, \rm K}}

\newcommand{\kpc}{{\rm kpc}}
\newcommand{\Mpc}{{\rm Mpc}}

\newcommand{\hMpc}{h^{-1}\,{\rm Mpc}}

\newcommand{\hMsun}{{h^{-1}\,{\rm M}_\odot}}

\newcommand{\nH}{{{\rm n}_{\rm H}}}
\newcommand{\mH}{{{m}_{\rm H}}}
\newcommand{\nHs}{{{\rm n}_{\rm H}^*}}

\newcommand{\ion}[2]{\hbox{#1\,{\sc #2}}}
\newcommand{\ionsubscript}[2]{\hbox{\scriptsize #1\,{\tiny #2}}}

\newcommand{\HI}{\ion{H}{i}}

\newcommand{\OVI}{\ion{O}{vi}}

\newcommand{\lya}{Ly$\alpha$}

\newcommand{\bea}{\begin{eqnarray}}
\newcommand{\eea}{\end{eqnarray}}
\newcommand{\beq}{\begin{equation}}
\newcommand{\eeq}{\end{equation}}
\newcommand{\bit}{\begin{itemize}}
\newcommand{\eit}{\end{itemize}}
\newcommand{\ben}{\begin{enumerate}}
\newcommand{\een}{\end{enumerate}}
\newcommand{\los}{sightline}
\newcommand{\loss}{sightlines}

\newcommand{\NHI}{{\rm N_{\ionsubscript{H}{I}}}}

\newcommand{\NH}{{\rm N_{\rm H}}}

\newcommand{\fHI}{{\rm f_{\ionsubscript{H}{I}}}}

\newcommand{\fhi}{({\rm n}_{\ionsubscript{H}{I}}/{\rm n}_{\rm H})}

\newcommand{\deltawhi}{\Delta_{\ionsubscript{H}{I}}}

\newcommand{\tempwhi}{T_{\ionsubscript{H}{I}}}

\newcommand{\metalwhi}{Z_{\ionsubscript{H}{I}}}

\newcommand{\fhiw}{({\rm n}_{\ionsubscript{H}{I}}/{\rm n}_{\rm H})_{\ionsubscript{H}{I}}}

\newcommand{\nhwhi}{\left({\rm n_H}\right)_{\ionsubscript{H}{I}}}

\newcommand{\bhi}{b_{\ionsubscript{H}{I}}}

\title[Broad \HI\ \lya\ Absorbers]{Absorption signatures of warm-hot gas at low redshift: Broad \HI\ \lya\ Absorbers}

\author[Tepper-Garc\'\i{}a et al.]{%
Thorsten Tepper-Garc\'\i{}a,$^{1}$\thanks{E-mail: tepper@astro.physik.uni-potsdam.de}
Philipp Richter,$^{1}$ 
Joop Schaye,$^{2}$ 
Craig M. Booth,$^{2,3,4}$
\newauthor
Claudio Dalla Vecchia,$^{2,5}$
Tom Theuns$^{6,7}$ 
\\
$^{1}$Institut f\"ur Physik und Astronomy, Universit\"at Potsdam, Karl-Liebknecht-Str. 24/25, 14476 Potsdam, Germany\\
$^{2}$Leiden Observatory, Leiden University, P.O. Box 9513, 2300 RA Leiden, The Netherlands\\
$^{3}$Department of Astronomy \& Astrophysics, The University of Chicago, Chicago, IL 60637, USA\\
$^{4}$Kavli Institute for Cosmological Physics and Enrico Fermi Institute, The University of Chicago, Chicago, IL 60637, USA\\
$^{5}$Max Planck Institut f\"ur Extraterrestrische Physik, Giessenbachstra\ss{}e 1, 85748 Garching, Germany\\
$^{6}$Institute for Computational Cosmology, Department of Physics, University of Durham, South Road, Durham, DH1 3LE, UK\\
$^{7}$Department of Physics, University of Antwerp, Groenenborgerlaan 171, B-2020 Antwerpen, Belgium
}
\begin{document}

\date{Accepted 2012 June 14. Received 2012 May 14; in original form 2012 January 26}

\pagerange{\pageref{firstpage}--\pageref{lastpage}} \pubyear{----}

\maketitle

\label{firstpage}

\begin{abstract}

We investigate the physical state of \HI\ absorbing gas at low redshift (\mbox{$z = 0.25$}) using a subset of cosmological, hydrodynamic simulations from the OWLS project, focusing in particular on broad (\mbox{$\bhi \geq 40 \kms$}) \HI\ \lya\ absorbers (BLAs), which are believed to originate in shock-heated gas in the warm-hot intergalactic medium (WHIM). Our fiducial model, which includes radiative cooling by heavy elements and feedback by supernovae and active galactic nuclei, predicts that by \mbox{$z = 0.25$} nearly 60 per cent of the gas mass ends up at densities and temperatures characteristic of the WHIM and we find that half of this fraction is due to outflows. The standard \HI\ observables (distribution of \HI\ column densities $\NHI$, distribution of Doppler parameters $\bhi$, \mbox{$\bhi - \NHI$} correlation) and the BLA line number density predicted by our simulations are in remarkably good agreement with observations.

BLAs arise in gas that is hotter, more highly ionised and more enriched than the gas giving rise to typical \lya\ forest absorbers. The majority of the BLAs arise in warm-hot (\mbox{$\log \left( T / \K \right) \sim 5$}) gas at low (\mbox{$\log \Delta < 1.5$}) overdensities. On average, thermal broadening accounts for at least 60 per cent of the BLA  line width, which in turn can be used as a rough indicator of the thermal state of the gas. Detectable BLAs account for only a small fraction of the true baryon content of the WHIM at low redshift. In order to detect the bulk of the mass in this gas phase, a sensitivity at least one order of magnitude better than achieved by current ultraviolet spectrographs is required. We argue that BLAs mostly trace gas that has been shock-heated and enriched by outflows and that they therefore provide an important window on a poorly understood feedback process.

\end{abstract}


\section{Introduction} \label{sec:intro}

The analysis of intervening \HI\ \lya\ absorption in the spectra of distant quasars (QSO) has become an extremely powerful tool to study the spatial distribution of the diffuse intergalactic medium (IGM) that follows the large-scale distribution of cosmological filaments, and to constrain the baryon content of the IGM as a function of redshift. At redshifts \mbox{$z > 3$}, more than 95 per cent of the baryonic matter resides in the form of photo-ionised, diffuse gas giving rise to the ``\lya\ forest'' in the spectra of distant QSOs \citep[e.g.][]{rau97a}. As a consequence of expansion, the \lya\ forest thins out, and at \mbox{$z\approx0$} the contribution of the \lya\ forest to the total baryon budget has decreased to \mbox{$\sim 20$} per cent \citep[e.g.][]{pen04a,leh07a}. At the same time, the formation of galactic structures and the gravitational heating of the IGM by collapsing large-scale filaments lead to a gradually increasing amount of shock-heated intergalactic gas at temperatures \mbox{$T \gtrsim 10^5 \K$}, which is referred to as the warm-hot intergalactic medium \citep[WHIM,][]{cen99a,the98b,dav01a,ber08a}.

Since collisional ionisation determines the ionisation state of the shock-heated IGM, the neutral gas fraction in the WHIM is significantly lower, by at least one order of magnitude, than in the photo-ionised IGM of the same density \citep[e.g.][]{ric08b}. Because of this very small neutral hydrogen fraction in the WHIM, most of the recent observational campaigns to study warm-hot intergalactic gas at low redshift have concentrated on intervening absorption by highly ionised metals in ultraviolet (UV) spectra of bright QSOs. In particular, five-times ionised oxygen (\OVI) has been used extensively to trace shock-heated intergalactic gas at low redshift and to constrain the baryon content of the WHIM \citep[e.g.][]{tri00a,ric04a,dan06a,dan08b,tho08a,tri08b,dan10a}. However, because \OVI\ predominantly traces metal-enriched gas in a critical (in terms of ionisation balance) temperature regime at \mbox{$T\approx 3\times10^5 \K$},  and because the metals may well be poorly mixed on small scales \citep[][]{sch07a} the interpretation of intervening \OVI\ absorbers is still controversial \citep[e.g.][]{opp09b,tep11a,smi11a}. In particular, it is not yet clear whether \OVI\ absorbers predominantly arise in photo-ionised \citep[e.g.][]{tho08b} or collisionally ionised gas \citep[e.g.][]{dan08b}, or in complex absorbing structures with cool gas intermingled with warm-hot gas \citep[][]{tri08b}.

An alternative to highly ionised metals as tracers of warm-hot gas is offered by \HI\ absorption. Due to the low neutral hydrogen fraction expected from collisional ionisation at temperatures \mbox{$T \gtrsim 10^5 \K$}, \lya\ absorption from shock-heated WHIM filaments is expected to be very weak. In addition, \HI\ absorption lines arising in gas at temperatures \mbox{$T>10^5 \K$} are expected to be relatively broad because of the effect of thermal  broadening. Such broad (\mbox{$\bhi \geq 40 \kms$}) and shallow (\mbox{$\left[\NHI / \bhi \right] \sim 10^{11} \psc {\rm km^{-1}s}$ or \mbox{$\tau_0(\HI) \sim 0.1$}}) \lya\ absorption features, the so-called Broad \lya\ Absorbers \citep[BLAs; ][]{ric06a}, are hence difficult to identify in the UV spectra of QSOs because of the limited signal-to-noise (S/N) and the low resolution of spectral data obtained with current space-based UV spectrographs.

In spite of being observationally challenging, directly detecting the small amounts of neutral hydrogen in the WHIM in absorption is a feasible task. The first systematic studies of BLAs at low redshift have been conducted using high-resolution {\em Hubble Space Telescope} (HST) {\em Space Telescope Imaging Spectrograph} (STIS) spectra of bright QSOs \citep[][]{ric04a,sem04a,ric06a,wil06a,leh07a,dan10a}. These studies indicate that BLAs may indeed account for a substantial fraction of the baryons in the WHIM at \mbox{$z\approx0$}. They also show, however, that identification and interpretation of broad spectral features in UV spectra with limited data quality is afflicted with large systematic uncertainties. In particular, the effects of non-thermal broadening and unresolved velocity-structure in the lines lead to the occurrence of broad spectral features that do not necessarily arise in gas at high temperatures. The {\em Cosmic Origins Spectrograph} \citep[COS;][]{gre12a}, a new UV spectrograph which has recently been installed on HST, is expected to substantially increase the number of BLA candidates at low redshift. Due to the limited spectral resolution of COS \mbox{$(\sim17\kms)$}, the systematic uncertainties in identifying thermally broadened \HI\ lines in the WHIM temperature range will nevertheless remain.

To investigate the physical properties and spectral signatures of BLAs at low redshift, \citet[][]{ric06b} have studied broad \HI\ absorption features using a cosmological simulation based on a grid-based adaptive mesh refinement (AMR) method \citep[][]{nor99a}. Their simulation reproduces the observed BLA number density and supports the idea that BLAs trace (at least in a statistical sense) a substantial fraction of shock-heated gas in the WHIM at temperatures \mbox{$T \sim 10^5-10^6 \K$}. However, since this (early) simulation ignored several important physical processes that are expected to affect the thermal state of this gas phase (i.e. energetic feedback, radiative heating and cooling by hydrogen and metals), it is important to  re-assess the frequency and physical properties of BLAs using state-of-the-art cosmological simulations with more realistic gas physics.

In this paper, we present a systematic study of BLAs at low redshift based on a set of cosmological simulations from the {\it OverWhelmingly Large Simulations} (OWLS) project \citep[][]{sch10a}. This work complements our previous study on intervening \OVI\ absorbers and their relation to the WHIM based on a slightly different set of OWLS simulations \citep[][henceforth Paper I]{tep11a}. The main features of the simulations we use are briefly described in Sec.~\ref{sec:sims}. As we have done in Paper I for the case of low redshift \OVI\ absorbers, we compare the predictions from our fiducial model to a set of standard \HI\ observables, and discuss various physical properties of the general \HI\ absorber population in Sec.~\ref{sec:obs}. Given the dependence of the WHIM mass fraction predicted by simulations on the particular implementation of the relevant physical processes reported in the past \citep[e.g.][]{cen06a}, we investigate the impact of different physical models on the thermal state of the various gas phases in our simulations in Sec.~\ref{sec:whim}. In this section we also present and discuss the results on the physical properties of the absorbing gas traced by BLAs. Finally, we summarise our main findings in Sec.~\ref{sec:sum}. In the Appendix we include: a full description of our fitting algorithm (Appendix \ref{sec:fit}); a detailed calculation of the observability of \HI\ absorbing gas in terms of optical depth as a function of density and temperature (Appendix \ref{sec:tau0_vs}); a discussion of the convergence of our results with respect to the adopted physical model (Appendix \ref{sec:cnvg_phys}), and with respect to the adopted resolution and simulation box size (Appendix \ref{sec:cnvg_num}).

\section[]{Simulations} \label{sec:sims}

The simulations used in this work are part of a large set of cosmological simulations that together comprise the OWLS project, described in detail in \citet[][and references therein]{sch10a}. Briefly, the simulations were performed with a significantly extended version of the $N$-Body, Tree-PM, Smoothed Particle Hydrodynamics (SPH) code \textsc{gadget iii} -- which is a modified version of \textsc{gadget ii} \citep[last described in][]{spr05b} --, a Lagrangian code used to calculate gravitational and hydrodynamic forces on a system of particles. The initial conditions were generated from an initial glass-like state \citep[][]{whi96a} with \textsc{cmbfast} \citep[version 4.1; ][]{sel96a} and evolved to redshift \mbox{$z = 127$} using the \citet[][]{zel70a} approximation.

The reference model, dubbed \mbox{\em REF}, in the OWLS framework adopts a flat $\Lambda$CDM cosmology characterised by the set of parameters $\{\Omega_{\rm m}, \, \Omega_{\rm b}, \, \Omega_{\Lambda}, \, \sigma_{8}, \, n_{\rm s}, \, h\} = \{ 0.238, \, 0.0418, \, 0.762, 0.74, \, 0.95, \, 0.73 \}$ as derived from the Wilkinson Microwave Anisotropy Probe (WMAP) 3-year data%
\footnote{These parameter values are largely consistent with the WMAP 7-year results \citep[][]{jar11a}, the largest difference being the value of $\sigma_{\,8}$, which is $2 \, \sigma$ lower in the WMAP 3-year data than allowed by the WMAP 7-year data.%
} \citep[][]{spe07a}. This model includes star formation following \citet[][]{sch08e}, metal production and timed release of mass and heavy elements by intermediate mass stars, i.e. asymptotic giant-branch (AGB) stars and supernovae of Type Ia (SNIa), and by core-collapse supernovae (SNIIe) as described by \citet[][]{wie09b}. It further incorporates kinetic feedback by SNIIe based on the method of \citet[][]{dal08b}, as well as thermal feedback by SNIa \citep[][]{wie09b}. Radiative cooling by hydrogen, helium and heavy elements is included following the method of \citet[][]{wie09a}. The ionisation balance for each SPH particle is computed as a function of redshift, density, and temperature using pre-computed tables obtained with the photoionisation package \textsc{cloudy} \citep[version 07.02.00 of the code last described by][]{fer98a}, assuming the gas to be optically thin and exposed to the \citet[][]{haa01a} model for the X-Ray/UV background radiation from galaxies and quasars. It is worth noting that a simulation run that adopts the \mbox{\em REF} model, although with a slightly different set of values for the cosmological parameters (from WMAP7), has been shown to reproduce the \HI\ absorption observed at \mbox{$z=3$} in great detail \citep[][]{alt11a}.

\begin{table} 
\begin{center}
\caption{Overview of the simulations used in this study. All model variations are relative to \mbox{\em REF}.
}  
\label{tbl:sims}
\begin{tabular}{rl}
\hline
 Model 					& Description							\\
\hline
 \mbox{\em NOSN\_NOZCOOL} 	& neglects SNe energy feedback and		\\ 
			 			& cooling assumes primordial abundances	\\ 
 \mbox{\em NOZCOOL} 			&  cooling assumes primordial abundances	\\ 
 \mbox{\em REF} 				& OWLS reference model (see text for details)	\\ 
 \mbox{\em AGN} 				& includes feedback by AGN (fiducial model)	\\ 
\hline
\end{tabular}
\end{center}
\end{table}

Along with \mbox{\em REF}, we consider three further models from the OWLS suite respectively referred to as \mbox{\em NOSN\_NOZCOOL}, \mbox{\em NOZCOOL}, and \mbox{\em AGN}. All these models differ from the reference model in one or more respects.  \mbox{\em NOSN\_NOZCOOL} neglects kinetic feedback by SNIIe, and the calculation of radiative cooling assumes primordial abundances. It is the most simple model in terms of input physics, and it is similar (and hence useful for comparison) to the simulation used by \citet[][]{ric06b}. The model \mbox{\em NOZCOOL} assumes primordial abundances when computing radiative cooling, and the model \mbox{\em AGN} includes feedback by active galactic nuclei (AGN) based on the model of black hole growth developed by \citet[][see also \citealt{spr05a}]{boo09a}. 

All these simulations were run in a cubic box of \mbox{$100 h^{-1}$} co-moving Mpc on a side, containing $512^3$ dark matter (DM) particles and equally many baryonic particles. The initial mass resolution is \mbox{$4.1\times10^8 \hMsun$} (DM) and \mbox{$8.7\times10^7 \hMsun$} (baryonic). The gravitational softening is set to $8 ~h^{-1}$ co-moving kpc and is fixed at \mbox{$2 h^{-1}$} proper kpc below \mbox{$z=3$}.


In this study, we choose \mbox{\em AGN} as our fiducial model since it is the most complete model in terms of input physics. In addition to reproducing various standard \HI\ statistics (see Appendix \ref{sec:cnvg_phys}), this model has been shown to reproduce: the observed mass density in black holes at \mbox{$z=0$}; the black hole scaling relations \citep[][]{boo09a} and their evolution \citep[][]{boo11a}; the observed optical and X-ray properties, stellar-mass fractions, star-formation rates (SFRs), stellar-age distributions and the thermodynamic profiles of groups of galaxies \citep[][]{mcc10a}; and the steep decline in the cosmic star formation rate below \mbox{$z=2$} \citep[][]{sch10a,van11c}. Note that, while the \HI\ statistics predicted by the \mbox{\em AGN} model are very similar to the predictions of the other models considered here (see Appendix \ref{sec:cnvg_phys}), there are notable differences in the temperatures of the gas traced by BLAs (see Fig.~\ref{fig:temp_dist}). We will address this point in more detail in Sec.~\ref{sec:obs_phys}. Table \ref{tbl:sims} briefly summarises the relevant features of the models described above. For a more detailed description of these (and other) models that are part of the OWLS project, see \citet[][]{sch10a}.

\section{The general \HI\ absorber population} \label{sec:obs}

In this section we test the predictions of our fiducial model (\mbox{\em AGN}) against observations using a set of well-measured \HI\ observables: the \HI\ column density distribution function (CDDF), the distribution of \HI\ line widths, and the correlation between \HI\ column density and line width.

\subsection{Synthetic spectra} \label{sec:spec}

For a meaningful comparison to existing data, we generate 5000 random \loss\ (1000 at five redshifts spanning the range \mbox{$0 \leq z \leq 0.5$} with step ${\rm d}z = 0.125$) through the simulation box covering a total redshift path \mbox{$\Delta z = 189$}, corresponding to an absorption path length \mbox{$\Delta \chi = 275$}.

We use the package \textsc{specwizard} written by Schaye, Booth, \& Theuns to generate a synthetic spectrum for each \los\ containing absorption by \HI\ \lya\ only. Briefly, we draw a random physical \los\ across the simulation box of size $L$, which is simply defined as the line between a given point on opposite faces of the box, and the collection of SPH particles with projected distances to this line smaller than their smoothing length. Next, we calculate the ionisation balance for each SPH particle as a function of redshift, density, and temperature, which we do using precomputed tables obtained with the photoionisation package \textsc{cloudy} \citep[version 07.02 of the code last described by][]{fer98a}, assuming the gas to be optically thin and exposed to the \citet[][]{haa01a} model for the X-Ray/UV background radiation from galaxies and quasars. We divide the physical \los\ into $N_{\,\rm pix} = \left[a(z) \, L/h \right] / \Delta x$ pixels of constant width $\Delta x$, where $h$ and $a(z)$ are the Hubble constant in units of \mbox{$100 \kms  \Mpc^{-1}$} and the expansion factor at the box's redshift $z$, respectively, and compute the smoothed ion density ${\rm n}_{\rm ion}$, the ion density-weighted gas temperature, and the ion density-weighted peculiar velocity at each pixel. Proper distance bins of width $\Delta x$ along the \los\ are transformed into velocity bins of width \mbox{$\Delta v = H(z) \Delta x$}, where $H(z)$ is the Hubble parameter at redshift $z$; ion number densities are transformed into ion column densities via \mbox{${\rm N}_{\rm ion} = {\rm n}_{\rm ion} \Delta x$}, and gas temperatures into Doppler parameters using \mbox{$b_T  = \sqrt{2 k T/ {\rm m}_{\rm ion}}$}., where $k$ is Boltzmann's constant and ${\rm m}_{\rm ion}$ is the ion's mass. The \HI\ optical depth $\tau(v)$ at each pixel is computed assuming a thermal (i.e. Gaussian) profile, taking peculiar velocities into account, as described by \citet[][their Appendix 4]{the98b}. Finally, the optical depth spectrum is transformed into a continuum-normalised flux via $F(v) = \exp [-\tau(v)]$.

We convolve our spectra with a Gaussian line-spread function (LSF) with a full width at half-maximum \mbox{(FWHM)  = 7 $\kms$} and re-sample our spectra onto \mbox{$3.5 \kms$} pixels. We add Gaussian noise to each spectrum assuming a flux dependent root-mean-square (rms) amplitude given by \mbox{$\left({\rm S/N } \right)^{-1} F(v)$}, where S/N is the adopted signal-to-noise ratio. We assume a {\em minimum}, i.e. flux-independent noise level \mbox{$\sigma_{\rm min} = 10^{-4}$}. This implies that our algorithm will underestimate the true column density of absorption features with a flux of the order of (or  lower than) \mbox{$\sigma_{\rm min}$}, which corresponds to a logarithmic central optical depth \mbox{$\log \tau_0 \sim 1$} (see Appendix \ref{sec:fit}). Our choice of a perhaps unrealistically low value for \mbox{$\sigma_{\rm min}$} thus allows us to reduce the gap between the true and the fitted column density of saturated lines.

We generate three sets of spectra, adopting S/N=10, S/N=30, and S/N=50, respectively. The spectra with S/N=10 and S/N=30 thus closely match the properties of the large sample thus far obtained with {\em HST/STIS}; these will be used in Secs.~\ref{sec:cddf}, \ref{sec:bvd}, and \ref{sec:bnc} to test the predicted \HI\ observables against observations; the synthetic spectra with S/N=50 are intended to investigate the physical properties of the \HI\ absorbing gas following a statistical approach, in the remaining sections of the paper.

Fitting of our 5000 synthetic spectra using the procedure described in Appendix \ref{sec:fit} yields a total of 93430, 66705, and 28649 components for S/N =50,  30, and 10, respectively. The resulting line-number densities and their corresponding Poisson uncertainties  are \mbox{$\left({\rm d} N / {\rm d} z\right) = 494 \pm 22$} (S/N=50), $353 \pm 19$ (S/N=30), and $152 \pm 12$ (S/N=10). For reference, the sample of 341 \lya\ absorbers at \mbox{$z \lesssim 0.4$} identified in seven FUSE+STIS spectra with average \mbox{S/N $\gtrsim 10$} by \citet[][]{leh07a} along an unblocked redshift path \mbox{$\Delta z = 2.064$} yields \mbox{$\left({\rm d} N / {\rm d} z\right) = 165 \pm 13$} at \mbox{${\rm S/N} \approx 10$}, which agrees (within the Poisson uncertainties) with our result at a similar S/N.

\subsection{Column-density distribution function} \label{sec:cddf}

\begin{figure}
\resizebox{\colwidth}{!}{\includegraphics{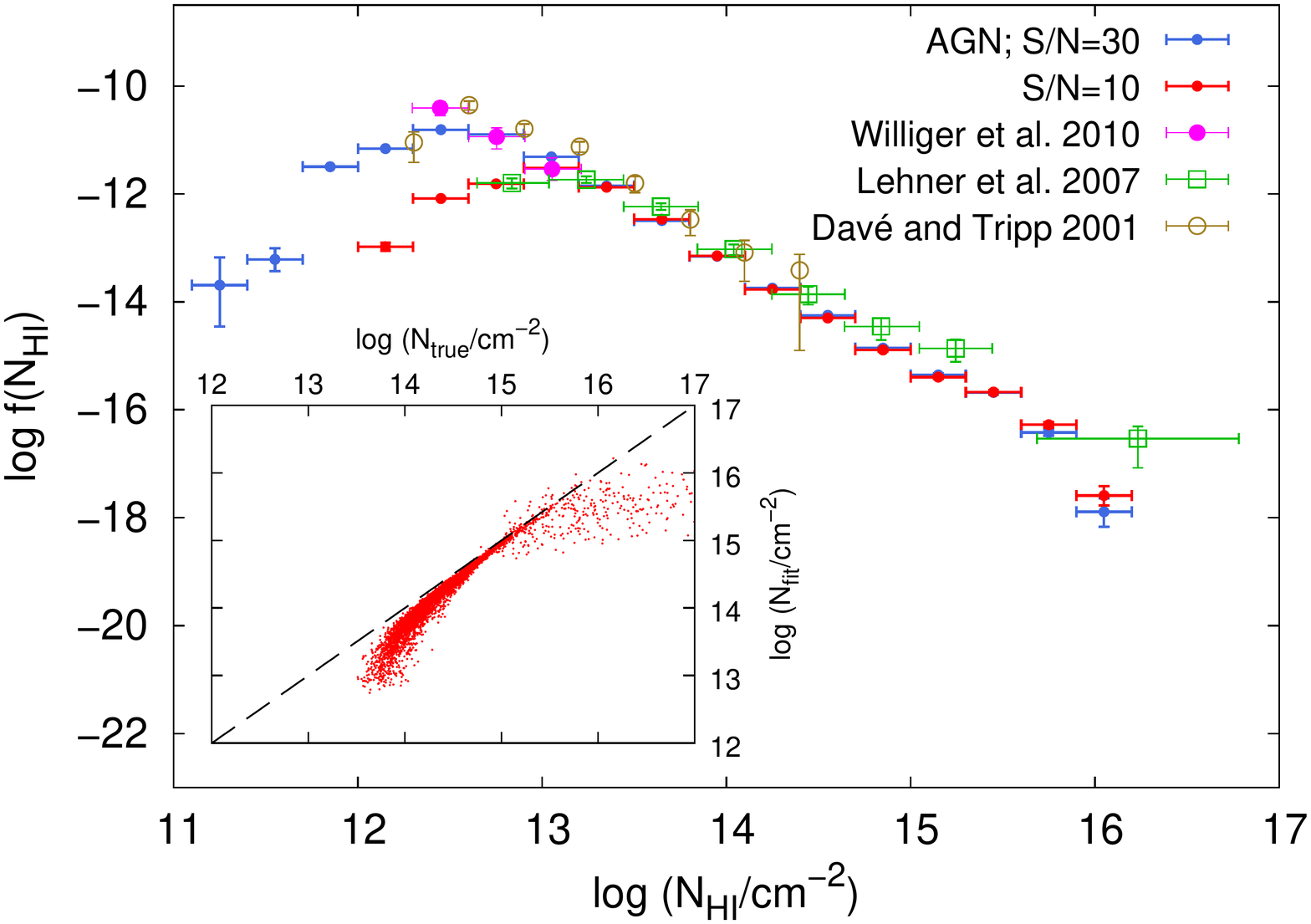}}
\caption[]{Column-density distribution function,  $f(\NHI)$, obtained from observations and from 5000 simulated spectra with different S/N values spanning the redshift range \mbox{$[0,0.5]$}. Error bars along the $y$-axis show Poisson uncertainties computed using the tables by \citet[][]{geh86a}. Assuming  \mbox{$f(\NHI) \propto \NHI^{-\beta}$} for logarithmic column densities in the range \mbox{$[13.0 , \, 15.2 ] $} (\mbox{$[12.5 , \, 15.2] $}), we find \mbox{$\beta = 1.916 \pm 0.044$} (\mbox{$\beta = 1.917 \pm 0.032$}) for S/N=10 (S/N=30). The inset shows the comparison between the true ($x$-axis) and the fitted ($y$-axis) \HI\ column density integrated along each individual \los. For clarity, only the result for S/N=10 is shown. The dashed line corresponds to a perfect match (see text for discussion).}
\label{fig:obs_1}
\end{figure}

In Fig.~\ref{fig:obs_1} we show the column-density distribution function (CDDF), \mbox{$f(\NHI)$}, obtained from our spectra with S/N=10 (red) and S/N=30 (blue) spanning the redshift range \mbox{$[0,0.5]$}, together with results from different observations at similar redshifts using spectra with comparable (average) S/N values. Assuming that the CDDF can be parametrized in the form of a single power-law,  \mbox{$f(\NHI) \propto \NHI^{-\beta}$}, we find \mbox{$\beta = 1.916 \pm 0.044$} (\mbox{$\beta = 1.917 \pm 0.032$}) for S/N=10 (S/N=30) for logarithmic column densities in the range \mbox{$[13.0, \, 15.2]$} (\mbox{$[12.5, \, 15.2] $}). The lower limit in \mbox{$\log \NHI$} approximately corresponds in each case to the completeness limit as given by eq.~\eqref{eq:sl_N}, while the upper limit roughly indicates the column density above which our fitting algorithm underestimates the true \HI\ column density due to the minimum noise-level adopted (see Appendix \ref{sec:fit}).

The slope we obtain is in fairly good agreement with the slope measured from different observations. For \mbox{$z \lesssim 0.4$} \citet[][their table 7]{leh07a} measure a range of values \mbox{$\beta = 1.52 - 1.92$} for absorbers in selected column-density intervals between \mbox{$\log\ (\NHI / \psc ) = 13.2 $} and $16.5$, and line widths \mbox{$\bhi \leq 40 \kms$} or \mbox{$\bhi \leq 80 \kms$}. If we extend the fitted column density range to \mbox{$\log\ (\NHI / \psc ) = 16.5 $}, we find \mbox{$\beta = 1.90 \pm 0.06$} and \mbox{$\beta = 1.95 \pm 0.06$} for S/N=10 and S/N=30, respectively. \citet[][]{wil10a} use a subsample from the \citet[][]{leh07a} data and their own data at \mbox{$ \log\ (\NHI/\cm^{-2} ) \leq 12.3 $}, and find \mbox{$\beta = 1.79 \pm 0.1$}. \citet[][]{dav01a} measure \mbox{$\beta = 2.04 \pm 0.23$} for absorbers with column densities \mbox{$ \log\ (\NHI/\cm^{-2} ) \geq 12.9 $} at a median redshift \mbox{$\widetilde{z} = 0.17$}. Note, however, that a significantly shallower slope is found by \citet[][]{pen04a} who identify 109 \lya\ absorbers at \mbox{$z < 0.069$} along 15 STIS spectra with \mbox{$\rm S/N \gtrsim 20$}, and  measure \mbox{$\beta = 1.65 \pm 0.07$} for logarithmic \HI\ column densities in the range \mbox{$[12.5, \,14.5 ] $}.

The amplitude of the CDDF resulting from the analysis of our synthetic spectra adopting different S/N is also in remarkable agreement with the observations. Note that the amplitude comes out naturally from our simulation, i.e. the CDDF has not been normalised to match the data in any way (even though that could have been justified because of uncertainties in the intensity of the UV background). At column-densities \mbox{$\log\ (\NHI/\psc ) \lesssim 15 $}, our predicted amplitude agrees well with the data all the way down to the lowest column densities measured, \mbox{$\log\ (\NHI/\psc ) = 12.3 $}. At  \mbox{$\log\ (\NHI/\psc ) > 15 $}, the amplitude of our predicted CDDF appears slightly lower (or its slope is steeper) than the result by \citet[][]{leh07a}. Note, however, that their data point at highest measured column-density bin has a rather large uncertainty. On the other hand, it is very likely that our choice of fitting parameters leads us to underestimate the amplitude of the predicted CDDF at \mbox{$\log\ (\NHI/\psc ) \gtrsim 14.5 $} by underestimating the true column density of {\em saturated} lines, as explained in Appendix \ref{sec:fit}. A comparison between the true and the fitted \HI\ column densities integrated along each \los\ reveals that our fitting procedure indeed yields integrated \HI\ column densities which are systematically lower than the true total column density, in particular for \mbox{$\log\ (\NHI/\psc ) \gtrsim 15 $} (see inset in Fig.~\ref{fig:obs_1}). This could explain the difference between our predicted CDDF and the result by \citet[][]{leh07a} at the high-$\NHI$ end. 

\subsection{Line width distribution} \label{sec:bvd}

\begin{figure}
\resizebox{\colwidth}{!}{\includegraphics{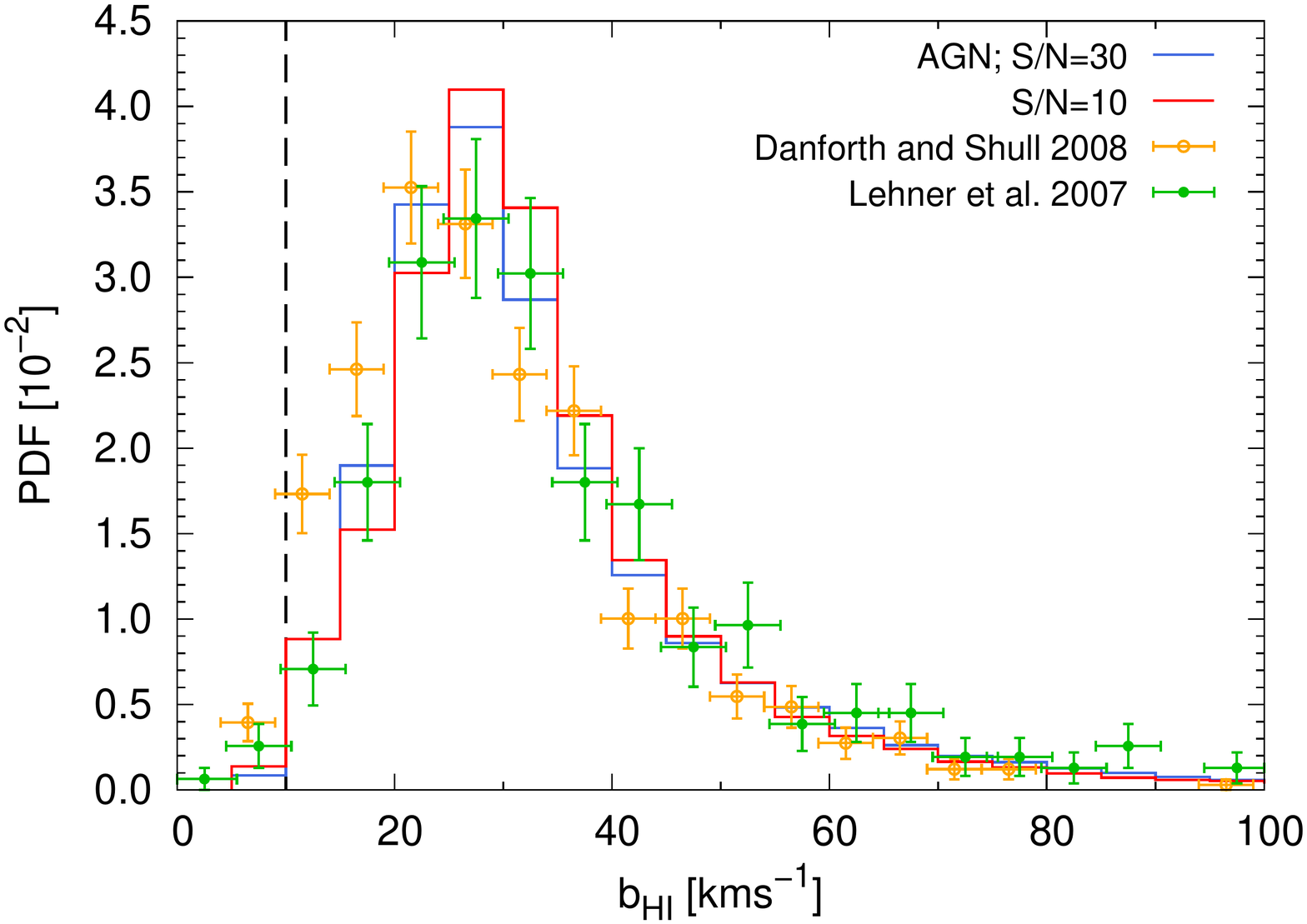}}
\caption[]{Distribution of $\bhi$-values from observations and from 5000 simulated spectra with S/N=10 (red), and S/N=30 (blue), spanning the redshift range \mbox{$[0,0.5]$}. Data points from \citet[][green]{leh07a} and \citet[][orange]{dan08b} with $y$-error bars showing Poisson uncertainties. The dashed vertical line indicates our adopted minimum allowed $b$-value. The distributions from both data and simulations have been binned using \mbox{$\Delta \bhi = 5 \kms$}.}
\label{fig:obs_2}
\end{figure}

Fig.~\ref{fig:obs_2} shows the distribution of Doppler parameters, $\bhi$, obtained from our synthetic spectra for S/N=10 and 30 spanning the redshift range \mbox{$[0,0.5]$}, together with the line width distributions obtained from data with comparable S/N values and redshifts by \citet[][green data points]{leh07a} and \citet[][orange data points]{dan08b}. The median values of our predicted distributions are \mbox{$\widetilde{\bhi} \approx 30.4 \kms$},  \mbox{$29.8 \kms$}, and \mbox{$29.4 \kms$} for \mbox{S/N = 10}, 30, and 50 (not shown), respectively. All of these agree well with the median value found by \citet[][]{hea02a}, \mbox{$\widetilde{\bhi} = 27 ~\kms$}, by \citet[][]{shu00b}, \mbox{$\widetilde{\bhi} = 28 ~\kms$}, and with the median value \mbox{$\widetilde{\bhi} = 31 \kms$} for the full \citet[][]{leh07a} sample. Note that all of these values are significantly larger than the median value \mbox{$\widetilde{\bhi} = 21 ~\kms$} measured by \citet[][]{dav01b}. Our simulation shows a lower fraction of broad (\mbox{$\bhi > 40 \kms$}) absorbers when compared to the \citet[][]{leh07a} $b$-value distribution, but our results compare well to the line width distribution from \citet[][]{dan08b}.

The predicted median $\bhi$-values indicate that a lower S/N value systematically shifts the line width distribution to slightly larger values. Yet, the number of components with \mbox{$\bhi \geq 40 \kms$} relative to the total number of components identified in each case decreases from $\sim26$ to $\sim23$ per cent when the adopted S/N value decreases from 50 to 10. Here, two competing mechanisms are at work: On the one hand, a low S/N value results in a stronger blending of narrow components into (artificial) broad features. On the other hand, since broader lines are shallower (at a given column density), and thus more difficult to detect at low S/N, the number of broad components detected decreases with decreasing S/N. Compared to a higher S/N value, the net effect of a low S/N value is to yield a smaller number (both relative and absolute) of broad absorption features (at a given resolution and sensitivity).

\subsection{The $\bhi - \NHI$ distribution} \label{sec:bnc}

\begin{figure}
\resizebox{\colwidth}{!}{\includegraphics{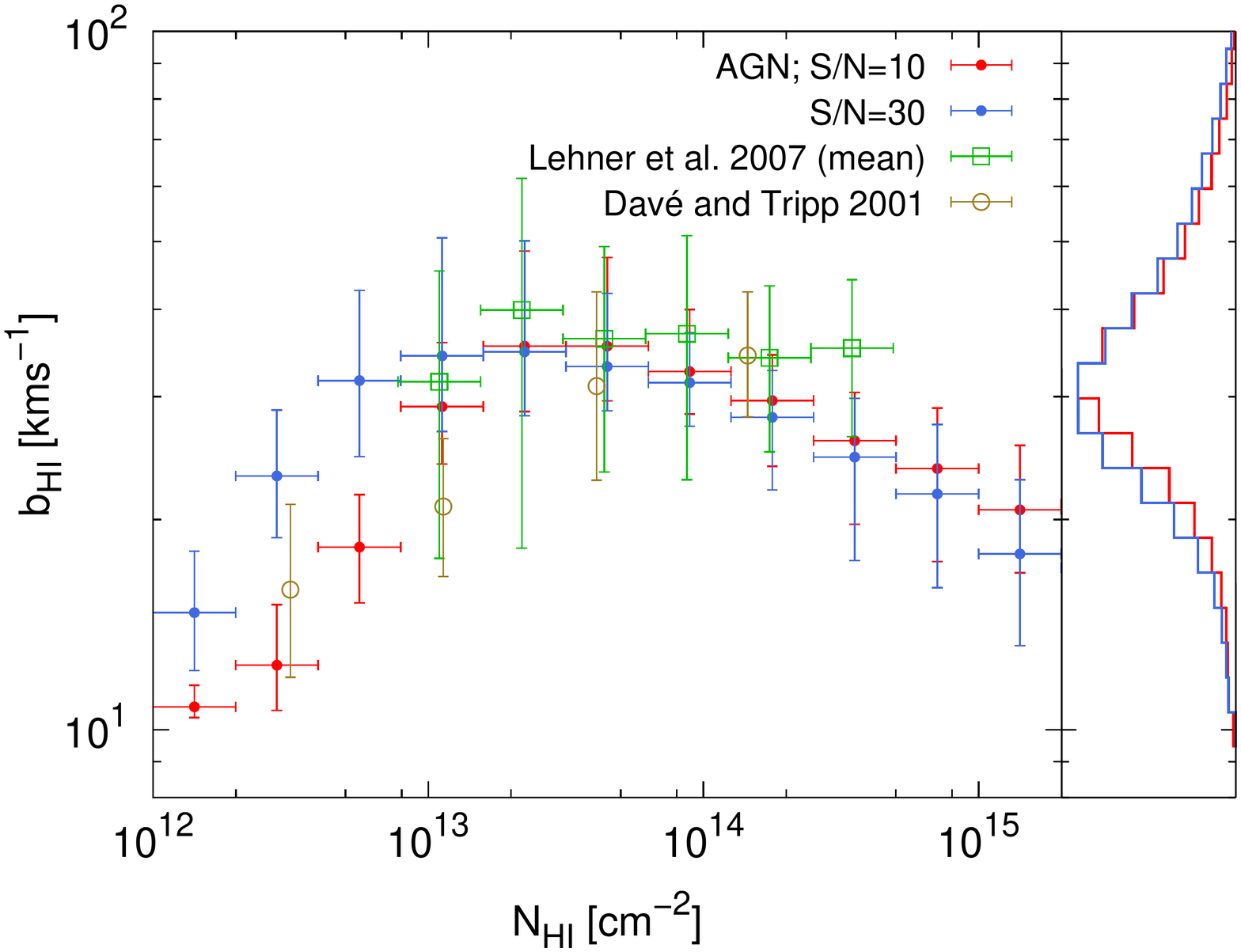}}
\caption[]{\mbox{$\bhi - \NHI$} distribution obtained from 5000 simulated spectra with S/N=10 (red), and S/N=30 (blue), spanning the redshift range \mbox{$[0,0.5]$}, and measurements by \citet[][green data points from their Table 3 for $\bhi > 0 \kms$]{leh07a} and \citet[][olive data points]{dav01b}. Points show the median $\bhi$-value in each bin of size \mbox{$\Delta \log (\NHI/\cm^{-2})  = 0.3 $}, while the error bars parallel the $y$-axis correspond to 25 and 75 percentiles, respectively. Note that the red (blue) histogram on the right sub-panel corresponds to the red (blue) histogram in Fig.~\ref{fig:obs_2}, but with a different binning and on a different scale.}
\label{fig:obs_3}
\end{figure}

Last, we compare the \mbox{$\bhi - \NHI$} distribution obtained from our simulated spectra with S/N=30 and S/N=10 to two different sets of observations used for the comparison of our predicted CDDF and the line width distribution discussed in the last sections. To this end, we bin the lines from observations and from our synthetic spectra in $\NHI$ using \mbox{$\Delta \log (\NHI/\cm^{-2})  = 0.3 $}, and compute the median $\bhi$-value, and 25-/75-percentiles in each bin. The result is shown in Fig.~\ref{fig:obs_3}. The \mbox{$\bhi -\NHI$} distribution from our simulated spectra matches the observations well within the uncertainties. Even the drop in $\bhi$ observed at low $\NHI$ in the \citet[][]{dav01b} is well reproduced by our simulation. Note that lines with \mbox{$\log (\NHI / \psc ) < 13.4 $} identified in spectra with S/N=30  generally have larger widths. This is a consequence of the fact that, at a fixed column density, lines with a given width are shallower with respect to narrower lines, and they can only be detected if the S/N is high enough. \\

Summarising, we conclude that the \HI\ observables predicted by our fiducial model are in excellent agreement with observations. This agreement may be surprising in view of the uncertainty in the input physics used in our simulation. However, in Appendix \ref{sec:cnvg_phys} we show that these results are quite robust against the model variations with respect to our fiducial model considered here (see Sec.~\ref{sec:mod_var}). We now proceed with the analysis of the physical conditions in low-$z$ \HI\ absorbers.

\subsection{Physical state of the \HI\ absorbing gas} \label{sec:phys}

In this section we present and discuss the physical properties of the gas detected via \HI\ absorption in our fiducial model (\mbox{\em AGN}; see Tab.~\ref{tbl:sims}). The method we use is similar to the method described in Paper I, in which we used optical-depth weighted quantities. Briefly, to compute the desired \HI\ optical-depth weighted quantity (e.g. density) associated with a given absorption line, we first compute the optical-depth weighted density in redshift space along the \los\ as in \citet[][]{sch99a}. Next, we compute the average of the optical-depth weighted density over the line profile, weighted again by the optical depth in each pixel and assign this last weighted average to the line. In concordance with Paper I, in the following we shall denote quantities weighted by \HI\ optical depth by adding a corresponding subscript; thus, for example, the \HI\ optical-depth weighted temperature is denoted by $\tempwhi$. We refer the reader to Sec.~5.1 of Paper I for a more detailed description about our method for computing optical depth-weighted quantities.

For simplicity, we obtain a new line sample of \HI\ absorbers identified in synthetic spectra with S/N=50 generated from 5000 \loss\ across a simulation box at a single redshift
\footnote{Note that our chosen redshift is slightly higher than the median redshift of most \HI\ absorption-line studies at low redshift \citep[e.g. \mbox{$\widetilde{z} \approx 0.17$} in ][]{leh07a}. Although {\em some} evolution does take place from $z=0.5 \to 0.0$, we do not expect the choice of this particular redshift to affect our conclusions in any significant way.
}, $z = 0.25$, spanning a total redshift path \mbox{$\Delta z = 187.5$},  which corresponds to an absorption path length \mbox{$\Delta \chi = 270$}. These spectra have been fitted following the method described in Appendix \ref{sec:fit}.

We restrict our analysis to ``simple'', i.e. single-component, absorbers, unless stated otherwise, We define an absorber $i$ as `simple' if the velocity distance from its centre to any other component $j$ along the same \los\ satisfies \mbox{$\Delta v > 2 \sigma_b$},  where \mbox{$\sigma_b^2 \equiv 0.5 [\bhi^2(i)+\bhi^2(j) ]$}. Absorption lines that do not satisfy this condition are referred to as `complex'.

Table \ref{tbl:div} contains various statistical and physical quantities resulting from the analysis of these new line sample, such as the relative number of identified components, the relative number of simple absorbers, the line-number density, $\left({\rm d} N / {\rm d} z\right)$, the total baryon content%
\footnote{The total baryon content in \HI\ is computed via
\beq \label{eq:omi}
	\Omega_{\ionsubscript{H}{I}} = \frac{\mH }{\rho_{\rm c}} ~\left( \frac{c}{H_0} \sum_{i=1}^{N_{\rm LOS}} \Delta \chi_{i} \right)^{-1} \sum_{i=1}^{N_{\rm LOS}} \sum_{j=1}^{N_{\rm ab s}} \left( \NHI \right)_{ij} \, . \notag
\eeq
} in \HI, $\Omega_{\ionsubscript{H}{I}}$, and the total baryon content in gas traced by \HI, $\Omega_{\rm b}(\HI)$ (see also Sec.~\ref{sec:bar_whim}). Note that the statistical and physical properties of this new sample are very similar to the corresponding properties of the line sample discussed in Secs.~\ref{sec:cddf} -- \ref{sec:bnc}.

\begin{table} 
\begin{center}
\caption{Line-number density, $\left({\rm d} N / {\rm d} z\right)$, total baryon content in \HI, $\Omega_{\ionsubscript{H}{I}}$, and total baryon content in gas traced by \HI, $\Omega_{\rm b}(\HI)$, related to simple \HI\ absorbers identified in 5000 spectra with S/N=50, 30, and 10 at \mbox{$z=0.25$}.
}  
\label{tbl:div}
\begin{tabular}{lccc}
\hline
  											& S/N=50	& S/N=30	& S/N=10	\\
\hline
components (rel. to S/N=50)						& 1				& 0.72			& 0.31			\\
simple absorbers (rel. to total)							& 0.47			& 0.53			& 0.65			\\
$\left({\rm d} N / {\rm d} z\right)$ $^{\, \rm a}$			& $460 \pm 22$	& $332 \pm 18$	& $144 \pm 12$	\\
$\Omega_{\ionsubscript{H}{I}}  ~[10^{-7}]$ $^{\, \rm b}$	& 1.20			& 1.15 			& 1.12	 		\\
$\Omega_{\rm b}(\HI) / \Omega_{\rm b}$				& 0.57			& 0.47			& 0.29			\\
\hline
\end{tabular}
\end{center}
\begin{list}{}{}
	\item[$^{\, \rm a}$] Quoted uncertainties are purely Poissonian. For comparison, \citet[][]{leh07a} obtain \mbox{$\left({\rm d} N / {\rm d} z\right) = 165 \pm 13$} at \mbox{${\rm S/N} \approx 10$}.
	\item[$^{\,\rm b}$] Total baryon content in \HI\ obtained by adding the column densities of all identified \HI\ components. The true  total baryon content in \HI\ along the fitted \loss\ at $z= 0.25$ is \mbox{$\Omega_{\ionsubscript{H}{I}}  = 2.11\times10^{-5}$}.
\end{list}
\end{table}

\subsubsection{Physical density and absorber strength} \label{sec:odvsN}

As previously noted by several studies \citep[e.g.][]{sch99a,dav99a}, there exists a tight correlation between \HI\ column-density, $\NHI$, and overdensity%
\footnote{The mean baryonic density in our model is \mbox{$\left< \rho_b \right>  = 4.18 \times10^{-31} ~\left( h / 0.73 \right)^{2} ~\left( 1+z \right)^3 ~\g~\pcc$}.
}, \mbox{$\Delta \equiv \rho_b / \left< \rho_b \right>$}, of the absorbing gas usually parametrized in the form of a power-law, \mbox{$\Delta / \Delta_0  = \left( \NHI / {\rm N}_0 \right)^{a}$}.
Due to variations in the (local) ionising radiation field, the influence of other heating mechanism (shocks), and other factors such as the geometry of the absorbing structures, etc., this relation has an intrinsic scatter, which decreases with increasing redshift \citep[][]{dav99a}.

The relation between overdensity and \HI\ column density for the diffuse IGM has been derived analytically by \citet[][]{sch01a}, who assuming local hydrostatic equilibrium\footnote{The assumption of `local hydrostatic equilibrium' implies that the size of a self-gravitating gas cloud is of the order of the local Jeans length.} and optically thin gas finds
\begin{equation} \label{eq:deltan}
	\log \Delta \propto \frac{2}{3 + \alpha ~(1-2b)} \left(\log \NHI - \frac{9}{2} \log (1+z) \right) \, .
\end{equation}
In the above equation, $\alpha$ is the slope of the temperature-density relation, \mbox{$T = T_0 ~\Delta^{\alpha}$}, which results from the balance between photo-heating and adiabatic cooling \citep[][]{hui97a}, and $-b$ is the power of the temperature in the expression for the \HI\ recombination rate coefficient which behaves as \mbox{$\propto T^{-b}$}. If re-ionisation of the IGM takes place at sufficiently high redshifts, its imprints on the thermal state of the IGM are eventually washed out, and the slope of temperature-density relation is expected to reach an asymptotic limit determined by the temperature dependence of the \HI\ recombination rate. More specifically, at low redshift \mbox{$\alpha \to 1/(1+b)$}. We find\footnote{We compute the recombination rate coefficient for recombination case A numerically, and fit a power law in the given  temperature range.} \mbox{$b = 0.755 \pm 0.001$} in the temperature range \mbox{$[10^3, \, 5\times10^4] \K$}, and hence \mbox{$\alpha \to 0.570$}. Inserting this value into eq.~\eqref{eq:deltan} gives \mbox{$\Delta \propto \NHI^{0.738} \cdot (1+z)^{-3.36} $}. Thus, the value of the amplitude $\Delta_0$ in the \mbox{$\Delta - \NHI$} relation decreases with redshift, implying that absorbers of a given column density trace gas at higher overdensities at lower redshift.

\begin{figure}
\resizebox{\colwidth}{!}{\includegraphics{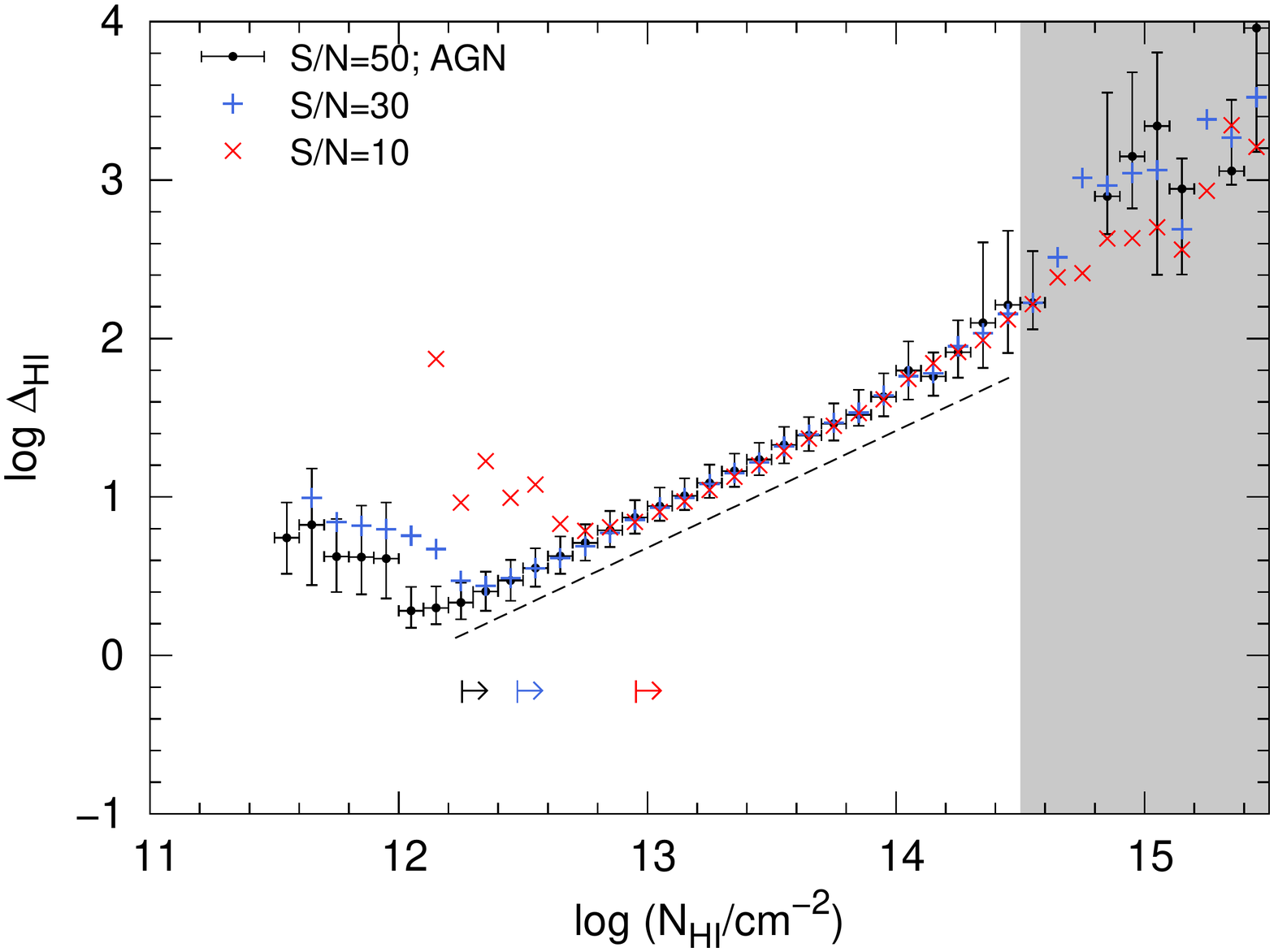}}
\caption[]{\mbox{$\deltawhi - \NHI$} relation for simple \HI\ absorbers identified in spectra with different S/N values at \mbox{$z=0.25$} for. Points show the median overdensity in each bin of size \mbox{$\Delta \log (\NHI/\cm^{-2})  = 0.1 $}. Error bars along the $y$-direction correspond to 25 and 75 percentiles in each bin. For clarity, only the error bars for the S/N=50 result are shown, but they are similar for S/N=30 and S/N=10.  The horizontal arrows indicate the formal completeness limit for each adopted S/N value as given by eq.~\eqref{eq:sl_N}. Note the deviation of the $\deltawhi-\NHI$ relation from a single power-law for column densities below the formal completeness limit (at a given S/N) and in the column density range for which the \HI\ \lya\ line generally saturates (shaded area). A power-law with the theoretical expected slope 0.738 \citep[but with arbitrary amplitude;][]{sch01a} has been included to guide the eye (black, dashed line).}
\label{fig:odvsN}
\end{figure}

\begin{figure}
\resizebox{\colwidth}{!}{\includegraphics{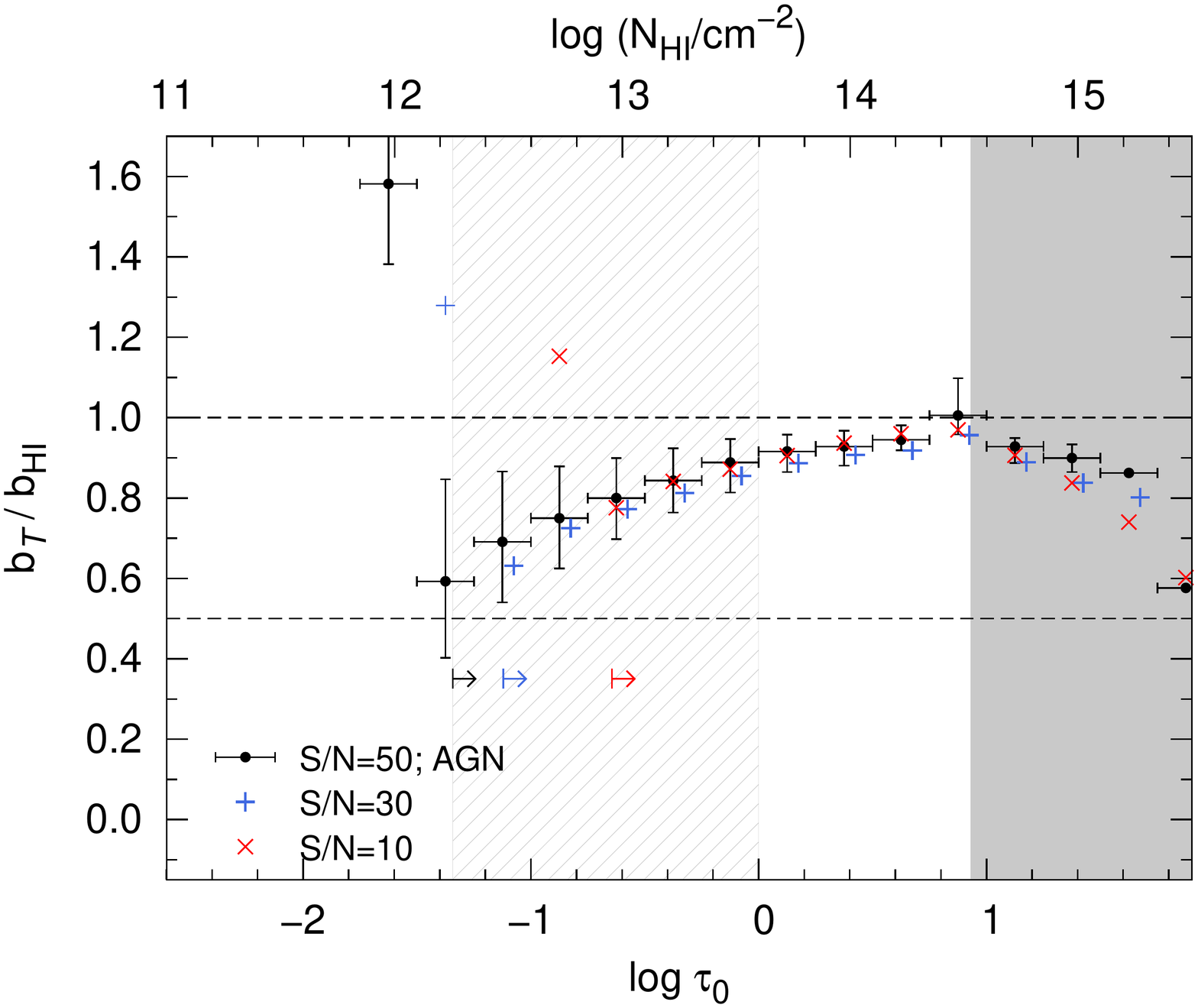}}
\caption[ ]{Ratio of \HI\ line thermal width, \mbox{$( b_T / \kms ) = 12.9 \sqrt{\tempwhi/ 10^4 \K}$}, to total line width, $\bhi$, as a function of the optical depth at line centre, $\tau_0$, inferred from Voigt profile fits, for simple \HI\ absorbers identified in spectra at \mbox{$z=0.25$} adopting different S/N values. The top axis indicates the corresponding \HI\ column density assuming \mbox{$\bhi = 30 \kms$}, which corresponds to the median value of the Doppler parameter distribution (see Sec.~\ref{sec:bvd}). The points show the median  \mbox{$\left( b_T / \bhi \right)$} value in each bin of size \mbox{$\Delta \log \tau_0 = 0.25$} indicated by the error bars parallel to the $x$-axis; the lower and upper error bars parallel to the $y$-axis correspond to the 25 and 75 percentiles in each bin, respectively. For clarity, only the error bars for the S/N=50 result are shown, but they are similar for S/N=30 and S/N=10. Note that the values corresponding to S/N=30 (blue crosses) have been slightly shifted for display purposes. The horizontal arrows indicate the formal completeness limit for each adopted S/N value as given by eq.~\eqref{eq:sl_N}. The dashed horizontal lines enclose the range \mbox{$0.5 \bhi \leq b_T \leq \bhi$}, and have been included to guide the eye. The gray, shaded area indicates the central optical depth (or column density) range for which the \HI\ \lya\ line generally saturates. The hatched area indicates the typical range in $\tau_0$ for BLAs, \mbox{$-1.34 < \log \tau_0 < 0$}, detected in spectra with S/N=50 (see Sec.~\ref{sec:sens}).}
\label{fig:bth_b}
\end{figure}

In Fig.~\ref{fig:odvsN} we show the \mbox{$\deltawhi - \NHI$} relation resulting from our fiducial model for all simple \HI\ absorbers identified in spectra with different S/N values at \mbox{$z=0.25$}. The turn-over in overdensity at column densities below the sensitivity limit (eq.~\ref{eq:sl_N}) for each adopted S/N is caused by errors in the measured $\NHI$ below this limit. Note also the deviation of the \mbox{$\deltawhi - \NHI$} relation from a single power-law at column densities \mbox{$\log (\NHI/\cm^{-2}) \gtrsim 14.5$} (indicated by the shaded area), which corresponds approximately to the column density for which the \lya\ line saturates. This is in part due to the inability of our algorithm to properly fit saturated lines.

Performing a least-square, error-weighted fit to the S/N=50 result for column densities above the corresponding sensitivity limit (\mbox{$\log (\NHI/\cm^{-2}) = 12.3$}; see eq.~\ref{eq:sl_N}) and restricted to \mbox{$\log (\NHI/\cm^{-2}) \leq 14.5 $}, we find \mbox{$a = 0.786 \pm 0.010$} and \mbox{$\Delta_0 = 48.3$}, normalised to \mbox{${\rm N}_0 = 10^{14} \psc$}. The resulting slope for S/N=30 (S/N=10) is \mbox{$a = 0.798 \pm 0.010$} (\mbox{$a = 0.847 \pm 0.014$}) in the logarithmic column density range \mbox{$[12.5, \, 14.5] $} (\mbox{$[13.0, \, 14.5]$}), where the lower column density limit is given by eq.~\eqref{eq:sl_N}. A power-law with the theoretical expected slope 0.738 and arbitrary amplitude has been included in this figure for reference (black, dashed line).

For comparison, \citet[][their equation 3]{dav10a} find \mbox{$a = 0.741\pm0.003$} and \mbox{$\Delta_0 = 38.9$} at \mbox{$z=0.25$}, for absorbers arising in gas with temperatures \mbox{$\log \left(T / \K \right)  < 4.5$} in their simulation. If we restrict our sample to single-component absorbers with \mbox{$\log \left(\tempwhi / \K \right)  < 4.5$}, we find \mbox{$a = 0.786 \pm 0.014$} and \mbox{$ \Delta_0 = 45.1$}. We note that we do not rescale the amplitude of the UV background in our simulation, while \citet[][]{dav10a} adjust its amplitude by a factor 3/2 to bring their predicted evolution of the \HI\ optical depth into better agreement with observations. 

\subsubsection{Gas temperature and line width} \label{sec:bth_b}

A matter of interest is to which extent the measured \HI\ line width can be used to estimate the temperature of the \HI\ absorbing gas. We explore this by comparing the \HI\ thermal line width, \mbox{$b_T  = 12.9 \sqrt{\tempwhi/ 10^4 \K}$}, computed from the optical depth-weighted gas temperature, $\tempwhi$, to the total \HI\ line width, $\bhi$, as a function of the line strength as given by the optical depth at the line centre, $\tau_0$. The optical depth at the line centre is computed using the \HI\ column density and the \HI\ line width inferred from a Voigt profile fit to the line (see Appendix \ref{sec:fit}). We bin the ratio \mbox{$\left( b_T / \bhi \right)$} in $\tau_0$, and plot  in Fig.~\ref{fig:bth_b} the median value and the 25-/75-percentiles in each bin as a function of $\tau_0$, for all single component absorbers identified in our synthetic spectra with different S/N values at \mbox{$z=0.25$}. 

We see that thermal broadening becomes increasingly important with increasing line strength (or \HI\ column density), and it contributes with at least 50 per cent to the total line width, i.e. \mbox{$b_T \gtrsim 0.5 ~\bhi$}, irrespective of the line strength and the adopted S/N value. The temperature of gas giving rise to absorption lines with central optical depths in the range \mbox{$1 < \tau_0 < 8.5$} (corresponding to strong lines) on average contributes with at least 90 per cent to the total line width, i.e. \mbox{$b_T \gtrsim 0.9 ~\bhi$}. The lower \mbox{$\left( b_T / \bhi \right)$} value shown by highly saturated lines, i.e. lines with \mbox{$\tau_0 > 8.5$} (gray, shaded area) is most probably due to the uncertainty in the fit parameters of such lines. These results suggest, in view of the tight \mbox{$\deltawhi - \NHI$} relation discussed above, that absorption arising in low density gas is subject to more significant non-thermal (i.e. Hubble) broadening than gas at higher density. This is consistent with the idea that low density, unbound gaseous structures are subject to the universal expansion, while gas at higher densities residing closer to galaxies may have detached from the overall expansion. We find (not shown) that the line width correlates well with the gas temperature for \mbox{$\log \left(\NHI / \psc \right) \gtrsim 13$}, but that it is a poor indicator of the thermal state of the gas for lower column densities.

Fig.~\ref{fig:bth_b} shows also that lines with central optical depths corresponding to \HI\ column densities below the formal completeness limit for each adopted S/N value (indicated by the arrows) as given by eq.~\eqref{eq:sl_N} have, on average, \mbox{$b_T > \bhi$}, which is un-physical. These lines correspond to absorption by gas at high temperatures, which gives rise to very shallow and extremely broad features that are (incorrectly) fitted with several components, thus yielding line widths that are narrower than allowed by the gas temperature.

In the central optical depth range typical for BLAs detected in spectra with S/N=50, \mbox{$-1.34 < \log \tau_0 < 0$} (see Sec.~\ref{sec:sens}) indicated by the hatched area, the contribution of thermal broadening to the line width amounts to 60 to 90 per cent. If non-thermal processes (e.g. turbulence) contribute to the line broadening in such a way that the total line width is given by \mbox{$\bhi^2 = b_T^2 + b_{\rm nt}^2$}, where $b_{\rm nt}$ is the non-thermal broadening (as would be the case for a purely Gaussian turbulence field), then the ratio of non-thermal broadening to total line width can be important, even though the thermal contribution is substantial. Take as an example the maximum, average thermal broadening to total line ratio for BLAs \mbox{$b_T / \bhi = 0.9$}; this value together with \mbox{$\bhi^2 = b_T^2 + b_{\rm nt}^2$} implies \mbox{$b_{\rm nt} / \bhi = 0.4$}.

\subsubsection{The $\left(\NHI,~\bhi \right)$-plane} \label{sec:b_vs_N}

\begin{figure*}
\resizebox{\textwidth}{!}{\includegraphics{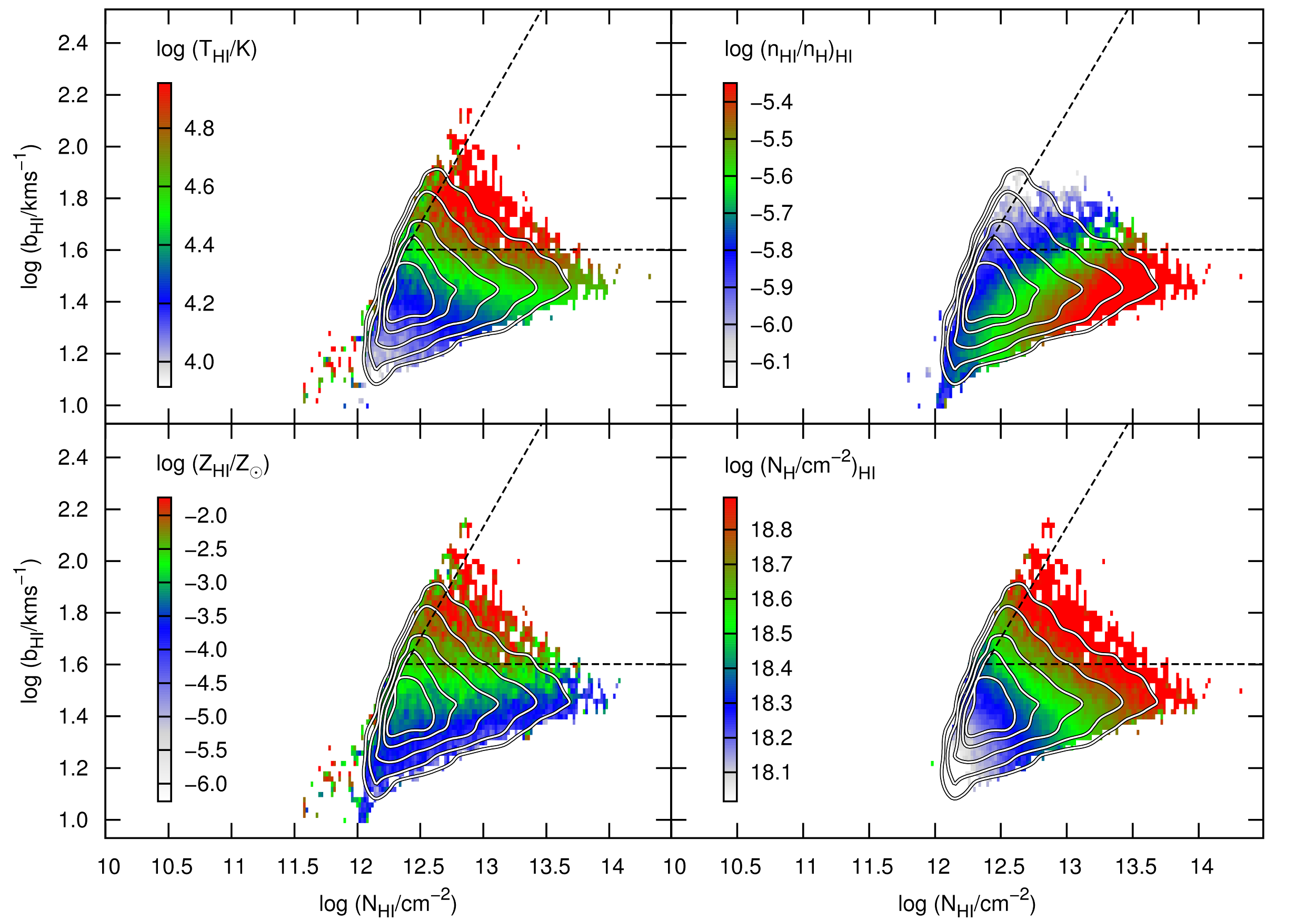}}
\caption[ ]{Temperature (top-left), neutral hydrogen fraction (top-right), metallicity (bottom-left), and total hydrogen column density $\rm N_{H}$ (bottom-right) of the gas at \mbox{$z=0.25$} traced by simple \HI\ absorbers identified in spectra with S/N=50, as a function of the line parameters $\NHI$ and $\bhi$. The colour coding shows the median value in each cell of the corresponding physical quantity. The white contours show the distribution (by number) of the sample of simple \HI\ absorbers, and they are identical in all panels. These contours enclose, starting from the innermost, 20, 40, 60, 80, and 90 per cent of the total number of single-component \HI\ absorbers. The horizontal and diagonal dashed lines indicate the usual BLA selection criteria \mbox{$\bhi \geq 40 \kms$} and \mbox{$\left(\NHI / \bhi \right) = 3\times10^{12} \psc {\rm km^{-1}s} \left({\rm S/N}\right)^{-1}$} \citep[e.g. ][]{ric06a}, respectively. Absorbers within this boundaries typically trace gas at \mbox{$\log (T / \K) \gtrsim 4.7$}, have total hydrogen column densities \mbox{$18 \lesssim \log ({\rm N_{H}}/\psc) \lesssim 18.9$}, and low ionisation fractions \mbox{$\log \fhiw \lesssim -5.5$}. Also, the (local) metallicity of the gas traced by these absorbers is typically \mbox{$\log (\metalwhi / Z_{\odot}) \gtrsim -2.5$}.}
\label{fig:bvsN}
\end{figure*}

A deeper insight into the physical state of gas giving rise to \HI\ absorption identified in real QSO spectra can be gained by studying the relation between selected physical quantities and the line observables, $\NHI$ and $\bhi$, simultaneously. We have followed such an approach in Paper I in order to study the physical conditions of \OVI\ absorbing gas, and now apply it to study gas traced via \HI\ absorption. We focus on four quantities: gas temperature $\tempwhi$, neutral hydrogen fraction $\fhiw$, total hydrogen column density $\NH$, and gas metallicity $Z$. Note that gas temperature and metallicity are `true' optical depth-weighted quantities, while total hydrogen column-density and ionisation fraction are `derived' quantities. For instance, the neutral fraction is computed using the optical depth-weighted hydrogen particle density, $\nhwhi$, and the optical depth-weighted gas temperature, using pre-computed tables obtained with the photoionisation package \textsc{cloudy} \citep[version 07.02 of the code last described by][]{fer98a}, as described in Sec.~\ref{sec:sims}.

For each of these physical quantities we proceed as follows: First we compute the desired physical quantity, e.g. $\tempwhi$, for each of the simple absorbers in our S/N=50 line sample. We then divide the $\left(\NHI,~\bhi \right)$-plane into cells, and compute the median value for the desired quantity using the values of all absorbers with $\left(\NHI,~\bhi \right)$ values in that cell. Fig.~\ref{fig:bvsN} displays the result for temperature (top-left),  neutral hydrogen fraction (top-right), metallicity (bottom-left), and total hydrogen column density (bottom-right). The colour code indicates the median value of the corresponding physical quantity. For reference, we include contours (white solid curves) showing the distribution by number of the simple \HI\ absorbers on the $\left(\NHI,~\bhi \right)$-plane. These contours enclose, starting from the innermost, 20, 40, 60, 80, and 90 per cent of the total number of \HI\ components. The dashed horizontal and diagonal lines at the top-right corner of each panel indicate, respectively, the BLA selection criteria \mbox{$\bhi \geq 40 \kms$} and \mbox{$\left(\NHI / \bhi \right) = 3\times10^{12} \psc {\rm km^{-1}s} \left({\rm S/N}\right)^{-1}$} \citep[][]{ric06a} adopting S/N=50.

There are several interesting features in this figure. First, all four physical quantities appear to have a relatively simple dependence on $\NHI$ and $\bhi$. The temperature of the gas (top-left panel), for example, shows a positive correlation with the line width, which appears to be tighter for absorbers at a given $\NHI$ (range), in agreement with the results presented in Sec.~\ref{sec:bth_b}. In this respect, note the population of narrow (\mbox{$\bhi \sim 10 \kms$}), low-column density (\mbox{$\NHI < 10^{12} \psc$}) absorbers with high (\mbox{$\log \left( \tempwhi / \K \right) > 4.8$}) optical depth-weighted temperatures. These correspond to the absorbers with column densities below the formal completeness limit and with \mbox{$b_T > \bhi$}, previously discussed.

The neutral hydrogen fraction (top-right panel) increases with $\NHI$, but strongly decreases with $\bhi$. This can be interpreted as a temperature-dependence, given the positive correlation between $\tempwhi$ and $\bhi$. Correspondingly, the total hydrogen column density (bottom-right panel) increases with both $\NHI$ and $\bhi$. The bottom-left panel shows that the (local) metallicity of the gas is strongly correlated with the line width. Given the correlation between gas temperature and line width shown the top-left panel, this suggests that there is a correlation between gas temperature and (local) gas metallicity. This correlation is very likely a consequence of strong feedback. Indeed, high-temperature, high-metallicity absorbers could be tracing shock-heated, enriched outflows in the surroundings of galaxies that have not had enough time to mix with the surrounding gas and to cool down, whereas low-temperature, low-metallicity absorbers could be tracing both gas that has not yet been impacted by outflows, and wind material that has been ejected at redshifts high enough for it to cool down and to dilute its metal content in the ambient gas.

The BLA selection regime defined by the dashed lines in each panel reveals a population of \HI\ absorbers tracing highly ionised (\mbox{$\log \fhiw \sim -6$}) gas with median temperatures \mbox{$\log \left( \tempwhi / K \right) \gtrsim 4.7$}, median (local) metallicities \mbox{$\log (\metalwhi / Z_{\odot}) \gtrsim -2.5$}, and total hydrogen column densities \mbox{$\log ({\rm N_{H}}/\psc) \approx 18.7$}, which is almost an order of magnitude higher than the total hydrogen column density of typical \lya\ forest absorbers (see also Fig.~\ref{fig:bla_various}). According to our previous interpretation of the \mbox{$\tempwhi - \metalwhi$} correlation, these results suggests that BLAs may be tracing galactic outflows. We will come back to this point in Sec.~\ref{sec:bla_phys}. \\

The results presented in this section indicate that the \HI\ column density of unsaturated absorbers is a reliable tracer of the underlying physical density of the gas giving rise to the detected \HI\ \lya\ absorption. Moreover, the temperature of the absorbing gas may be roughly estimated from the measured line width, as suggested by the average contribution of thermal broadening to the total line width of these absorbers. Finally, \HI\ absorbers subject to the commonly adopted BLA selection criteria trace gas which appears to be physically distinct from the gas traced by typical \lya\ forest absorbers.

\section{The warm-hot diffuse gas} \label{sec:whim}

\begin{figure*}
\resizebox{\textwidth}{!}{\includegraphics{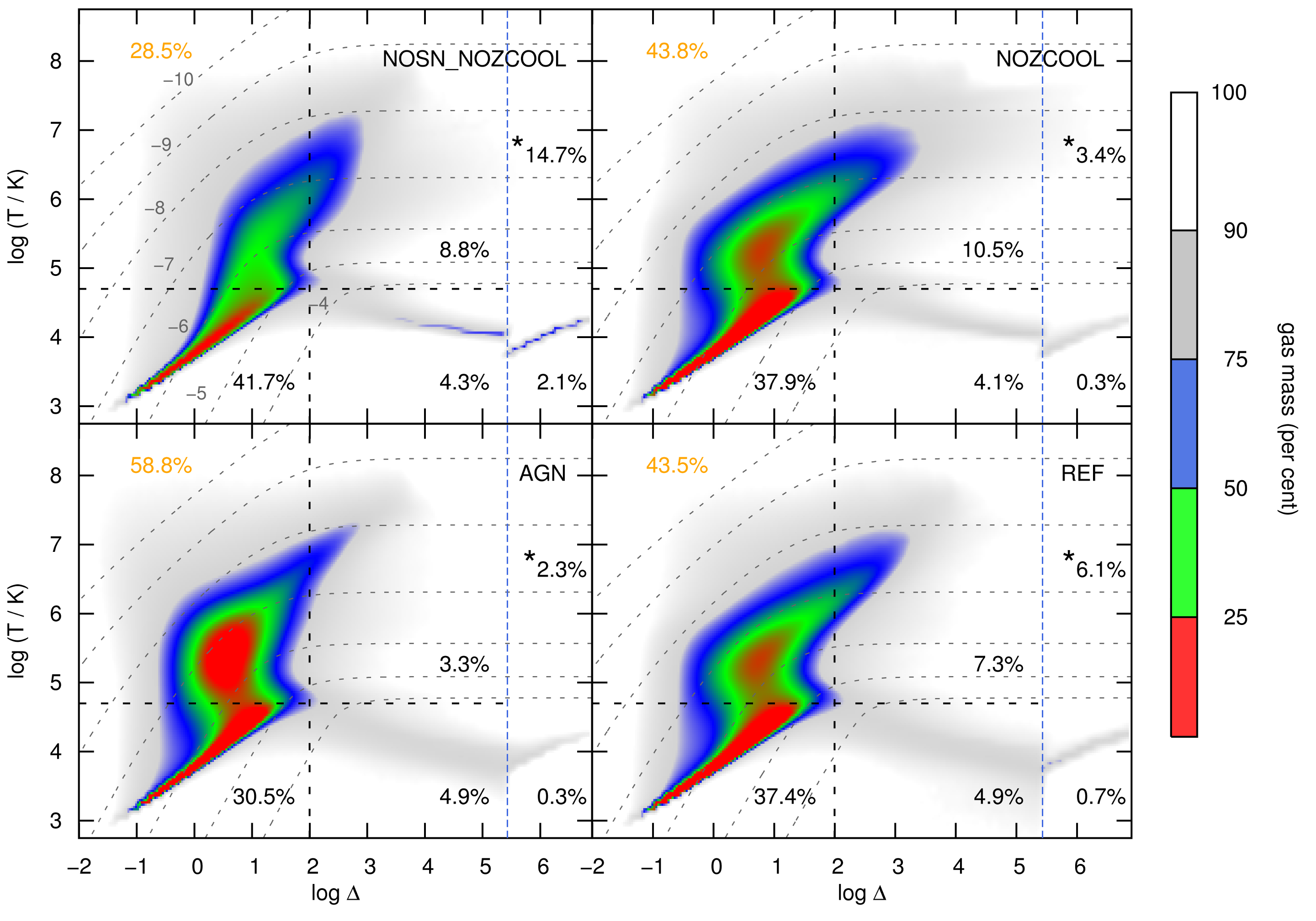}}
\caption[]{Distribution of gas mass over various phases predicted by different models, indicated on the top-right corner of each panel (see Tab.~\ref{tbl:sims}). The vertical (horizontal) dotted line indicates the density (temperature) threshold at $\Delta = 10^2$ ($T = 5\times10^4 \K$) that separates unbound (cool) from collapsed (warm-hot) gas. The region to the right of the blue dashed line shows the high-density, star-forming gas with physical densities \mbox{$\nHs \geq 0.1 \, \cm^{-3}$} (or \mbox{$\Delta \gtrsim 3\times10^{5}$} at \mbox{$z = 0.25$}). The coloured areas show the cumulative gas mass (in per cent) indicated by the colour bar on the right. The percentages in each panel show the baryon mass fraction in the corresponding phase defined by the temperature/density thresholds. We have highlighted the baryon mass fraction in the diffuse, warm-hot gas in orange. The starred percentage indicates in each case the baryonic mass fraction in stars. The dotted contours, which are identical in all panels, give the neutral hydrogen fraction $\fhi$, of gas in ionisation equilibrium as a function of $\Delta$ and $T$; the logarithmic $\fhi$-value is indicated next to the corresponding contour in the top-left panel. The left panels show the effect of our two most extreme scenarios, i.e. including both feedback by SNII and AGN with respect to neglecting feedback altogether; the right panels show the effect of neglecting radiative cooling by heavy elements. Clearly, feedback by SNII (and AGN) heats a significant amount of gas above temperatures \mbox{$T = 5\times10^4 \K$}, with the WHIM fraction increasing from 28.5 per cent (top-left) to 58.8 per cent (bottom-left). Interestingly, the IGM fraction only changes by \mbox{$\sim10$} per cent, indicating that feedback shifts a large fraction of the ISM from haloes into intergalactic space.}
\label{fig:gasmass_sims}
\end{figure*}

In the next sections we explore in detail the effect of feedback and metal-line cooling on the physical state (i.e. density and temperature) of the gas in our simulations. Also, we investigate the \HI\ absorption signatures of warm-hot diffuse gas, and the physical properties of the gas traced by broad \HI\ absorption features (BLAs) identified in synthetic QSO absorption spectra. For this purpose, we use the \HI\ sample from our fiducial model  \mbox{\em AGN} presented in Sec.~\ref{sec:phys}, and generate similar samples for all the other model runs. Comparison between model predictions and observations (whenever possible) are done exclusively for our fiducial run.

\begin{table} 
\begin{center}
\caption{Definition of the various gas phases we consider in terms of temperature and density thresholds. For example, we define the {\em warm-hot}, {\em diffuse} gas (WHIM) to have temperatures \mbox{$T \geq 5\times10^4 \K$} and overdensities  \mbox{$\Delta \leq 10^2$}.
}  
\label{tbl:phases}
\begin{tabular}{lcc}
\hline
 			& Overdensity ($\Delta$)	& Temperature ($T [\K]$)	\\
\hline
cool			& --					& $<5\times10^4$		\\ 
warm-hot		& --					& $\geq 5\times10^4$	\\ 
diffuse		& $\leq 10^2$			& --					\\ 
condensed	&  $> 10^2$			& --					\\ 
star-forming	&  $\gtrsim 3\times10^5$	& (EoS)$^{\, \rm a}$		\\ 
\hline
\end{tabular}
\end{center}
\begin{list}{}{}
\item[$^{\,\rm a}$] We consider `star-forming' the gas with physical densities that exceed our adopted star-formation threshold \mbox{$\nHs = 0.1 \, \cm^{-3}$} and which is allowed to form stars. The temperature of this gas phase is set by an imposed equation of state (EoS) of the form \mbox{$P \propto \rho^{4/3}$}. This gas phase can be thought of as the inter-stellar medium  (ISM). Note that, although this gas is cold, it is not included in the gas phase defined as `cool'.
\end{list}
\end{table}

\subsection{Model-dependence of the predicted warm-hot gas mass} \label{sec:mod_var}

On super-galactic scales, two mechanisms are able to shock-heat intergalactic gas to temperatures \mbox{$T \gtrsim 5\times10^4 \K$}: a) galactic outflows driven by SNII explosions and by AGN activity;  b) accretion shocks caused by infall onto the potential wells of dark matter halos. We have selected four model runs from the OWLS project, \mbox{\em NOSN\_NOZCOOL}, \mbox{\em NOZCOOL}, \mbox{\em REF}, and \mbox{\em AGN}, to investigate the effect of each of these mechanisms on the properties of diffuse gas and its imprints on simulated absorption spectra. Note that these models have already been described in Sec.~\ref{sec:sims} (see also Tab.~\ref{tbl:sims}).

We are interested in the predicted distribution of gas mass among the various (gas) phases --in particular the warm-hot diffuse phase-- defined in Tab.~\ref{tbl:phases}. We adopt a temperature threshold at \mbox{$\log \left( T / \K \right) = 4.7 $} (or \mbox{$T  \approx 5\times10^4 \K$}) and a density threshold at \mbox{$\Delta = 10^2$} to distinguish these gas phases. The density threshold has been chosen so as to roughly separate unbound gas from collapsed structures (at \mbox{$z = 0.25$}). The temperature threshold is motivated by the bi-modality in the gas mass distribution predicted by the models considered here (see below). Note that our value is somewhat below the `canonical' but to some extent arbitrary  \mbox{$T = 10^5 \K$} commonly adopted to distinguish cool from warm-hot intergalactic gas \citep[but see][]{wie10a}. 

The distribution of gas mass as a function of temperature and (over-)density predicted by the different models is presented in Fig.~\ref{fig:gasmass_sims}. The coloured areas show the cumulative gas mass (in per cent) indicated by the colour bar on the right. The vertical (horizontal) solid line in each panel indicates the density (temperature) threshold separating the various phases. Star-forming gas, which is defined as gas with physical densities $\nHs \geq 0.1 \, \cm^{-3}$ (or%
\footnote{The relation between hydrogen particle density $\nH$ and (baryonic) overdensity $\Delta$ is given by eq.~\eqref{eq:nHtoD} in Appendix \ref{sec:tau0_vs}.
} $\Delta \approx 3\times10^{5}$ at \mbox{$z = 0.25$}), is shown to the right of the blue, vertical dashed line in each panel. The percentage included in each separate region indicates the baryonic mass in the corresponding (gas) phase relative to the total baryonic mass. In particular, the number (orange) in the top-left corner of each panel gives the mass fraction of gas with \mbox{$T \geq5\times10^4 \K$} and \mbox{$\Delta \leq 10^2$}, i.e. warm-hot diffuse gas. The starred percentage indicates in each case the baryonic mass confined in stars. Note that our adopted temperature threshold seems appropriate to separate cool, photo-ionised from shock-heated gas; the gas mass distribution at \mbox{$\Delta \leq 10^2$} is clearly bimodal, with two phases having significant gas mass fractions above and below \mbox{$\log \left( T / \K \right) = 4.7$}. The dotted contours indicate the neutral hydrogen fraction, $\fhi$, as function of density and temperature at \mbox{$z = 0.25$}; it has been computed using pre-computed \textsc{cloudy} tables as described in Sec.~\ref{sec:sims}. The logarithmic $\fhi$-value is indicated next to the corresponding contour in the top-left panel. 

The sequence of models given by moving clock-wise from the top-left panel is essentially a sequence of increasing feedback strength (and model complexity). The mass fraction in warm-hot diffuse gas in the model \mbox{\em NOSN\_NOZCOOL} (top-left panel) indicates that by \mbox{$z=0.25$} roughly 30 per cent of the gas mass is shock-heated to temperatures \mbox{$\log \left( T / \K \right) > 4.7 $} by gravity alone. In the absence of any feedback on galactic scales, a large fraction of the gas that is accreted via gravitational infall at higher redshifts is able to cool and fuel star formation, with nearly 15 per cent of the gas mass ending up in stars by \mbox{$z=0.25$}. The cool, photo-ionised diffuse gas at  \mbox{$\log \left( T / \K \right) \lesssim 4.7 $} and \mbox{$\Delta \leq 10^2$} in this model contains roughly 40 per cent of the total gas mass.

Moving on to the top-right panel (model \mbox{\em NOZCOOL}), we see that nearly 45 per cent of the baryonic mass in the simulation is in the form of warm-hot diffuse gas as a consequence of the kinetic energy released by supernova explosions. This corresponds to an absolute increase in mass of 15 per cent in this gas phase compared to \mbox{\em NOSN\_NOZCOOL}. At the same time, the mass fraction in the cool diffuse IGM predicted by \mbox{\em NOZCOOL} decreases with respect to \mbox{\em NOSN\_NOZCOOL}, but by a far smaller amount ($\sim4$ per cent in absolute terms). Thus, about 10 per cent of the gas mass that ends up in the WHIM by \mbox{$z=0.25$} must be removed from a gas phase other than the cool diffuse IGM. The significantly lower mass in stars in the \mbox{\em NOSN\_NOZCOOL} model compared to the \mbox{\em NOZCOOL} model strongly suggests that at higher redshifts supernova feedback shock-heats and blows a significant fraction of the ISM out of halos, which ends up in the WHIM by \mbox{$z=0.25$}.

Including radiative cooling by heavy elements (model \mbox{\em REF}; bottom-right panel) has a negligible effect on the WHIM and the cool diffuse IGM, suggesting that metal-line cooling in the these gas phases is inefficient, either because their density is low, or the metals they contain are not yet well mixed, or perhaps because the level of enrichment is low, or a combination of them all. Interestingly, the \mbox{\em REF} model predicts a much higher mass fraction in stars and star-forming gas with respect to  \mbox{\em NOZCOOL}, in spite of including SN feedback. The corresponding decrease in mass in warm-hot gas at high densities (which can be considered as the intra-group and intra-cluster medium; ICM), suggests that some of the gas in this phase is accreted and fuels star formation. However, the exact evolutionary path of the gas in $T -\Delta$ phase space might be more complex than this.

Perhaps the most remarkable result is the fact that feedback from AGN has a very strong impact on the thermal state of the diffuse gas. Indeed, comparison of the bottom panels shows that an additional \mbox{$\sim15$} per cent of the total gas mass in the simulation is shock-heated to temperatures above \mbox{$\log \left( T / \K \right) = 4.7 $} and pushed into regions of low density ($\Delta < 10^2$), such that by \mbox{$z=0.25$} nearly 60 per cent of the gas mass ends up in the WHIM. A comparison of the mass distributions among the different phases predicted by \mbox{\em REF} and \mbox{\em AGN} suggests that half of the additional WHIM mass, i.e. \mbox{$\sim7$} per cent, at \mbox{$z=0.25$} is removed at higher redshifts mainly from the ISM (thus reducing the mass in stars at \mbox{$z=0.25$} by a factor \mbox{$\sim3$}), and from the ICM. The remaining \mbox{$\sim7$} per cent of the WHIM mass apparently comes from the IGM. Comparison of the gas mass fractions in the warm-hot diffuse phase between the models \mbox{\em NOZCOOL} and \mbox{\em AGN} suggests that SN and AGN contribute roughly a similar amount to the baryon content of the WHIM. Equally important, the gas mass in this gas phase predicted by the models \mbox{\em NOSN\_NOZCOOL} ($\sim 30$ per cent) and \mbox{\em AGN} ($\sim 60$ per cent) indicates that (strong) feedback (both by SN and AGN) may be able to shock-heat an amount of gas comparable to the gas shock-heated via gravitational infall. This results thus indicate that it is crucial to understand feedback processes on super-galactic scales before being able to make any reliable predictions about the baryon content of warm-hot gas in the Universe.

Consider finally the hydrogen neutral fraction $\fhi$ indicated by the dotted contours. The logarithmic value of $\fhi$ is indicated next to the corresponding contour in the top-left panel, and they are identical in all the other panels. At a fixed temperature, the neutral hydrogen fraction increases with density, since the ionisation state of the gas is dominated by photo-ionisation. However, at sufficiently high densities, i.e. \mbox{$\Delta \gtrsim 10^2$} at \mbox{$z=0.25$}, collisional ionisation dominates and the neutral hydrogen fraction depends only on the gas temperature, resulting in contours running parallel to the $\Delta$-axis. In either case, the neutral hydrogen fraction steeply decreases with temperature at all densities \citep[see also][]{ric08b,dan10a}. As a consequence, the gas at densities and temperatures characteristic of the warm-hot diffuse gas is expected to be highly ionised. In particular, the model {\em AGN} predicts that the vast majority of the gas in the WHIM contains a neutral hydrogen fraction \mbox{$\fhi \lesssim 10^{-6}$}. This has important implications for the detectability of this gas phase via \HI\ absorption which will be discussed in detail in Sec.~\ref{sec:sens}.

\begin{figure}
{\resizebox{\colwidth}{!}{\includegraphics{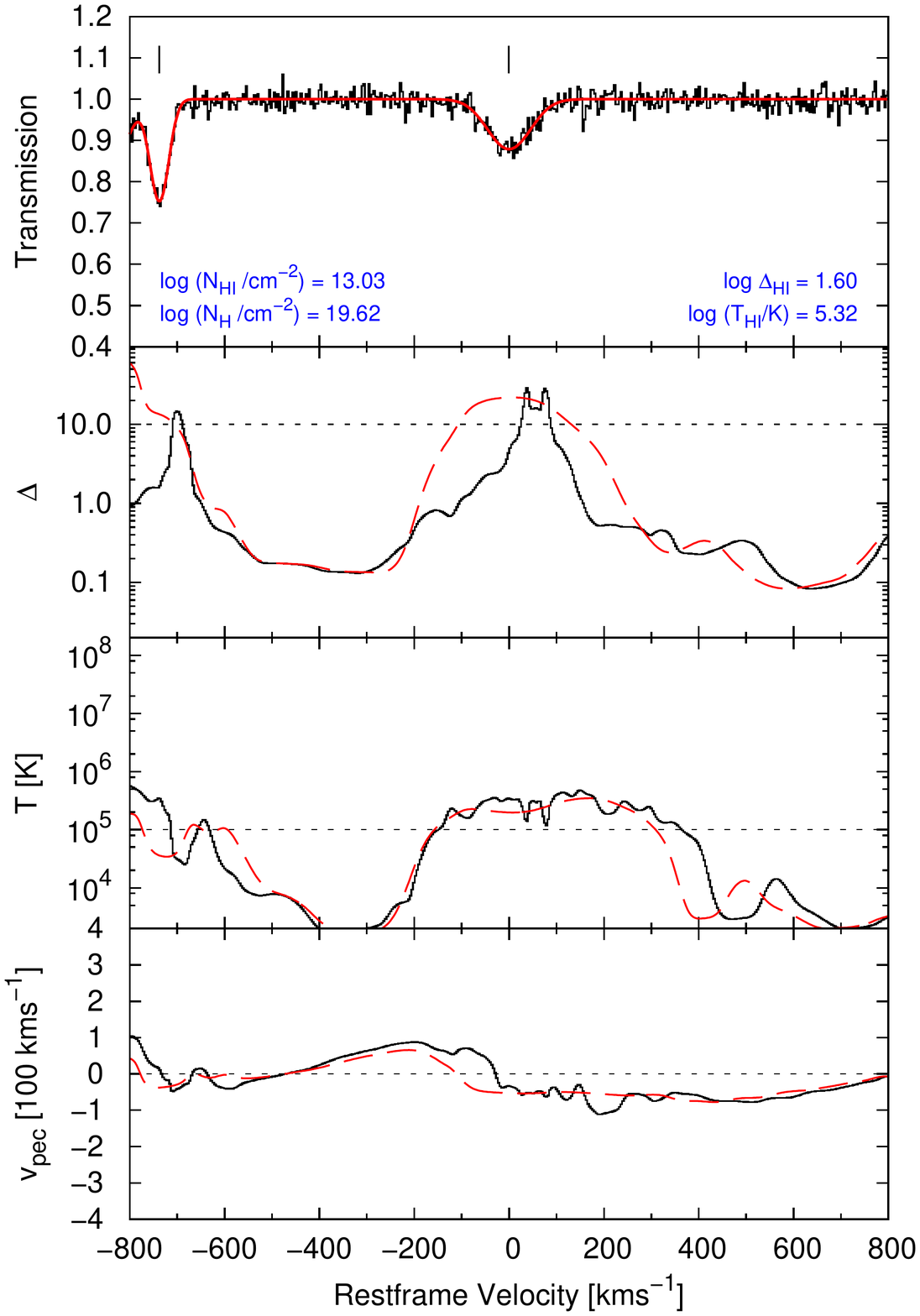}}}
\caption[]{Example of a \los\ through a simulation box at \mbox{$z=0.25$} from a run adopting our fiducial model (\mbox{\em AGN}). The top panel shows a synthetic spectrum generated by \textsc{specwizard} using S/N=50 (black) and the corresponding fit (red) centred on a simple BLA with \mbox{$\bhi \approx 63 \kms$}, and showing an additional narrow \HI\ \lya\ absorption line (vertical marks flag the line centres). The subsequent panels show, from top to bottom, the gas overdensity, the gas temperature, and the peculiar velocity along the \los. Black, solid lines indicate smoothed quantities; red, dashed lines \HI\ optical-depth weighted quantities. The dotted horizontal line in the second panel (from the top) indicates the detection limit at S/N=50 (see eq.~\ref{eq:sl_N}) expressed in terms of $\Delta$ using the \mbox{$\NHI(\Delta)$} relation discussed in Sec.~\ref{sec:odvsN}. The values in the lower-left and lower-right corner of the top panel give, respectively, the \HI\ column density and the \HI\ optical depth-weighted total hydrogen column density, and the overdensity and the temperature of the gas traced by the BLA. Note that \mbox{$\log \left( T / \K\right) = 5.32 $} corresponds to a thermal width \mbox{$b_T = 59.0 \kms$}, implying that the width of this BLA is dominated by thermal broadening.}
\label{fig:spec_los}
\end{figure}

\subsection{Observability of (warm-hot) gas using (broad) \HI\ absorption} \label{sec:detect}

In this section we explore to what extent the actual gas mass (distribution) in the cool and warm-hot diffuse phases are traced by the \HI\ detected in absorption. In particular, we investigate the thermal state of the gas traced by absorbers selected in terms of their line width, albeit only on a statistical basis. At the same time, we invert the approach and inquire about the spectral signatures (and physical properties) of \HI\ absorbers arising in warm-hot gas. Even though we are interested primarily in broad absorbers, we include narrow absorbers in our analysis as well. This allows for a more robust interpretation of our results. We define the following classes of \HI\ absorbers in terms of their spectral and/or physical properties:
\ben
	\item \mbox{\em NLA:} \HI\ absorber components with Doppler parameters \mbox{$\bhi < 40 ~\kms$}. We adopt the notation introduced by \citet[][]{leh07a}.
	\item {\em BLA:} \HI\ absorber components with Doppler parameters \mbox{$\bhi \geq 40 ~\kms$} that satisfy the sensitivity limit introduced by \citet[][]{ric06a} and adopted in other studies \citep[][]{dan10a,wil10a}
\beq \notag
	\left( \frac{ \NHI / \psc }{ \bhi / \kms } \right) \geq 3\times10^{12} \left({\rm S/N}\right)^{-1} \, .
\eeq
	This limit is equivalent to a \HI\ \lya\ central optical depth
\beq \label{eq:sl_tau0}
	\tau_0 \geq 2.27 \left({\rm S/N}\right)^{-1} \, .
\eeq
	We feel that using a detection limit in terms of $\tau_0$ is more intuitive than the limit in terms of $\left( \NHI / \bhi \right)$, in particular for small values of $\tau_0$, since in this limit \mbox{$\tau(v) \approx 1 - F(v)$}. Henceforth, we will use eq.~\eqref{eq:sl_tau0} instead of the limit in terms of \mbox{$\left( \NHI / \bhi \right)$} as our second BLA selection criterion; also, all corresponding results will be expressed in terms of $\tau_0$ rather than \mbox{$\left( \NHI / \bhi \right)$}.
	\item {\em hot-BLA:} BLAs with optical-depth weighted temperatures $\tempwhi \geq 5\times10^4 \K$. This class is defined in order to isolate BLAs genuinely tracing warm-hot gas. Note that our adopted temperature threshold is lower than the actual temperature implied by Doppler parameters \mbox{$\bhi \geq 40 \kms$} assuming pure thermal broadening%
\footnote{The thermal width of an \HI\ line arising in gas at \mbox{$T \gtrsim 5\times10^4 \K$} is \mbox{$\bhi \gtrsim 30 \kms$}.
}. This is, however, not an issue since as we have shown ins Sec.~\ref{sec:bth_b}, the line width of BLAs is never entirely set by thermal broadening.
\een

As in previous sections, all the results presented here refer to {\em simple} absorbers as defined in Sec.~\ref{sec:phys}, unless stated otherwise. Tab.~\ref{tbl:abs_single} gives an overview of the statistics of simple absorbers in each of the classes defined above, for all four models considered here. An example of such a simple (broad) \HI\ absorber is shown in Fig.~\ref{fig:spec_los}. The top panel shows the spectrum (black) and corresponding fit (red) centred at a \HI\ \lya\ line with \mbox{$\bhi =63 \kms$} and \mbox{$\NHI \sim 10^{13} \psc$}. The next panel shows the smoothed overdensity (black) and the \HI\ optical depth-weighted overdensity (red dashed) along the \los. Note the relative simple density structure of the absorbing gas. The smoothed (black) and \HI\ optical depth-weighted (red dashed) temperatures are shown in the third panel (from the top). Clearly, the gas giving rise to the BLA shown in the top panel has a temperature \mbox{$T \gtrsim 10^5 \K$}. As a consequence of the high temperature, the absorbing gas is highly ionised, with a total \HI\ column density \mbox{$\NH \approx 4\times10^{19} \psc$}, and it represents thus a significant baryon reservoir. The slight off-set between the BLA and the density peak of the absorbing gas is due to its (small) peculiar velocity along the \los\ (bottom panel).

\begin{table} 
\begin{center}
\caption{Number fraction (in percent) of simple absorbers relative to the total number of \HI\ components in each class identified in synthetic spectra at \mbox{$z=0.25$} with S/N=50 for different models.}  
\label{tbl:abs_single}
\begin{tabular}{lcccc}
\hline
 				& \mbox{\em NOSN\_NOZCOOL}	& \mbox{\em NOZCOOL}	& \mbox{\em REF }		& \mbox{\em AGN}	\\
\hline
all$^{\,\rm a}$		& 43							& 44					& 44					& 47				\\
NLA$^{\,\rm b}$	& 37							& 37					& 37					& 38				\\
BLA 				& 4.5							& 5.5					& 5.6					& 7.1				\\
hot-BLA 			& 1.0							& 1.7					& 1.7					& 3.1				\\
\hline
\end{tabular}
\end{center}
\begin{list}{}{}
\item[$^{\,\rm a}$] Note that about half of the identified absorbers are single-component, irrespective of the model.
\item[$^{\,\rm b}$] Note that the fraction of NLA is similar for all models, suggesting that the gas traced by these absorbers is not significantly affected by feedback.
\end{list}
\end{table}

\subsubsection{Spectral sensitivity} \label{sec:sens}

\begin{figure*}
\resizebox{\textwidth}{!}{\includegraphics{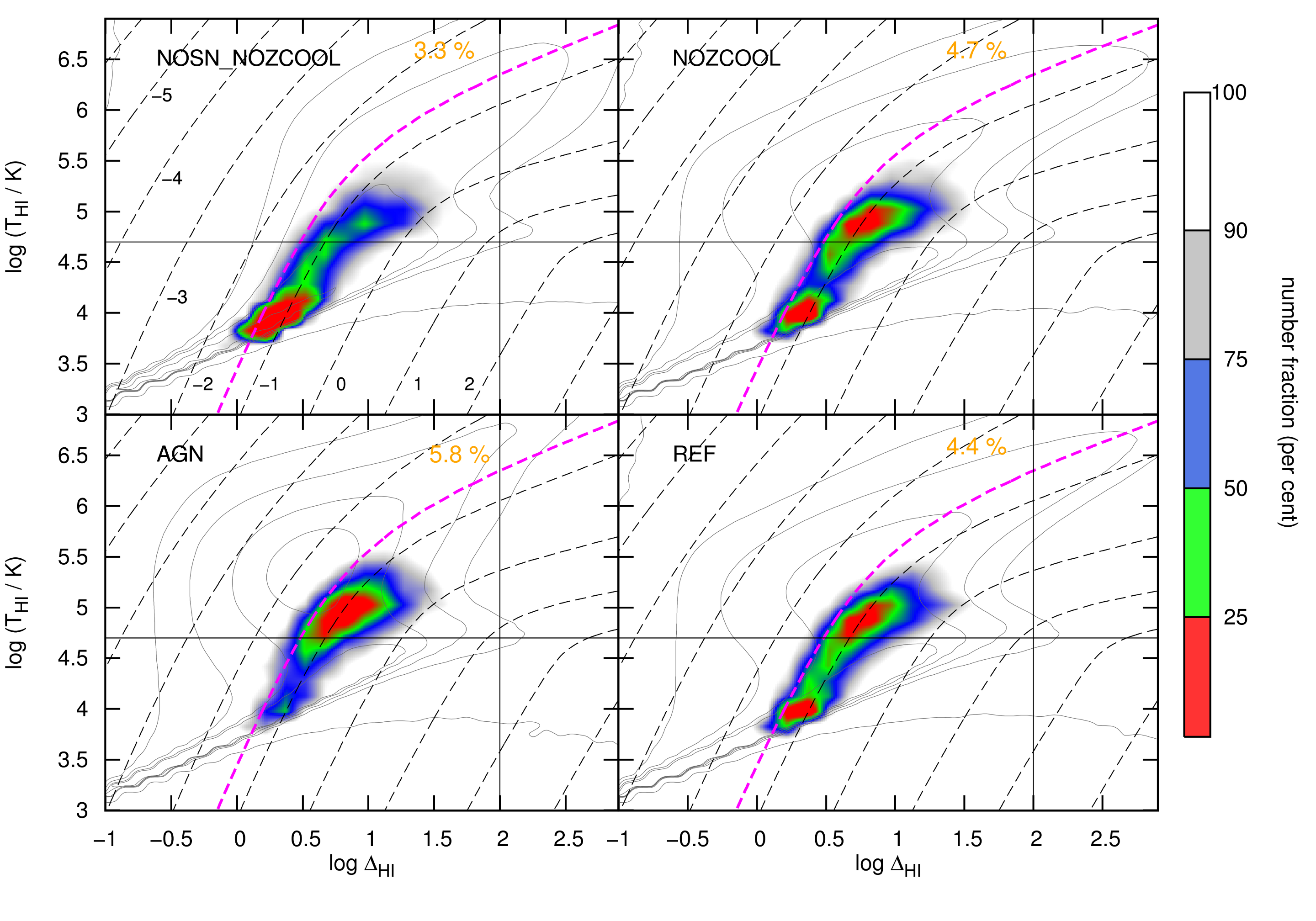}}
\caption[]{Distribution of temperature and overdensities of the gas traced by simple BLAs (coloured regions) identified in spectra with S/N=50 at \mbox{$z=0.25$} obtained from various models. In each panel, the coloured regions show the cumulative number fraction (in per cent) of BLAs. For instance, the red region contains 25 per cent of the total number of BLAs, while the green and red regions together contain half of them. The gray contours correspond to the gas mass distributions shown in Fig.~\ref{fig:gasmass_sims}. Note the change in scale on both the $x$- and $y$-axes with respect to  Fig.~\ref{fig:gasmass_sims}. The vertical (horizontal) solid line in each panel indicates the density (temperature) threshold adopted to distinguish the different gas phases (see Tab.~\ref{tbl:phases}). The dashed contours, which are identical in all panels, indicate the central optical depth, $\tau_0$, as a function of (over-)density and temperature (see Appendix \ref{sec:tau0_vs}, eq.~\ref{eq:tau0_jeans}). For clarity, the corresponding logarithmic value of $\tau_0$ is included next the each curve only in the top-left panel. The magenta dashed contour indicates our adopted sensitivity limit for S/N=50, \mbox{$\log \tau_0 = -1.34$} (eq.~\ref{eq:sl_tau0}). The percentage (orange) in each panel indicates the fraction of baryonic mass of the gas traced by simple BLAs with \mbox{$\tempwhi \geq 5\times10^4 \K$}, i.e. simple hot-BLAs. Reassuringly, the overwhelmingly majority of the identified BLAs lie below the thick magenta contour, irrespective of the model.  However, the large fraction of mass in gas at \mbox{$T \gtrsim 10^5 \K$} and \mbox{$\Delta \lesssim 10 $} (in particular for \mbox{\em AGN}) is not observable at our adopted sensitivity. The dashed contours indicate that at least \mbox{$\log \tau_0 \sim -2$} is required. BLAs selected in terms of their width and eq.~\eqref{eq:sl_tau0} thus only trace the low-temperature regime of the warm-hot diffuse gas, since the gas at higher temperatures is not observable, probably due it its high ionisation degree (see Fig.~\ref{fig:Noverb_vs_T}). Note that the estimated baryon content of the absorbing gas in each case corresponds to \mbox{$\sim10$} per cent of the actual baryonic mass in the this phase for the corresponding model (cf. Fig.~\ref{fig:gasmass_sims}).}
\label{fig:tvso_bla}
\end{figure*}

Before we investigate the physical and statistical properties of BLAs in our simulations, we need to assess how setting a fixed sensitivity limit as given by eq.~\eqref{eq:sl_tau0} may bias the detection of warm-hot gas. Under rather simple assumptions, it is possible to model the \HI\ \lya\ central optical depth, $\tau_0$, of \HI\ absorbing gas as a function of its temperature and density. This allows one to put constraints on the physical state of the gas phase traced using \HI\ absorption, given a set of instrumental limitations that lead to a minimum detection (or sensitivity) limit. Conversely, following this approach it is possible to estimate the sensitivity needed to detect gas at a given temperature and density. We describe the basic assumptions and give a detailed calculation of our model in Appendix \ref{sec:tau0_vs}. In particular, we show that it depends on the assumed size of the absorbing structure. With no better estimate at hand, we assume the absorbers to have a linear size%
\footnote{Note that the simulation runs we use do resolve the Jeans length, in particular at the relatively low densities consider here.
} of order the local Jeans length \citep[][see also eq.~\ref{eq:jeans}]{sch01a}. Note that we have already showed in Section \ref{sec:odvsN} that this assumption can account for the $\NHI(\Delta)$ relation predicted by the simulations. Also, our model neglects non-thermal broadening, implying that all sensitivities in terms of $\tau_0$ given henceforth are strict lower limits.

We now investigate the relation between the gas mass distribution in our simulations and the gas mass detected in \HI\ absorption. For each of the BLAs in the line sample obtained for each model considered here, we estimate $\deltawhi$ and $\tempwhi$, and plot the resulting distribution on the \mbox{$\tempwhi - \deltawhi$} plane. The result is shown in Fig.~\ref{fig:tvso_bla}. The coloured areas indicate the cumulative number fraction (in per cent) of BLAs at a given density and temperature. The gray solid contours correspond in each case to the gas mass distribution shown in Fig.~\ref{fig:gasmass_sims}. Note that for the contours the axes indicate the actual gas overdensity and gas temperature. We plot in each panel a series of black dashed contours which indicate the central optical depth, $\tau_0$, as a function of $\deltawhi$ and $\tempwhi$ as given by eq.~\eqref{eq:tau0_jeans}. The corresponding contour values are indicated next to each curve only in the top-left panel, but they are identical for all the other panels. Note in particular the thick dashed contour (magenta) which indicates our adopted sensitivity limit as given by eq.~\eqref{eq:sl_tau0} for S/N=50, i.e. \mbox{$\log \tau_0 = -1.34$} (or \mbox{$\log \left( \NHI / \bhi \right) = 10.8$}). 

One notable feature in this figure is the bi-modality of the distribution of gas traced by broad \HI\ absorbers, irrespective of the model. We see in each case a population of BLAs tracing gas at low temperatures (\mbox{$\tempwhi < 5\times10^4 \K$}) and overdensities \mbox{$\log \deltawhi < 0.5$}, and a second population tracing warm-hot gas at \mbox{$\tempwhi > 5\times10^4 \K$} and overdensities \mbox{$\log \deltawhi > 0.5$}. Note, however, that the amplitude of the distribution varies from model to model. Comparing the gray contours to the coloured distribution we see clearly that the \HI\ detected in absorption traces only a fraction of the gas mass in the simulations. In particular, the model \mbox{\em AGN} (bottom-left panel) shows a large fraction of gas mass at \mbox{$10^5 \K \lesssim \tempwhi \lesssim 3\times10^5 \K$} and \mbox{$\log \deltawhi \sim 0.5$} which is not detected in \HI\ absorption. The same is true for the models \mbox{\em NOZCOOL} and \mbox{\em REF}, although at slightly different temperatures and overdensities. Consideration of the thick dashed contour reveals that this is a selection effect, i.e. the gas is simply not detectable at our adopted sensitivity limit. As discussed above, this comes about because the gas at such high temperatures and relative low densities is extremely ionised, with neutral hydrogen fractions \mbox{$\log ~\fhi \lesssim -6 $}, and its absorption simply falls below our adopted detection threshold (cf. Fig.~\ref{fig:Noverb_vs_T}).

Thus, in our fiducial model BLAs detected in spectra with S/N=50 typically have \mbox{$-2 < \log \tau_0 < 0$}, while the bulk of the gas mass at \mbox{$10^5 \K \lesssim \tempwhi \lesssim 3\times10^5 \K$} and \mbox{$\log \deltawhi \sim 0.5$} is predicted to give rise to absorption with \mbox{$ \log \tau_0 < -2$}. The fact that we do detect in absorption some of the gas at temperatures and densities which correspond to sensitivities slightly smaller than our adopted value (i.e. to the left of the thick dashed contour) reflects the simplicity of the assumptions that go into modelling the absorption strength in terms of $\Delta$ and $T$. Nevertheless, the expected and actual detections are fairly consistent with each other. Based on this, we estimate that in order to detect most of the baryonic mass in the WHIM using thermally broadened \HI\ absorption, spectra with very high S/N are required that allow detection at the \mbox{$\log \tau_0 \sim -2$} level, which is roughly an order of magnitude lower than the typical sensitivities adopted in BLA studies \citep[][]{ric06a,dan10a,wil10a} .

\subsection{BLA number density} \label{sec:dndz_bla}

\begin{table} 
\begin{center}
\caption{Line-number density, $\left( dN/dz \right)$, and corresponding Poisson uncertainties for BLAs and NLAs identified in spectra with various S/N values from our fiducial run at \mbox{$z=0.25$}.}  
\label{tbl:bla_stats}
\begin{tabular}{l r@{$\pm$}l r@{$\pm$}l r@{$\pm$}l}
\hline
 			& \multicolumn{2}{c}{S/N=50}	& \multicolumn{2}{c}{S/N=30} 	& \multicolumn{2}{c}{S/N=10}	\\
\hline
 all $^{\rm a}$	& $460$&$21$ 			& $332$&$18$ 			& $144$&$12$	 			\\
 simple		& $214$&$15$ 			& $175$&$13$ 			& $94$&$10$ 		 		\\
 NLA (simple)	& $173$&$13$ 			& $140$&$12$ 			& $73$&$9$ $^{\rm b}$		\\ 
 NLA (all)		& $343$&$19$ 			& $248$&$16$ 			& $110$&$10$ $^{\rm b}$		\\ 
 BLA (simple) 	& $33$&$6$  				& $28$&$5$				& $15$&$4$				\\ 
 BLA (all)		& $95$&$10$  				& $66$&$8$ 				& $25$&$5$				\\ 
\hline
\end{tabular}
\end{center}
\begin{list}{}{}
	\item [$^{\rm a}$] This corresponds simple and complex \HI\ absorbers taken together.
 	\item [$^{\rm b}$] For reference, \citet[][]{leh07a} find a mean \mbox{$\left({\rm d} N / {\rm d} z \right)_{\rm NLA} = 66 \pm 17$} over 7 \loss\ for {\em all} NLAs in their data with an average \mbox{${\rm S/N} \approx 10$}, discarding lines with associated Voigt-profile parameter errors larger than 40 per cent.
\end{list}
\end{table}

\begin{figure}
\resizebox{\colwidth}{!}{\includegraphics{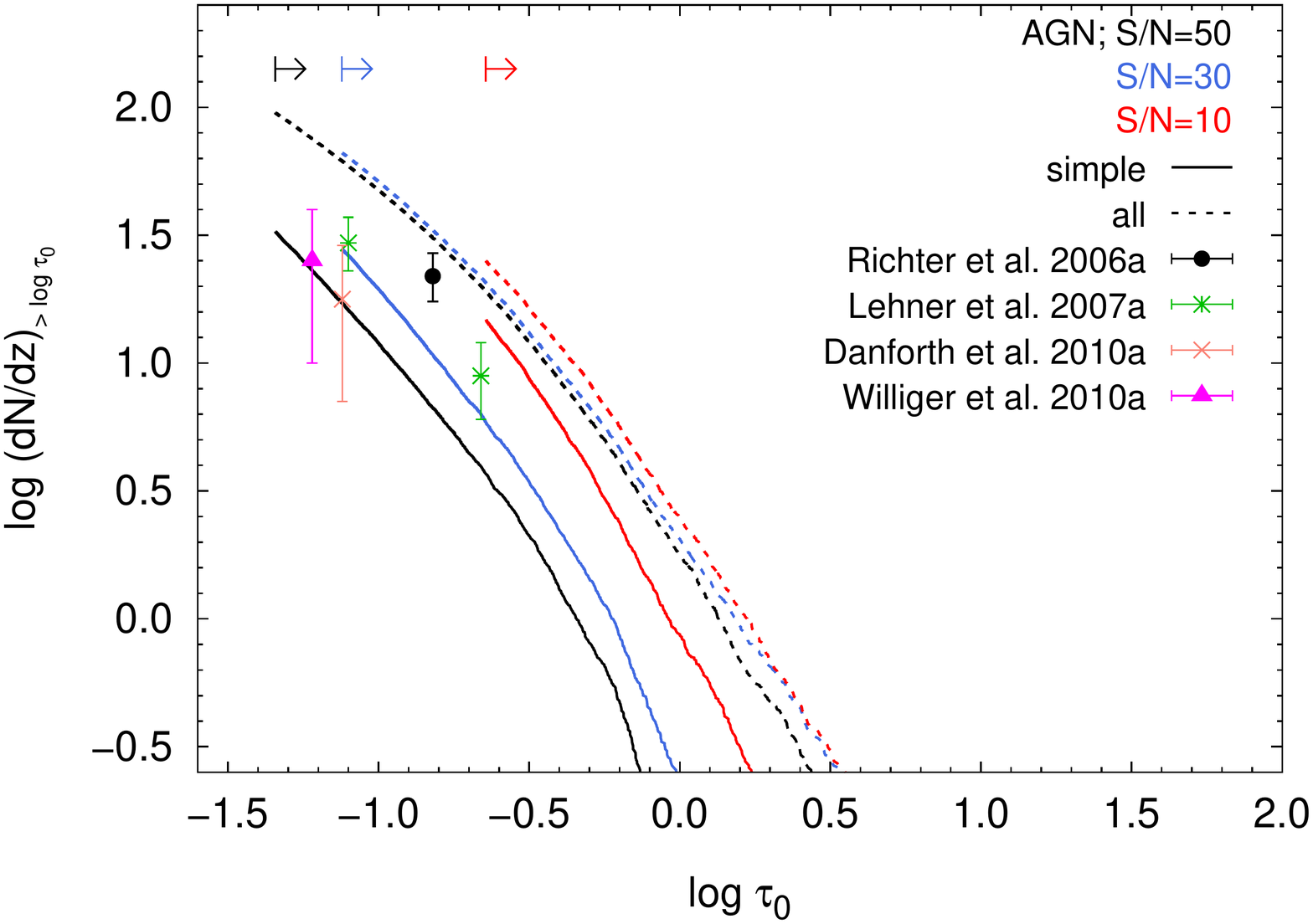}}
\caption[]{Cumulative number density of BLAs as a function of central optical depth, $\tau_0$. Measurements available in the literature are plotted along with the predictions of our fiducial run \mbox{\em AGN} at \mbox{$z=0.25$} (see text for details). The solid lines show the result for simple BLAs identified in 5000 spectra with S/N=50 (black), S/N=30 (blue), and S/N=10 (red). We include our results for all (i.e. simple {\em and} complex) BLA candidates as well (dashed lines). The arrows indicate the BLA sensitivity limit in terms of $\tau_0$ for the corresponding S/N value as given by eq.~\eqref{eq:sl_tau0}.}
\label{fig:dndz}
\end{figure}

Since our adopted sensitivity limit matches the value used to identify BLA candidates in real QSO spectra, we may directly compare the predicted and observed line frequencies. Applying the selection criteria described above (i.e. \mbox{$\bhi \geq 40 \kms$} and \mbox{$\tau_0 \geq 2.27 \left( {\rm S/N} \right)^{-1}$}) to our \mbox{\em AGN} \HI\ line sample obtained from spectra with S/N=50 at \mbox{$z=0.25$} results in 6120 BLA candidates, which corresponds to a line-number density \mbox{$\left({\rm d} N / {\rm d} z\right)_{\rm BLA} = 33 \pm 6$}, where the quoted uncertainty is pure Poissonian. The number densities of BLAs identified in spectra with S/N=30 and S/N=10 are given in Tab.~\ref{tbl:bla_stats}. For completeness, we also include in this table the corresponding numbers for NLAs. Note that the absolute number of BLAs decreases while the relative number of simple BLAs increases with decreasing S/N. This is due to a low S/N causing line blending and increasing the line widths, as discussed in Sec.~\ref{sec:bvd} \citep[see also][their figure 2]{ric06a}.

In Fig.~\ref{fig:dndz} we compare the cumulative BLA number density as a function of $\tau_0$ predicted by our fiducial run obtained from spectra with various S/N values to the available observational results obtained from QSO spectra at comparable redshifts and with similar (average) S/N. The arrows indicate the BLA sensitivity limit in terms of $\tau_0$ for the corresponding S/N value as given by eq.~\eqref{eq:sl_tau0}. The predictions for simple BLAs are indicated by the solid lines. Since our definition of simple absorbers is somewhat arbitrary, we include the corresponding predictions for all, i.e. simple and complex, BLA candidates as well (dashed lines). These two sets of lines thus span our predicted line-frequency range for \mbox{${\rm S/N} \in [10,50]$}.

It is noteworthy that our predictions are broadly consistent with the observed range of BLA number densities. For example, our prediction for S/N=30 (blue solid) agrees with the results by \citet[][green data points]{leh07a}. These authors find in their data with an average \mbox{${\rm S/N} \approx 15$}, discarding lines with associated line-parameter errors larger than 40 per cent, a fraction of single-component BLAs close to 30 per cent and a mean (averaged over 7 \loss) line-number density \mbox{$\left({\rm d} N / {\rm d} z\right)_{\rm BLA} = 30 \pm 7$} at  \mbox{$\log \tau_0 \gtrsim -1.10$} (equivalent to \mbox{$\log \left( \NHI / \bhi \right) \gtrsim 11.02 $}), and \mbox{$\left({\rm d} N / {\rm d} z\right)_{\rm BLA} = 9 \pm 3$} at \mbox{$\log \tau_0 \gtrsim -0.66$} (or \mbox{$\log \left( \NHI / \bhi \right) \gtrsim 11.46 $} ). Both our S/N=50 (black) and S/N=30 (blue) predictions match the result by \citet[][orange data point]{dan10a}, who find \mbox{$\left({\rm d} N / {\rm d} z \right)_{\rm BLA} =  18 \pm 11$} at \mbox{$\log \tau_0 \gtrsim -1.12$} (\mbox{$\log \left( \NHI / \bhi \right) \gtrsim 11.0 $}) along seven \loss\ at \mbox{$z \lesssim 0.5$}, spanning a total redshift path \mbox{$\Delta z =2.193$}, with \mbox{${\rm S/N} \geq 5$}. Their sample consists of 15 single-component BLAs and 48 BLAs with uncertain velocity structure. \citet[][magenta data point]{wil10a} estimate their detection limit at \mbox{$\log \tau_0 \gtrsim -1.22$} (or \mbox{$\log \left( \NHI / \bhi \right) \gtrsim 10.9 $}) and obtain \mbox{$\left({\rm d} N / {\rm d} z \right)_{\rm BLA} = 25 \pm 15$} and \mbox{$\left({\rm d} N / {\rm d} z \right)_{\rm BLA} = 8.5 \pm 8.5$} for their full and single-component BLA samples, respectively, along a single \los\  ($\Delta z = 0.117$) and corresponding spectrum with \mbox{S/N=20 -- 30}. Finally, \citet[][black data point]{ric06a} measure \mbox{$\left({\rm d} N / {\rm d} z\right)_{\rm BLA} = 22$} using their reliable sample of single-component BLAs, detected along four \loss\ at \mbox{$z \lesssim  0.4$} (\mbox{$\Delta z =0.928$}), at a sensitivity \mbox{$\log \tau_0 \gtrsim -0.82$} (or \mbox{$\log \left( \NHI / \bhi \right) \gtrsim 11.3 $}), in spectra with an average \mbox{${\rm S/N} \approx 15$}.

\subsection{Physical properties of broad \HI\ absorbers} \label{sec:obs_phys}

In Sec.~\ref{sec:obs} we discussed the relation between the physical conditions of the gas traced by typical \HI\ absorbers and their line observables ($\NHI$, $\bhi$) using our fiducial model. We now focus on the physical conditions of the gas giving rise to \HI\ absorbers identified as simple BLAs in our synthetic spectra at \mbox{$z=0.25$}; these correspond to single-component \HI\ absorbers falling within the region defined by the polygon in Fig.~\ref{fig:bvsN} which is defined through two criteria: a) the line width  satisfies \mbox{$\bhi \geq 40 \kms$}; b) the central optical depth obeys eq.~\eqref{eq:sl_tau0}, i.e.\ \mbox{$\tau_0 \geq 2.27 \left( {\rm S/N} \right)^{-1}$}.

Given the importance of broad \HI\ absorbers as potential WHIM tracers, and the dependence of the predicted WHIM mass fraction on the adopted physical model, we next explore the relation between the measured line width and the temperature of the absorbing gas using the models introduced in Sec.~\ref{sec:sims}. We further use our fiducial model ({\em AGN}) to investigate the ionisation state, neutral hydrogen fraction, and metallicity of the gas traced by different classes of \HI\ absorbers. Finally, we estimate the baryon content of the gas traced by BLAs using all models, but compare our results to observations using only the model {\em AGN}.

\subsubsection{Temperature distribution of (broad) lines} \label{sec:bla_temp}

\begin{figure*}
\resizebox{1.1\textwidth}{!}{\includegraphics{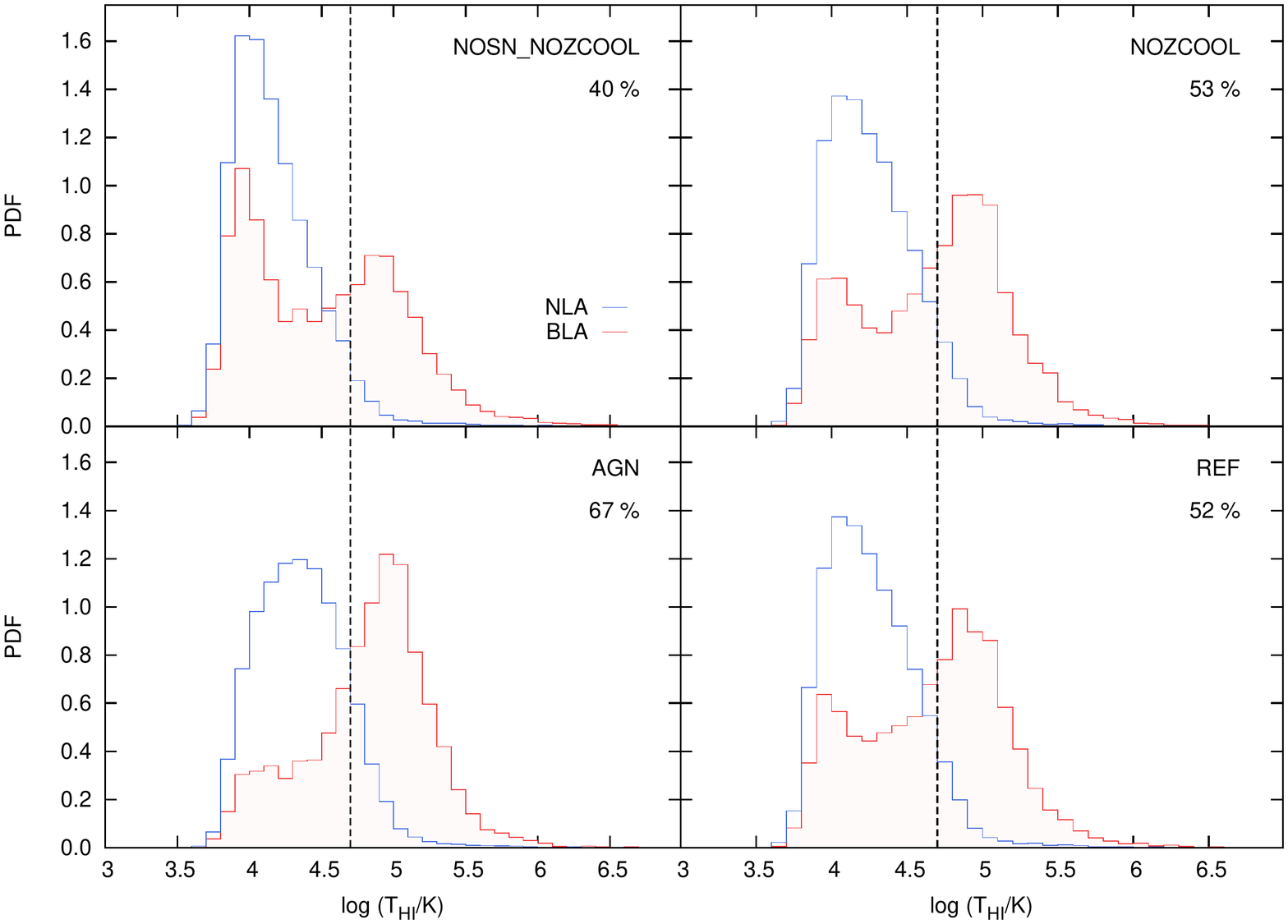}}
\caption[]{Temperature distribution of the gas traced by simple \HI\ absorbers identified in spectra at \mbox{$z=0.25$} with S/N=50 obtained from different models. The blue histograms show in each case the distribution for gas traced by narrow (i.e. $\bhi < 40 \kms$) absorbers, while the red histograms indicate the temperature distribution of gas traced by BLAs. The percentage shown in each panel indicates the number fraction of BLAs arising in gas at $\tempwhi \geq 5\times10^4 \K$, i.e. hot-BLAs. Note the correspondence between increasing mass fraction of warm-hot diffuse gas in Fig.~\ref{fig:gasmass_sims} and the increasing fraction of BLAs tracing warm-hot gas shown here. The distribution predicted by our fiducial run (\mbox{\em AGN}) suggests that, statistically, two out of three BLA candidates trace gas at \mbox{$T \gtrsim 5\times10^4 \K$}.}
\label{fig:temp_dist}
\end{figure*}

In Fig.~\ref{fig:temp_dist} we show the distribution of temperatures of the gas traced by both BLAs (red histograms) and NLAs (blue histograms) identified our spectra with \mbox{S/N = 50} at \mbox{$z = 0.25$} obtained from different models. The dashed, vertical line in each panel indicates the temperature threshold adopted to separate cool from warm-hot gas (see Tab.~\ref{tbl:phases}).

In all models, the temperature distribution of the gas traced by NLAs shows that lines with Doppler parameters \mbox{$\bhi < 40 \kms$} preferentially arise in gas at temperatures \mbox{$\tempwhi \lesssim 5\times10^4 \K$} with a peak at \mbox{$\tempwhi \sim10^4 \K$}, as expected from their width. The few lines which are narrower than allowed by the temperature of the absorbing gas (i.e. the section of the blue histogram to the right of the vertical, dashed line) are mostly weak lines that fall below the formal completeness limit, as discussed in Sec.~\ref{sec:bth_b}.  But some of these lines are real detections, which suggest that gas at different temperatures overlaps in velocity space (due to redshift-space distortions), and some of it may even overlap in position space, indicating the existence of multi-phase absorbing structures. In models with feedback ({\em NOZCOOL}, {\em REF}, {\em AGN}), the temperature distribution of the gas traced by NLAs is broader than in the model without feedback (\mbox{\em NOSN\_NOZCOOL}), and the fraction of lines with `un-physical' widths is higher. This can be explained as follows. Outflows driven by SNe and AGN follow the path of least resistance in space, thus escaping into the voids while leaving the cooler, denser filaments intact \citep[][]{the02a}. With increasing feedback strength, the cross-section of such high-temperature outflows increases as well, and so does the chance for a random \los\ to intersect both cool, dense filaments and shock-heated material, with their corresponding absorption overlapping from time to time along the spectrum.

Note that the NLA temperature distributions in the models \mbox{\em NOZCOOL} and \mbox{\em REF} are very similar to each other, and the same is true for the corresponding BLA temperature distributions. Moreover, the fraction of hot-BLAs is only slightly lower in \mbox{\em REF} than in \mbox{\em NOZCOOL}. This indicates that metal-line cooling is of secondary importance in setting the thermal state of the gas phase traced by (broad) \HI\ absorbers.

In contrast to narrow \HI\ absorbers, broad \HI\ lines trace gas around two different temperatures, \mbox{$\tempwhi \sim 10^4 \K$} and \mbox{$\tempwhi \sim 10^5 \K$}, irrespective of the model (red histograms). Clearly, BLAs arising in gas at low temperatures must be subject to substantial non-thermal broadening,  such as bulk flows and/or Hubble broadening%
\footnote{Note that our simulations lack the resolution to capture small-scale turbulence within the gas.
}. Since the line width of an \HI\ absorber with $\bhi = 40 \kms$ arising in gas at \mbox{$T \sim 10^4 \K$} is completely dominated by non-thermal broadening, its linear size (assuming the line width is entirely due to Hubble broadening) must be $\sim 500 ~\kpc$, which is consistent with the Jeans length of a filament with a mean density \mbox{$\nH \sim 10^{-6} \pcc$} and temperature \mbox{$T \sim 10^4 \K$} (eq.~\ref{eq:jeans}).

\begin{figure*}
{\resizebox{0.33\textwidth}{!}{\includegraphics{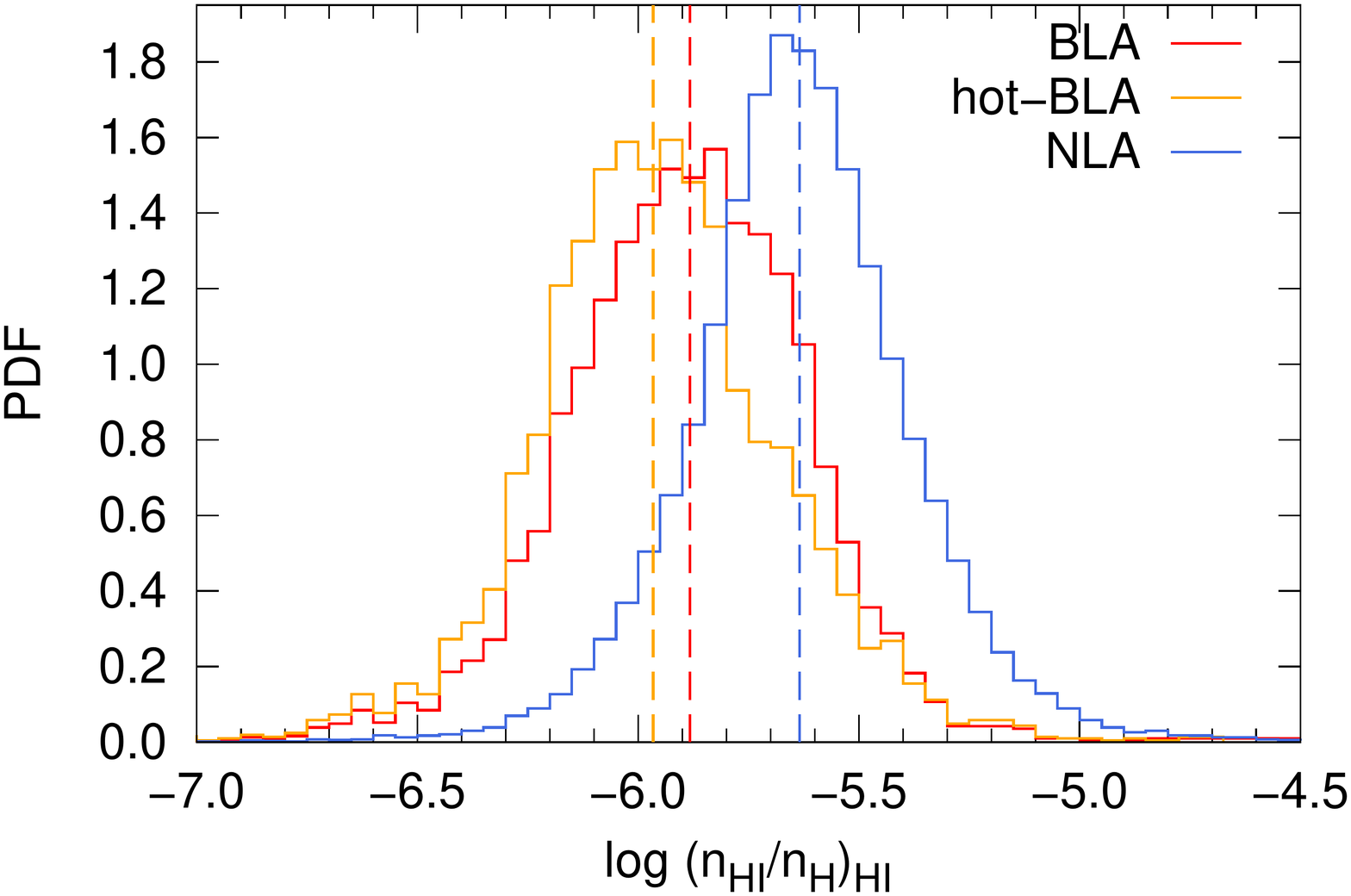}}}
{\resizebox{0.33\textwidth}{!}{\includegraphics{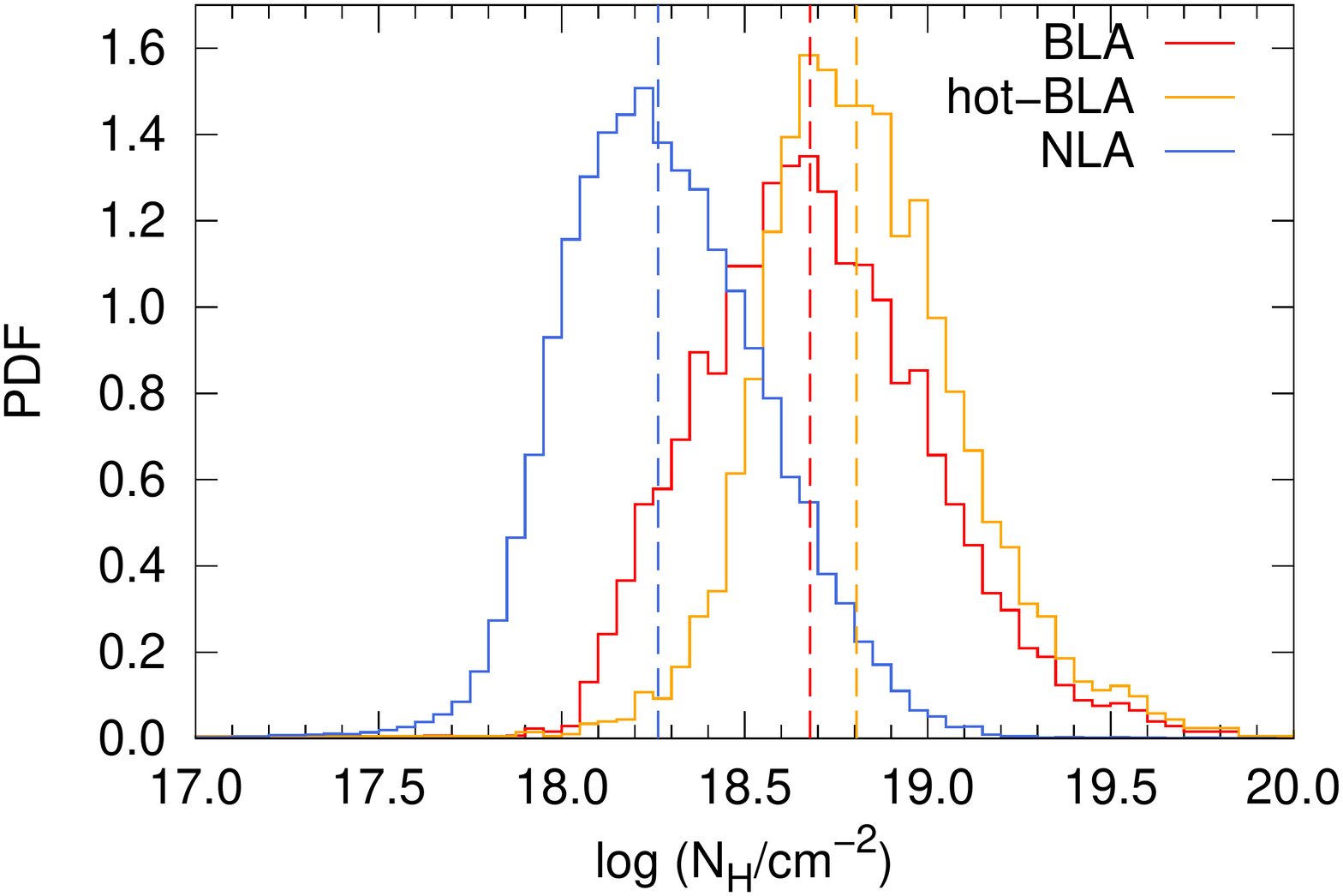}}}
{\resizebox{0.33\textwidth}{!}{\includegraphics{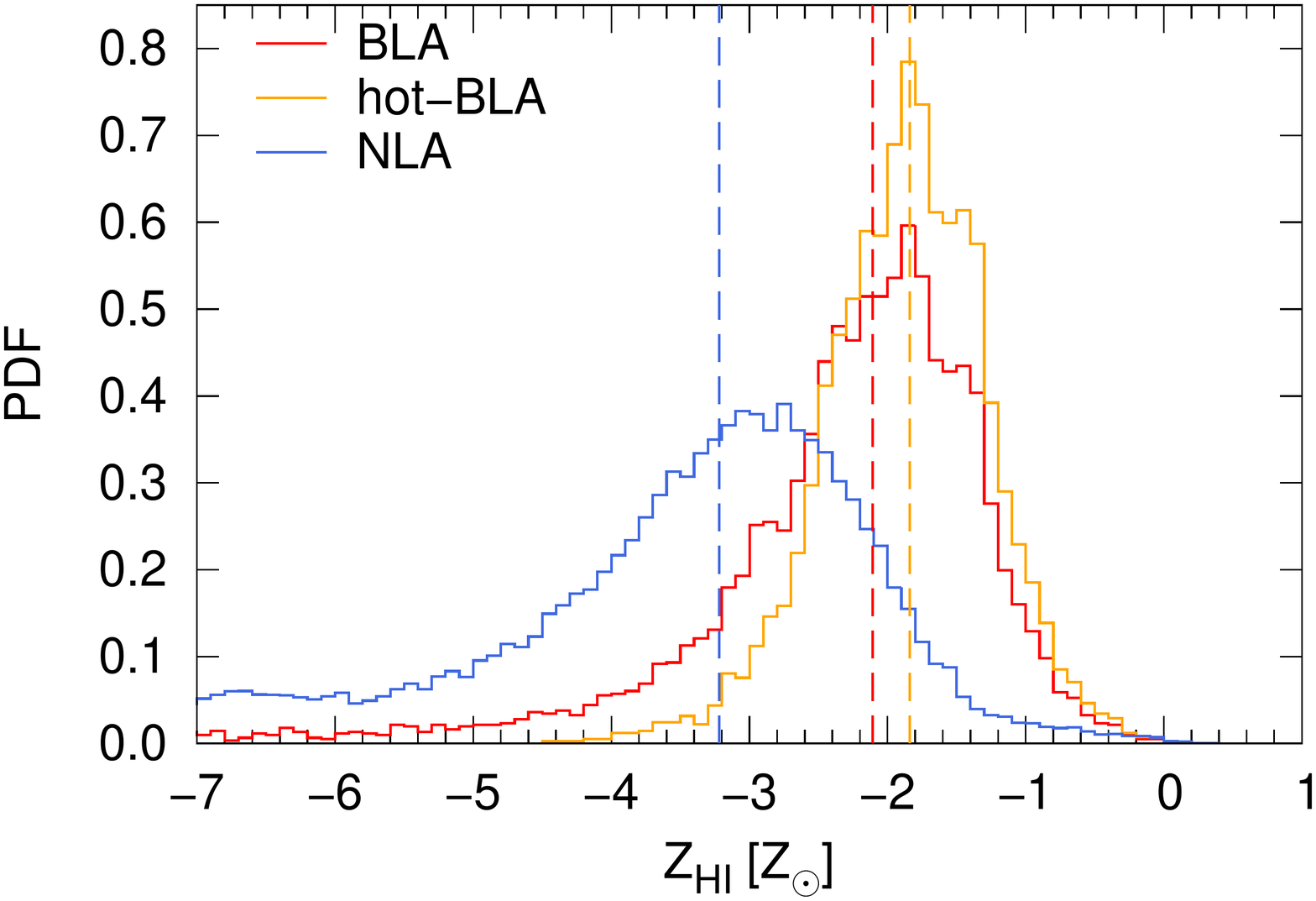}}}
\caption[]{Distribution of various physical quantities characterising the gas giving rise to single-component absorption features identified as NLA (blue), BLA  (red), and hot-BLA (orange) in spectra at \mbox{$z=0.25$} with S/N=50 obtained from our \mbox{\em AGN} model. {\em Left:} Neutral hydrogen fraction {\em Middle:} Total hydrogen column density. \mbox{\em Right:} Metallicity relative to the solar value (\mbox{$Z_{\odot} = 0.0127$}). The vertical dashed lines in each panel indicate the median values of the corresponding distributions. On average, NLAs and (hot-)BLAs appear to trace different gas phases.}
\label{fig:bla_various}
\end{figure*}

Interestingly, the fraction of BLAs tracing gas at low (\mbox{$\tempwhi \sim 10^4 \K$}) and high (\mbox{$\tempwhi \sim 10^5 \K$}) temperatures is very different in each model, as can be judged qualitatively by the shape of the corresponding histograms and quantitatively by the percentage included in each panel which gives the fraction of hot-BLAs, i.e. BLAs tracing gas with \mbox{$T \geq 5\times10^4 \K$}. In fact, the ratio of BLAs tracing cool gas to those tracing warm-hot gas appears to be very sensitive to feedback strength. In principle, this could be used to constrain feedback models observationally. The caveat is that a statistically significant sample of confirmed BLAs would be required for which the gas temperature can be measured reliably. In our fiducial model (\mbox{\em AGN}), which includes the strongest feedback, the majority of the BLAs trace gas at \mbox{$\tempwhi \sim 10^5 \K$}, with little contamination by non-thermally broadened lines. In fact, two out of three BLA candidates arise in gas at temperatures \mbox{$T \geq 5\times10^4 \K$}.

The bi-modal character of the gas temperature distributions for BLAs predicted by the model \mbox{\em NOSN\_NOZCOOL} is also consistent with previous results. \citet[][]{ric06b} find in a simulation that included a model similar to our \mbox{\em NOSN\_NOZCOOL} that \mbox{$\sim30$} per cent of the BLAs trace gas at \mbox{$T < 2\times10^4 \K$}, and a significant fraction gas at \mbox{$\tempwhi \gtrsim 5\times10^4 \K$}. The quantitative difference between theirs and our result for \mbox{\em NOSN\_NOZCOOL} is probably caused by the difference in the simulation methods used. Using a linear model to fit the thermal to total line width, \citet[][]{ric06b} find that \mbox{$b_{T} / \bhi = 0.91$} (no error quoted). We note that we do not find such a tight correlation between $b_{T}$ and $\bhi$, but if we perform a linear fit between these two quantities, we find\footnote{For reference, the corresponding results for S/N=30 and S/N=10 are \mbox{$b_{T} / \bhi = 0.622 \pm 0.04$} and \mbox{$b_{T} / \bhi = 0.556 \pm 0.006$}, respectively.} \mbox{$b_{T} / \bhi = 0.641 \pm 0.004$}. This is consistent with the result presented in Sec.~\ref{sec:bth_b} that thermal broadening on average contributes with (at least) 60 per cent to the total line width of BLAs.

\citet[][]{leh07a} argue that broad \HI\ lines may trace both cool and warm-hot gas, but that the majority of BLAs trace gas at \mbox{$T \sim 10^5 - 10^6 \K$} if their width is dominated by thermal broadening. As mentioned in the previous paragraph, thermal broadening accounts for a significant fraction to the total line width of single-component, broad \HI\ absorbers. Thus, our simulations are consistent with the result inferred from observations that these absorbers do preferentially trace gas at high temperatures, at least in models with (some type of) feedback.\\

In summary, we find that in the absence of feedback BLA samples are contaminated by a large fraction of non-thermally broadened lines. Conversely, the fraction of broad \HI\ absorption lines tracing gas at temperatures \mbox{$\tempwhi \sim 10^5 \K$} increases when feedback is included. For instance, our fiducial model predicts that, in a statistical significant sample, 67 per cent of the BLAs trace gas at \mbox{$\tempwhi \gtrsim 5\times10^4 \K$}. Our results thus strongly support the idea that reliable BLAs detected in real absorption spectra are genuine tracers of gas at such high temperatures.

\subsubsection{Neutral hydrogen fraction, total hydrogen column density, and metallicity} \label{sec:bla_phys}

In Fig.~\ref{fig:bla_various} we show the distribution of neutral hydrogen fraction (left panel), total hydrogen column density (middle panel), and (local) metallicity (right panel) of the gas traced by narrow absorbers (blue), BLAs (red), and hot-BLAs (orange) identified in spectra with S/N=50 obtained from our fiducial model. The vertical dashed line indicates in each case the corresponding median value.

In general terms, the physical properties of the gas traced by BLAs and hot-BLAs show similar distributions and comparable median values, but they are somewhat different from the corresponding properties of the gas traced by NLAs. For example, the median neutral hydrogen fraction of gas traced by NLAs is \mbox{$\log \fhi \sim -5.5 $}, which is slightly higher than the neutral hydrogen fraction of the gas traced by (hot-)BLAs, \mbox{$\log \fhi \sim -6 $}. This is expected since, as we have shown previously, the temperature of gas giving rise to broad \HI\ absorption is, on average, higher than the temperature of gas giving rise to narrow \HI\ absorbers. Furthermore, the median total hydrogen column density of the gas detected via (hot-)BLAs is \mbox{$\NH \sim 6\times10^{18} \psc$}, which is several times larger than the median total hydrogen column density of the gas traced by NLAs, \mbox{$\NH \sim 2\times10^{18} \psc$}, and its distribution extends out to significantly larger values, \mbox{$\NH \sim 10^{20} \psc$}, as compared to \mbox{$\NH \sim 10^{19} \psc$}. This is due to several factors. First, as shown in Figs.~\ref{fig:Noverb_vs_T} and \ref{fig:tvso_bla}, high-temperature gas detected via \HI\ absorption at a fixed sensitivity necessarily has a higher density with respect to gas at lower temperatures. Also, a higher density implies an average higher $\NHI$ as a consequence of the \mbox{$\deltawhi - \NHI$} correlation. Finally, gas at high temperature has a lower neutral hydrogen fraction, which in turn yields higher total hydrogen column densities for a given $\NHI$. The high(-er) total hydrogen density of the gas traced by broad \HI\ absorbers implies that its baryon content is considerable. We will come back to this point in more detail in Sec.~\ref{sec:bar_whim}.

Quite interesting is the difference between the gas metallicity distributions. While the metallicity of the gas traced by NLAs shows a broad distribution with a tail extending to very low values and a median \mbox{$Z \sim 10^{-3} Z_{\odot}$}, the metallicity distribution of the gas traced by broad absorbers is narrow, with most values falling in the range \mbox{$(0.001, \, 1) ~Z_{\odot} $}, centred around \mbox{$Z \sim 10^{-2} Z_{\odot}$}. On average, the metallicity of the gas traced by (hot-)BLAs exceeds the metallicity of the gas traced by NLAs by an order of magnitude.

These results together indicate that broad \HI\ absorbers trace gas that is physical distinct from the gas traced by narrow \HI\ absorbers, as already mentioned in Sec.~\ref{sec:b_vs_N}. In particular, the relatively high level of enrichment is inconsistent with the idea that BLAs  trace primordial gas that is sinking along filaments towards the centre of high-density regions, as commonly assumed. Rather, our results suggest that broad \HI\ absorbers may be tracing recent (or on-going) galactic outflows, and/or gravitationally shock-heated gas that has been enriched by galactic ejecta at early epochs.

\subsection{Baryon content of \HI\ absorbing gas} \label{sec:bar_whim}

\begin{figure}
\resizebox{\colwidth}{!}{\includegraphics{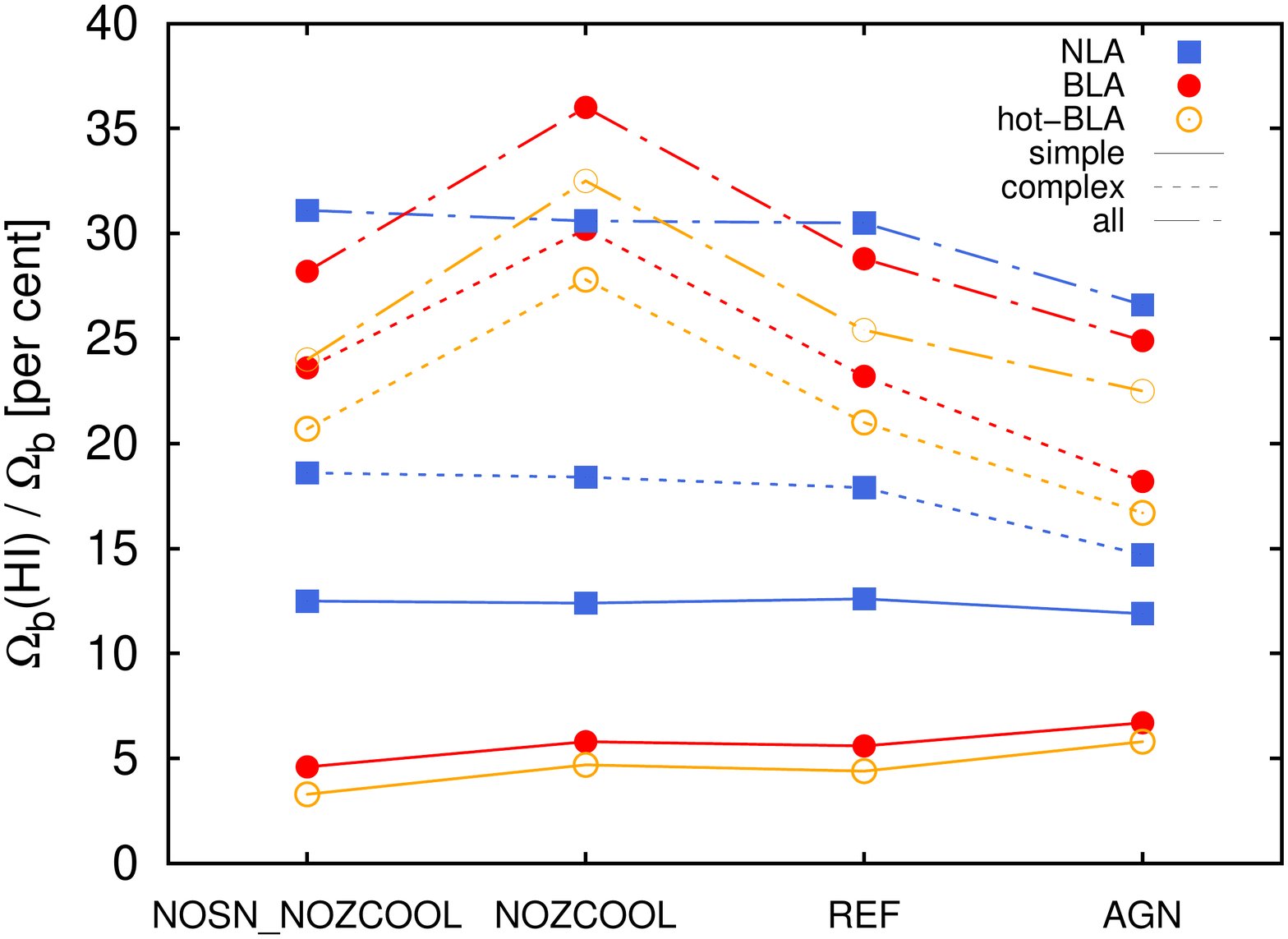}}
\resizebox{\colwidth}{!}{\includegraphics{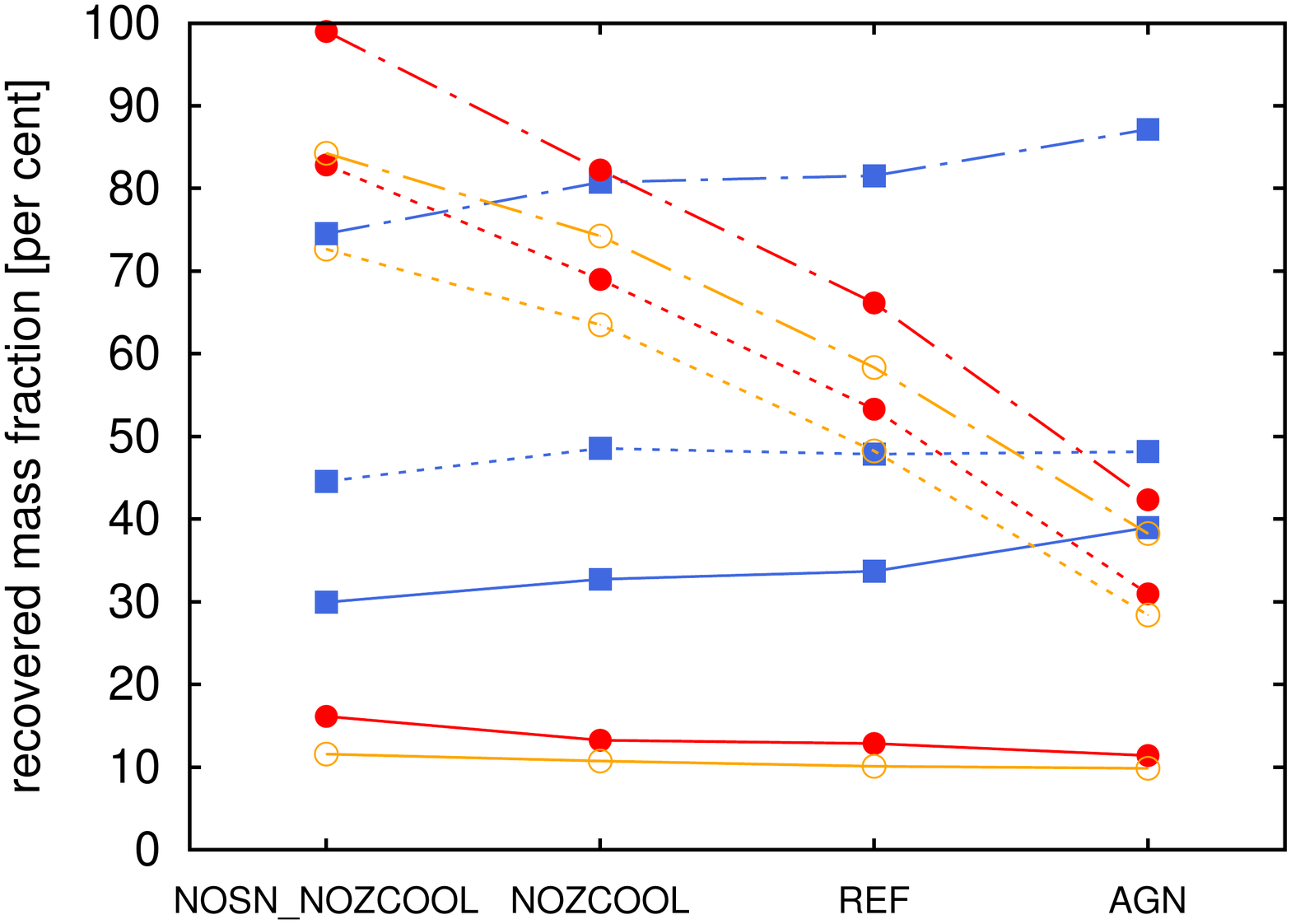}}
\caption[]{{\em Top:} Baryon content (in per cent) relative to the cosmic value \mbox{$\Omega_{\rm b} = 0.0418$} of the gas traced by \HI\ absorbers identified in synthetic spectra with S/N=50 for different model runs. The \HI\ absorber sample has been dissected into NLA (blue squares), BLAs (red filled circles), and hot-BLAs (orange open circles); these classes have in turn been divided into single-component absorbers (solid lines) and complex systems (dotted lines). The black, dot-dashed lines indicate in each case the result for all (i.e. simple and complex) absorbers of a given class.  {\em Bottom:} Gas mass traced by a given \HI\ absorber type relative to the actual gas mass in the phase expected to be traced by that particular type (see text for details). Symbols, lines, and colours as in the top panel.}
\label{fig:bar_fracs}
\end{figure}

In this section we briefly investigate the dependency of the predicted baryon fraction of the gas traced by \HI\ absorbers on the adopted physical model. As we have done in Paper I for \OVI\ absorbers, we estimate the baryon fraction, i.e. the baryon density relative to the critical  density $\rho_{\rm c}$, in \HI\ absorbers using
\beq \label{eq:omb}
	\Omega_{\rm b}(\HI) = \frac{\mH }{\rho_{\rm c}} ~\left( \frac{c}{H_0} \sum_{i=1}^{N_{\rm LOS}} \Delta \chi_{i} \right)^{-1} \sum_{i=1}^{N_{\rm LOS}} \sum_{j=1}^{N_{\rm ab s}} \frac{\NHI}{ \left({\rm X_{H}} \right)_{\ionsubscript{H}{I}} ~\fhiw} \, ,
\eeq
where $\mH$ is the hydrogen mass, and ${\rm X_{H}}$ and \mbox{$\fhiw$} are the optical-depth weighted hydrogen mass fraction and neutral hydrogen fraction, respectively. Note that $\NHI$, $\left({\rm X_{H}} \right)_{\ionsubscript{H}{I}}$, and $\fhiw$ are computed for {\em individual} absorbing components along each \los, but we have omitted the running indices for simplicity.

The top panel of Fig.~\ref{fig:bar_fracs} shows the baryonic mass fractions, \mbox{$\Omega_{\rm b}(\HI) / \Omega_{\rm b}$}, predicted by various models in different types of absorbers: NLAs (blue squares), BLAs (red filled circles), and hot-BLAs (orange open circles), where each of these classes have been sub-divided into simple (solid lines) and complex (dotted lines) absorbers (see Secs.~\ref{sec:phys} and \ref{sec:detect}). Note that we consider the baryonic mass fraction of the gas traced by both simple and complex absorbers, since our adopted criterium to define simple absorbers is somehow arbitrary. The net baryon fractions of simple and complex absorbers (of a given class) taken together are indicated by the dot-dashed lines.

The baryonic mass fraction in NLAs (simple or complex) is very similar in all models, being only slightly lower in our fiducial run \mbox{\em AGN}. This is consistent with the lower gas mass fraction in the cool diffuse gas (which is expected to be traced by NLAs) in this model compared to all other models (see Fig.~\ref{fig:gasmass_sims}). The small difference in \mbox{$\Omega_{\rm b}(\rm NLA) / \Omega_{\rm b}$} between all the models indicates that feedback has a negligible impact on the gas phase typically traced by NLAs. This in turn is consistent with the fact that the predicted \HI\ statistics (which are dominated by these absorbers) are rather insensitive to variations in the feedback model (see Appendix \ref{sec:cnvg_phys}).

The baryonic mass fraction of gas traced by simple BLAs is relatively low, and varies significantly between the models, from \mbox{$\sim3$} per cent (\mbox{\em NOSN\_NOZCOOL}) to \mbox{$\sim7$} per cent (\mbox{\em AGN}). The baryonic mass fractions in complex BLAs are much higher than the baryonic mass fractions in simple BLAs, and they are also very different in each model. For instance, the baryon fraction of complex BLAs in the model \mbox{\em NOZCOOL} is higher ($\sim5$ per cent) than in the model \mbox{\em NOSN\_NOZCOOL}; this is consistent with the fact that SN feedback significantly increases the mass in warm-hot diffuse gas, as has been shown previously (see Fig.~\ref{fig:gasmass_sims} and corresponding text), together with the idea that BLAs preferentially trace this gas phase. In contrast, the baryon fraction in complex BLAs predicted by the models \mbox{\em REF} and \mbox{\em AGN}, is lower (by $\sim5$ and $\sim10$ per cent, respectively) compared to predictions of the model \mbox{\em NOZCOOL}.

The lower baryon fraction in complex BLAs predicted by the model \mbox{\em REF} with respect to \mbox{\em NOZCOOL} can be explained as follows. Complex absorbers trace kinematically disturbed gas, most probably SN-driven outflows. These ejecta carry heavy elements with them, which allow a significant fraction of the gas to cool down radiatively, thus reducing the number of thermally broadened lines and their net baryonic mass. However, the lower baryonic mass content of complex BLAs in the model \mbox{\em AGN} with respect to all other models is in contrast with the actual total mass fraction in the warm-hot phase predicted by this model, which is higher compared to all other models (see bottom-left panel of Fig.~\ref{fig:gasmass_sims}). The discrepancy between the predicted mass fraction in the warm-hot phase and the baryon content of the gas traced by BLAs in the \mbox{\em AGN} model can be understood as consequence of the limited sensitivity. As already discussed, AGN feedback shifts a significant fraction of gas into the warm-hot phase; however, most of this mass ends up at temperatures and densities which lead to a \HI\ fraction and corresponding absorption signal that is beyond our adopted detection limit (see Fig.~\ref{fig:tvso_bla}).

Note that the baryon fractions traced by hot-BLAs are only sightly lower than for BLAs, irrespective of the model. This is important because it implies that the contamination of the BLA sample by non-thermally broadened lines does not significantly affect the inferred baryon fraction of the WHIM. In other words, the baryon fraction in warm-hot gas is dominated by the absorbers arising in gas at the highest temperatures. This is a direct consequence of the steep decline of the hydrogen neutral fraction with temperature.

The bottom panel of Fig.~\ref{fig:bar_fracs} shows the {\em recovered mass fraction} (in per cent) of a given gas phase. This quantity is defined as the total baryonic mass in a given absorber class relative to the actual gas mass in the phase (Fig.~\ref{fig:gasmass_sims}) {\em expected} to be traced by that particular absorber class. So, for example, the gas mass recovered from hot-BLAs is given in each model as the total baryonic mass in these absorbers (orange, filled squares in the top panel of Fig.~\ref{fig:bar_fracs}) divided by the gas mass in the warm-hot diffuse phase (orange percentages in the top-left sections of Fig.~\ref{fig:gasmass_sims}). The gas mass recovered from NLAs is correspondingly given as the total baryonic mass in these absorbers (blue squares in the top panel of Fig.~\ref{fig:bar_fracs}) divided by the gas mass in the cool diffuse phase (percentages indicated in the bottom-left sections of Fig.~\ref{fig:gasmass_sims}).

Apparently, none of the absorber classes traces the total gas mass in its corresponding (expected) phase, with exception perhaps of the full (i.e. simple and complex) BLA sample (red filled circles) in the model \mbox{\em NOSN\_NOZCOOL}. The gas mass fraction traced by both simple and complex NLAs is very similar in all models, and taken together these absorbers trace between \mbox{$\sim70$} per cent (\mbox{\em NOSN\_NOZCOOL}) and \mbox{$\sim90$} per cent (\mbox{\em AGN}) of the gas mass in cool diffuse gas. This again is consistent with our previous statement that the gas in this phase is left almost intact by feedback mechanisms such as SN-driven winds and AGN outflows,

Note that simple (hot-)BLAs trace roughly 10 to 15 per cent of the true baryonic mass in warm-hot gas, irrespective of the model. This suggests that baryonic mass estimates based on this type of absorbers are robust. In contrast, the gas mass traced by {\em complex} (hot-)BLAs is very different in each model. As already mentioned above, in the model \mbox{\em NOSN\_NOZCOOL} the full BLA sample traces practically all the mass contained in warm-hot gas, with complex BLAs contributing with more than 80 per cent to the recovered gas mass. This suggests that the bulk of gas shock-heated by gravity (which is the only possible heating mechanism in the model \mbox{\em NOSN\_NOZCOOL}) can be fully accounted for using (simple and complex) BLAs, at our adopted sensitivity. The recovered WHIM mass is, however, systematically lower in models that include SN and AGN feedback, and metal-line cooling.\\

Taking the result from our fiducial run \mbox{\em AGN} at face value, we estimate the total baryon content in gas traced by \HI\ in our simulation at \mbox{$z=0.25$} to be \mbox{$\Omega_{\rm b}(\HI) / \Omega_{\rm b} = 0.57$} (S/N=50), 0.48 (S/N=30), and 0.29 (S/N=10). The last two values are in remarkable agreement with the results from observations at comparable sensitivity. Assuming a simple ionisation model, \citet*[][]{pen04a} measure%
\footnote{Relative to \mbox{$\Omega_{\rm b} = 0.0418$}, rather than \mbox{$\Omega_{\rm b} = 0.047$} assumed by \citet[][]{pen04a}, and re-scaled to \mbox{$h = 0.73$}.
} \mbox{$\Omega_{\rm b}(\HI) / \Omega_{\rm b} = 0.31\pm0.04$} at \mbox{$z \approx 0$} for absorbers with column densities \mbox{$12.5  \leq \log\left( \NHI / \psc \right) \leq 17.5 $} and \mbox{$\bhi \lesssim 100 \kms$}. Similarly, assuming the gas to be isothermal and photo-ionised, \citet[][]{leh07a} obtain%
\footnote{Relative to \mbox{$\Omega_{\rm b} = 0.0418$}, rather than \mbox{$\Omega_{\rm b} = 0.044$}, and re-scaled to \mbox{$h = 0.73$} rather than \mbox{$h=0.7$}.
} \mbox{$\Omega(\HI)/\Omega_{\rm b}  = 0.40$} from their data with an average \mbox{${\rm S/N} \approx 15$} and for \mbox{$12.4  \leq \log \left( \NHI / \psc \right) \leq 16.5 $} and \mbox{$\bhi \lesssim 150 \kms$}.

\subsubsection{Baryon content of warm-hot gas at  low $z$}

Estimates of baryonic mass contained in the WHIM based on broad \HI\ absorbers detected in real QSO spectra are very uncertain, even with a reliable sample of BLA candidates at hand, since they are highly sensitive to the ionisation state of the absorbing gas (see eq.~\ref{eq:omb}),  which is probably dominated by collisions between ions and electrons in the plasma. In this case, the neutral hydrogen fraction is a steeply decreasing function of temperature, and an accurate estimate of the WHIM baryon content thus relies on a precise measurement of the temperature of the absorbing gas. As we have shown in Sec.~\ref{sec:bth_b}, temperature estimates from the line width of broad \HI\ absorbers may yield values that are uncertain by, at least, factors of a few. 

The first attempt to measure the baryon content of the WHIM using BLAs was undertaken by \citet[][]{ric04a}, who found \mbox{$\Omega_{\rm b}({\rm BLA}) \leq 3.2 \times 10^{-3} ~\left( h / 0.73 \right)^{-1}$} assuming CIE, which represents less than 8 per cent of the cosmic baryon budget. In a follow-up study, \citet[][]{ric06a} measured \mbox{$\Omega_{\rm b}({\rm BLA}) \geq 2.6 \times 10^{-3} ~\left( h / 0.73 \right)^{-1}$}, corresponding to at least 6 per cent of the baryons in the Universe. These authors also assumed CIE, but recognised the potential importance of photo-ionisation (PI) in determining the ionisation state of the WHIM, and concluded that their baryon content measurement could be underestimated by 15 - 50 per cent.

Using a significantly larger sample than previous studies, \citet[][]{leh07a} report \mbox{$\Omega_{\rm b}({\rm BLA})/\Omega_{\rm b} = 0.08$} assuming the gas to be in collisional ionisation equilibrium (CIE) for absorbers with \mbox{$13.2  \leq \log \left( \NHI / \psc \right) \leq 16.5 $} and \mbox{$40 \kms < \bhi \lesssim 150 \kms$}. Using the same sample and assuming a hybrid model (including photo- and collisional ionisation) to compute the neutral hydrogen fraction, these authors find \mbox{$\Omega_{\rm b}({\rm BLA})/\Omega_{\rm b} = 0.20$}. Both estimates are based on a series of assumptions. First, in order to account for the possible contamination of their sample with lines broadened by unresolved velocity structure or any other non-thermal mechanism, these authors randomly discard one third of the BLAs in their sample. Moreover, they assume the thermal width to be 90 per cent of the observed line width, based on the results from previous simulations by \citet[][]{ric06b}. If, instead, the line width is dominated by thermal broadening, they get \mbox{$\Omega_{\rm b}({\rm BLA})/\Omega_{\rm b} = 0.13$} (CIE) and \mbox{$\Omega_{\rm b}({\rm BLA})/\Omega_{\rm b} = 0.32$} (PI+CIE).

In a more recent study%
\footnote{We note that there is an error in the computation of the total absorption path length in \citet[][ their Table 1]{dan10a}, which might have affected their reported estimate of the baryon content in BLAs (C.Danforth, {\em private communication}).
}, \citet[][]{dan10a} report \mbox{$\Omega_{\rm b}({\rm BLA}) = 6.0^{+1.1}_{-0.8} \times 10^{-3} ~\left( h / 0.73 \right)^{-1}$}, equivalent to \mbox{$\Omega_{\rm b}({\rm BLA})/\Omega_{\rm b} = 0.14^{+0.03}_{-0.02}$}. These authors analyse in detail the systematic uncertainties that afflict their (and others') baryon estimates, such as unresolved velocity structure, sample completeness, ionisation corrections, and the assumed relation between line width and gas temperature, and find their estimate to vary between \mbox{$\Omega_{\rm b}({\rm BLA}) = 2.3 \times 10^{-3} ~\left( h / 0.73 \right)^{-1}$ }and \mbox{$\Omega_{\rm b}({\rm BLA}) = 15.2 \times 10^{-3} ~\left( h / 0.73 \right)^{-1}$}, i.e. between \mbox{$\sim 6$} and \mbox{$\sim 36$} per cent of the cosmic baryon budget. Clearly, there is still a high uncertainty in the estimate of the baryonic mass traced by observed BLAs.

The results from our fiducial model are as follows. If we take simple and complex BLAs together, we find that they trace \mbox{$\sim25$} (S/N=50), \mbox{$\sim20$} (S/N=30), and \mbox{$\sim10$} (S/N=10) per cent of the total baryon budget in our simulation. For comparison, BLAs (simple and complex) that arise in gas at \mbox{$T \geq 5\times10^4 \K$} yield \mbox{$\sim24$} (S/N=50), \mbox{$\sim18$} (S/N=30), and  \mbox{$\sim9$} (S/N=10) per cent, which are very close the values obtained from the whole BLA sample. If we restrict the BLAs to be single-component, the resulting baryon fractions in these absorbers are \mbox{$\sim7$} (S/N=50), \mbox{$\sim6$} (S/N=30), and \mbox{$\sim5$} (S/N=10) per cent. This confirms that contamination of the BLA sample by non-thermally broadened lines does not significantly affect the inferred baryon fraction of the WHIM. Incidentally, this suggests that there should be little overlap between the estimates of $\Omega_{\rm b}({\rm NLA})$ and $\Omega_{\rm b}({\rm BLA})$.\\

One final important remark. We have demonstrated that the broad \HI\ absorbers trace only a fraction of the total mass in the WHIM phase. Thus, even in the case that one could accurately estimate the baryon fraction in a given sample of absorbers, there is still a large gap between the observed and true mass contained in this gas phase. Although our simulation suggests that the baryonic masses estimated from observations represent 1/10 to 1/3 of the true baryonic mass in the WHIM, it is not clear how to use the measured baryonic mass to infer the true, total amount of baryons in this gas phase.

\section{Summary} \label{sec:sum}

In this paper, we have used a set of cosmological simulations from the {\em OverWhelmingly Large Simulations} (OWLS) project \citep[][]{sch10a} to study the physical conditions of the gas traced by Broad \HI-\lya\ Absorbers (BLAs) with low and moderate column densities (\mbox{$\log \left( \NHI / \psc \right) \lesssim 15$}) observed in QSO spectra. We have chosen the \mbox{\em AGN} model of the OWLS suite to test the predictions of our simulations against a set of \HI\ observables. We have investigated the impact of metal-line cooling, kinetic feedback by supernovae (SNe) explosions, and feedback by active galactic nuclei (AGN) on the distribution of the gas mass over different phases such as the photo-ionised intergalactic medium (IGM) and the shock-heated, warm-hot intergalactic medium (WHIM). Finally, we have explored the relation between the physical state and the baryon content of these gas phases and both narrow \HI\ absorbers (NLAs) and BLAs.\\

Our detailed results can be summarised as follows:
\begin{itemize}
	
	\item Accretion shocks due to gravitational infall into the potential wells of dark matter halos heat \mbox{$\sim30$} per cent of the total gas mass to temperatures \mbox{$T \geq 5\times10^4 \K$} by \mbox{$z=0.25$} (Sec.~\ref{sec:mod_var}).

	\item Feedback by SNe and AGN each remove a similar amount of gas from the ISM in haloes at early epochs and displace it to the warm-hot diffuse phase, increasing its total mass fraction by another $\sim30$ per cent by \mbox{$z=0.25$} (Sec.~\ref{sec:mod_var}); in other words, roughly half of of the gas mass predicted to be in the WHIM at low redshift ($\sim60$ per cent) has been heated by accretion shocks while the other half is due to strong feedback.
	
	\item The predictions from our simulations are in excellent agreement with standard \HI\ observables (CDDF, line-width distribution, \mbox{$\bhi -  \NHI$} correlation; Secs.~\ref{sec:cddf} -- \ref{sec:bnc}); these observables are rather insensitive to feedback and/or metal-line cooling (Appendix \ref{sec:cnvg_phys}).
	
	\item The line-number density of narrow (\mbox{$\bhi \geq 40 \kms$}; NLA) and broad (\mbox{$\bhi \geq 40 \kms$}; BLA) \lya\ absorbers predicted by our fiducial run \mbox{\em AGN} are in broad agreement with the corresponding observed line-frequencies (Sec.~\ref{sec:dndz_bla}).
	
	\item The density of the \HI\ absorbing gas shows a tight correlation with the \HI\ column density, which agrees well with the analytic prediction of \citet[][Sec.~\ref{sec:odvsN}]{sch01a}; this implies that our simulations are consistent with the assumption that typical \HI\ absorbers are self-gravitating clouds in hydrostatic equilibrium with linear sizes of the order of the local Jeans length.
	
	\item The temperature of the \HI\ absorbing gas correlates well with the \HI\ line width for \mbox{$\log \left( \NHI / \psc \right) \gtrsim 13$}, but it is a poor indicator of the thermal state of the gas for lower column densities; thermal broadening contributes, on average, with at least 60 per cent to the line width of BLAs (Sec.~\ref{sec:bth_b}).
	
	\item	The overwhelming majority of NLAs are found to trace gas at \mbox{$T \sim 10^4 \K$}; their number, temperature distribution and baryon content is very similar in models with/without feedback, thus strongly suggesting that feedback has a negligible impact on the cool, diffuse gas (i.e. the IGM; Sec.~\ref{sec:obs_phys}).
	
	\item BLAs trace gas both at \mbox{$T \sim 10^4 \K$} and at \mbox{$T \sim 10^5 \K$}; our fiducial model, which includes feedback by both SNe and AGN, predicts that 2 out of 3 BLAs arise in gas at \mbox{$T \gtrsim 5\times10^4 \K$}; the number ratio of thermally to non-thermally broadened \HI\ absorbers is very sensitive to (the adopted) feedback (model), and could in principle be used as an indicator of feedback strength (Secs.~\ref{sec:detect}, \ref{sec:obs_phys}).
	
	\item The ionisation state, the total hydrogen content, and the level of enrichment of the gas traced by BLAs indicates that these absorbers arise in gas that is physically distinct from the gas traced by NLAs; we argue that BLAs mostly trace gas that has been recently shock-heated and enriched by outflows (Secs.\ref{sec:b_vs_N}, \ref{sec:obs_phys}).
	
	\item While models including SN and AGN feedback predict a higher fraction of gas mass to be in the warm-hot diffuse phase, the baryon fraction of the gas inferred from BLAs in these models is lower compared to a model without feedback; the reason is that much of the mass is displaced to temperatures and densities for which the \HI\ fraction is too low for the gas to be detectable at the adopted sensitivity (Sec.~\ref{sec:bar_whim}).

	\item The baryon fraction of the gas traced by both NLAs and BLAs predicted by our fiducial model shows broad agreement with corresponding measurements from observations (Sec.~\ref{sec:bar_whim}).
	
	\item Baryonic mass estimates using simple BLAs are robust; in contrast, the gas mass traced by complex BLAs is very sensitive to the adopted (feedback) model (Sec.~\ref{sec:bar_whim}).

	\item Our fiducial model predicts that roughly 6 per cent of the total gas mass is traced by single-component BLAs, which represents about 10 per cent of the total WHIM mass in this model; if the restriction that the absorbers be single-component is dropped, then around 25 per cent of the total gas mass (40 per cent of the WHIM mass) can be recovered from the detected 	broad \HI\ absorption (Sec.~\ref{sec:bar_whim}).

	\item Although some of the gas mass with temperatures \mbox{$T \geq 5\times10^4 \K$} and densities \mbox{$\Delta \lesssim 10^2$} (i.e. the diffuse warm-hot phase) can be traced using BLAs, a significant fraction remains undetected as a consequence of a minimum (instrumental) sensitivity limit. Detection of the bulk of warm-hot gas requires a sensitivity (in terms of the \HI\ central optical depth) of \mbox{$\log \tau_0 \lesssim -2$} (Secs.~\ref{sec:detect}, \ref{sec:bar_whim}).
	
\end{itemize}

\section*{Acknowledgments}
The simulations presented here were run on Stella, the LOFAR Blue Gene/L system in Groningen and on the Cosmology Machine at the Institute for Computational Cosmology in Durham as part of the Virgo Consortium research programme. This work was sponsored by the National Computing Facilities Foundation (NCF) for the use of supercomputer facilities, with financial support from the Netherlands Organisation for Scientific Research (NWO), an NWO VIDI grant, the Marie Curie Initial Training Network CosmoComp (PITN-GA-2009-238356), and the {\em Deutsche Forschungsgemeinschaft} (DFG) through Grant DFG-GZ: Ri 1124/5-1.\\

\bibliographystyle{mn2e} 


\label{lastpage}

\clearpage 
%

\appendix

\section{Line-fitting} \label{sec:fit}

We fit our spectra using a significantly modified version of \textsc{autovp} \citep[][]{dav97a}, assuming each absorption component to be described by a Voigt profile given by the analytic approximation of \citet[][]{tep06a}. \textsc{autovp} identifies absorption features using the equivalent-width significance criterion of \citet[][]{lan87a}. The spectrum is scanned using a window of width $n$ pixels in search of regions with significant absorption. A region is considered significant in absorption if its equivalent width satisfies \mbox{$W \geq N ~\sigma_{W}$}, where $N$ is the significance level, and \mbox{$\sigma_{W}$} is the uncertainty in the equivalent width, integrated over $n$ pixels, given by
\beq
	\sigma_{W} \approx   \sqrt{n}  \left(\frac{\Delta v}{c} \lambda_0 \right)  \cdot (1 + z) \cdot (S/N)^{-1} \, .
\eeq
Here, \mbox{$\Delta v$} is the pixel width, S/N is the adopted signal-to-noise ratio, $z$ is the (central) redshift of the absorption feature, $\lambda_0$ is the rest-frame wavelength of the transition considered (e.g. \HI\ \lya), and $c$ is the speed of light. We adopt \mbox{$N = 7$} and \mbox{$n = 25$} (corresponding to approx. \mbox{$88 \kms$} for our chosen resolution), and \mbox{$\Delta v = 3.5 \kms$}. With these values, the significance value translates into a {\em rest-frame} equivalent width
\beq \label{eq:sl_W}
	W_{r} \approx 500 \left({\rm S/N}\right)^{-1} ~{\rm m\AA} \, .
\eeq
Note that our adopted window width does not affect the parameters of the fitted line(s) in any way.

If fitted by a single component, this implies that our line sample is formally complete down to \HI\ column densities%
\footnote{The quoted value is valid only for absorption lines on the linear part of the curve-of-growth, which is the case for the majority of the components identified in our synthetic spectra. Also note that we do detect lines with column densities (and rest-frame equivalent widths) smaller than the quoted values, since a detected region can be fitted by more than one component.
}
\beq \label{eq:sl_N}
	\NHI \approx 9 \times10^{13} \left({\rm S/N}\right)^{-1} \psc  \, .
\eeq
For Doppler parameters in the range \mbox{$\bhi \geq 40 \kms$}, characteristic of BLAs, the above corresponds to a sensitivity limit in terms of absorption strength
\beq
	\left( \frac{ \NHI / \psc }{ \bhi / \kms } \right) \lesssim 2.3 \times10^{12} \left({\rm S/N}\right)^{-1} \, ,
\eeq
which is equivalent to a \HI\ \lya\ optical depth at the line centre%
\footnote{The relation between line-strength and central optical depth for the \lya\ line is given by eq.~\eqref{eq:tau_0} in Appendix \ref{sec:tau0_vs}.
} \mbox{$\tau_0 \gtrsim 1.74 \left({\rm S/N}\right)^{-1}$}. Note that the values implied by the above equation are below the value commonly adopted for the identification of BLA candidates in real QSO spectra (see eq.~\ref{eq:sl_tau0}). 

A spectrum is fitted in two steps. In the first step, an absorption component is fitted at the pixel with the minimum flux in each detection region, starting with the region with the overall minimum flux. The column density, $\NHI$, and the Doppler parameter, $\bhi$, of the line are both iteratively reduced by a factor 0.99 starting from large values (e.g. \mbox{$\NHI = 10^{20} \psc$} and \mbox{$\bhi = 300 \kms$}; see below) until the flux at that pixel is within $2 \sigma$ below the actual flux level, i.e. in the range \mbox{$[F - 2\sigma, \, F]$}, where $\sigma$ is the local noise. Further components are added and their parameters correspondingly adjusted, taking all previous fitted lines into account, until the residual flux (i.e. the difference between actual flux and model flux) across the detection region is below $2 \sigma$. This procedure is repeated for all detected regions. In a second step, the line parameters (velocity centroid $v_0$, column density $\NHI$, and Doppler parameter $\bhi$) of all lines are simultaneously adjusted using the Levenberg-Marquardt algorithm \citep[][]{lev44a,mar63a} as implemented in \citet[][]{num92a} until the reduced $\chi^2$-value, i.e. the $\chi^2$-value divided by the degrees of freedom, is below \mbox{$\chi^2_{\rm bad} \equiv 1.2$}. If convergence is not achieved, the \los\ is discarded. We note that the fraction of discarded \loss\ is vanishingly small, and it amounts to 1/5000 for our spectra at \mbox{$z=0.25$} and none for our spectra in the range \mbox{$0 \leq z \leq 0.5$}.

Since we do not take higher order \HI\ Lyman transitions into account, saturated \HI\ \lya\ lines deserve special attention. A pixel is considered saturated if the corresponding flux is below \mbox{$2\sigma$}. For \mbox{$\left({\rm S/N} \right) \gtrsim 10$}, this implies that the flux is of the order of, or lower than, \mbox{$2\sigma_{\rm min}$} (where \mbox{$\sigma_{\rm min} = 10^{-4}$}), which is equivalent to a \HI\ \lya\ central optical depth \mbox{$\tau_0 \approx 8.52$} or a \HI\ column density \mbox{$\log (\NHI / \psc) \approx 13.1  + \log (\bhi / \kms)$} (see eq.~\ref{eq:tau_0}). Assuming a Doppler parameter \mbox{$\bhi = 30 \kms$} (which approximately corresponds to the median b-value of our line sample, see Sec.~\ref{sec:bvd}), this corresponds to \mbox{$\log (\NHI / \psc) \approx 14.5$}. In order to prevent our algorithm from severely underestimating the true column density of such saturated lines, and at the same time to avoid fitting lines with unrealistically large column densities along a given \los, during the second fitting step we limit the column density of an individual absorption line%
\footnote{This particular value chosen is arbitrary, but has been found to give satisfactory results. Note that our algorithm may still underestimate the true column density of heavily saturated lines.
} to five times this value, e.g. \mbox{$\log (\NHI / \psc)_{\rm max} \approx 15.2$} for \mbox{$\bhi = 30 \kms$}. Note that this value is not a strict limit but may still vary (in particular, it can be larger) depending on the actual $b$-value of the saturated line. As a consequence, we highly underestimate (by up to two orders of magnitude; see Table \ref{tbl:div}) the actual baryon content in \HI\ along all fitted \loss, which is dominated by high \HI\ column density gas. Yet, the properties of the \HI\ absorbers, in particular the broad \HI\ absorbers that are relevant for the present study, are not affected, since these are dominated by the low column density population with \mbox{$\log (\NHI / \psc) \lesssim 14.5 $}.

During the second fitting step, we impose a minimum line width of \mbox{$\bhi\ = 10 \kms$}, corresponding to \mbox{$T \approx 6000 \K$} assuming pure thermal broadening. Lines narrower than this are discarded during the fitting process unless the new $\chi^2$-value increases above \mbox{$1.2 ~\chi^2_{\rm bad}$}. We note that this cut-off does not appreciably affect the resulting $b$-value distribution (see Fig.~\ref{fig:obs_2}). Indeed, observations indicate that narrow (i.e. \mbox{$\bhi \lesssim 15 \kms$}) \HI\ absorbers at low redshift are scarce \citep[][]{leh07a}. 

Since our synthetic spectra are continuum-normalised by construction, we do not fit a continuum prior to line identification. We limit the line width to a maximum value of \mbox{$\bhi = 300 \kms$}, although larger values are allowed if doing so reduces the $\chi^2$-value below \mbox{$1.2 ~\chi^2_{\rm bad}$}. As a consequence, we find a small number of very broad (\mbox{$\bhi > 200 \kms$}), very shallow absorption features, which become less numerous with decreasing S/N. Since most of these features are real, though scarce, we do not discard them but include them in our resulting line sample.

Finally, any candidate lines with formal relative errors in $\NHI$ or $\bhi$ larger than 50 per cent are sequentially discarded unless the $\chi^2$-value increases above \mbox{$1.2 ~\chi^2_{\rm bad}$}. Note that the final formal errors are typically much smaller than this, around 10 per cent for both $\NHI$ and $\bhi$. 

\section{Observability of \HI\ absorbing gas} \label{sec:tau0_vs}

The neutral hydrogen column density $\NHI$ is given by
\begin{equation}
	\NHI = \NH \cdot \fHI \, , \notag
\end{equation}
where $\NH$ and $\fHI \equiv \fhi$ are the total hydrogen column density and neutral hydrogen fraction, respectively. The total hydrogen column density can be written using the hydrogen particle density $\nH$ as
\begin{equation}
	\NH = \int_0^L \nH ~dl = \overline{ \nH } \cdot L\, , \notag
\end{equation}
where $L$ is the physical, linear extension of the absorbing structure along the \los, and \mbox{$\overline{ \nH }$} is the average hydrogen particle density.  In the following we will write \mbox{$\nH \equiv \overline{ \nH }$}, but the reader should keep the (slight) difference in mind.

The width of an \HI\ absorbing line as measured by the Doppler parameter $\bhi$ may be modelled as
\begin{equation}
	\bhi^2 = b_T^2 + b_H^2 + b_{nt}^2 \, . \notag
\end{equation}
The thermal width, i.e. the broadening due to the temperature $T$ of the absorbing gas is given by
\begin{equation}
	\left[ b_T / \kms \right] = 12.9 \sqrt{T / 10^4\K} \, , \notag
\end{equation}
and the Hubble broadening by \citep[see e.g.][]{sch01a}
\begin{equation}
	b_H \sim \frac{1}{2} ~H(z) \cdot L \, , \notag
\end{equation}
where $H(z)$ is the Hubble parameter (at the appropriate epoch), expressed as
\begin{equation}
	H(z) \equiv h(z) \cdot 10^2 \kms \Mpc^{-1} \, ,
\end{equation}
with
\begin{equation} \label{eq:lil_h}
	h(z) = h_{0} ~ [ \Omega_{\rm m} \left( 1 + z \right)^3 + \Omega_{\Lambda}]^{1/2} \, .
\end{equation}
We adopt the cosmological parameters $\{\Omega_{\rm m}, \, \Omega_{\Lambda}, \, h_{0}\} = \{ 0.238, \, 0.762, \, 0.73 \}$ as derived from the Wilkinson Microwave Anisotropy Probe (WMAP) 3-year data, and find, e.g. \mbox{$h(z=0.25) = 0.81$}.

The remaining term, $b_{nt}$, in the expression for $\bhi$ includes all other forms of non-thermal broadening and is less straightforward to model. It may include turbulence within the absorbing gas, peculiar motions of the absorbing structures, etc. Assuming that these are negligible compared to the thermal and Hubble components, the line width can be approximated by
\begin{equation}
	\left[ \bhi / \kms \right]^2 \approx 166 \cdot \left[ T / 10^4 \K \right] + \frac{1}{4} 10^4 \left[ h(z) \cdot L / \Mpc \right]^2 \, . \notag
\end{equation}
Putting all the above equations together and simplifying, we find that the \HI\ \lya\ absorption strength of the gas is given by
\begin{equation} \notag
	\left( \frac{\NHI / \psc}{ \bhi / \kms } \right) =  \frac{6.17\times10^{12} ~h(z)^{-1} \cdot  [ \nH /  10^{-5} \pcc ] \cdot [\fHI / 10^{-5} ]}{\sqrt{6.64\times10^{-2} \cdot [ T / 10^4 \K ]   \cdot [ h(z) \cdot L/ \Mpc ]^{-2} + 1}} \, .
\end{equation}
The central optical depth of the \HI\ \lya\ transition can be expressed in terms of \mbox{$\left( \NHI / \bhi \right)$} as
\begin{equation} \label{eq:tau_0}
	\tau_0  = \frac{\sqrt{\pi} e^2}{m_{e} c} f_{\text{\lya}} \lambda_{\text{\lya}} \left( \frac{\NHI}{\bhi} \right) = 7.56\times10^{-13} \left( \frac{\NHI / \psc}{\bhi/ \kms} \right) \, .
\end{equation}
Using the above equations we obtain an expression for $\tau_0$ in terms of $\nH$, $T$, and $L$:
\begin{equation}
	\tau_0 =  \frac{4.66 ~h(z)^{-1} \cdot  [ \nH /  10^{-5} \pcc ] \cdot [\fHI / 10^{-5} ]}{\sqrt{6.64\times10^{-2} \cdot [ T / 10^4 \K ] \cdot [ h(z) \cdot L/ \Mpc ]^{-2} + 1}} \, .
\end{equation}
Note that  \mbox{$\fHI = \fHI(\nH, T,z)$}, where the $z$-dependence comes about through the redshift dependence of the ionisation background included in the calculation of $\fHI$.

If we assume that the absorbers have linear sizes of the order of the local Jeans length \citep[][]{sch01a}
\begin{equation} \label{eq:jeans}
	L_{\rm J} = 0.169 ~\Mpc ~[\nH / 10^{-5} \pcc ]^{-1/2} \cdot [T / 10^4 \K ]^{1/2} \cdot [f_g / 0.168 ]^{1/2} \, ,
\end{equation}
we get
\begin{equation}  \label{eq:tau0_jeans}
	\tau_0 = \frac{4.66 ~h(z)^{-1} \cdot [ \nH /  10^{-5} \pcc ] \cdot [\fHI / 10^{-5} ]}{\sqrt{2.32 \cdot [\nH / 10^{-5} \pcc ] \cdot h(z)^{-2} + 1}} \, ,
\end{equation}
where we have assumed the fraction of mass in gas to be close to its universal value%
\footnote{The total mass density parameter is \mbox{$\Omega_m = \Omega_b + \Omega_c$}. The most recent measurements of the  baryonic and dark matter density parameters yield, respectively, \mbox{$\Omega_b = 0.0449\pm0.0028$} and \mbox{$\Omega_c = 0.222 \pm 0.026$} \citep[][]{jar11a}.}
\mbox{$f_g \equiv \Omega_b / \Omega_m = 0.168$}. Note that the above equation does no longer depend explicitly on the temperature, but it does depend {\em implicitly} on it through the dependence on $\fHI$.\\

Using
\begin{equation} \label{eq:nHtoD}
	\nH = \frac{\left<\rho_{\rm b}\right>}{\mH}~X_{\rm H}~(1+z)^{\,3} \Delta  \approx 1.88 \times 10^{\,-7} \cm^{\,-3}~\left( \frac{X_{\rm H}}{0.752} \right)~(1+z)^{\,3} \Delta \notag
\end{equation}
the above equations can all be expressed in terms of the overdensity $\Delta$ as well.

\begin{figure}
{\resizebox{\colwidth}{!}{\includegraphics{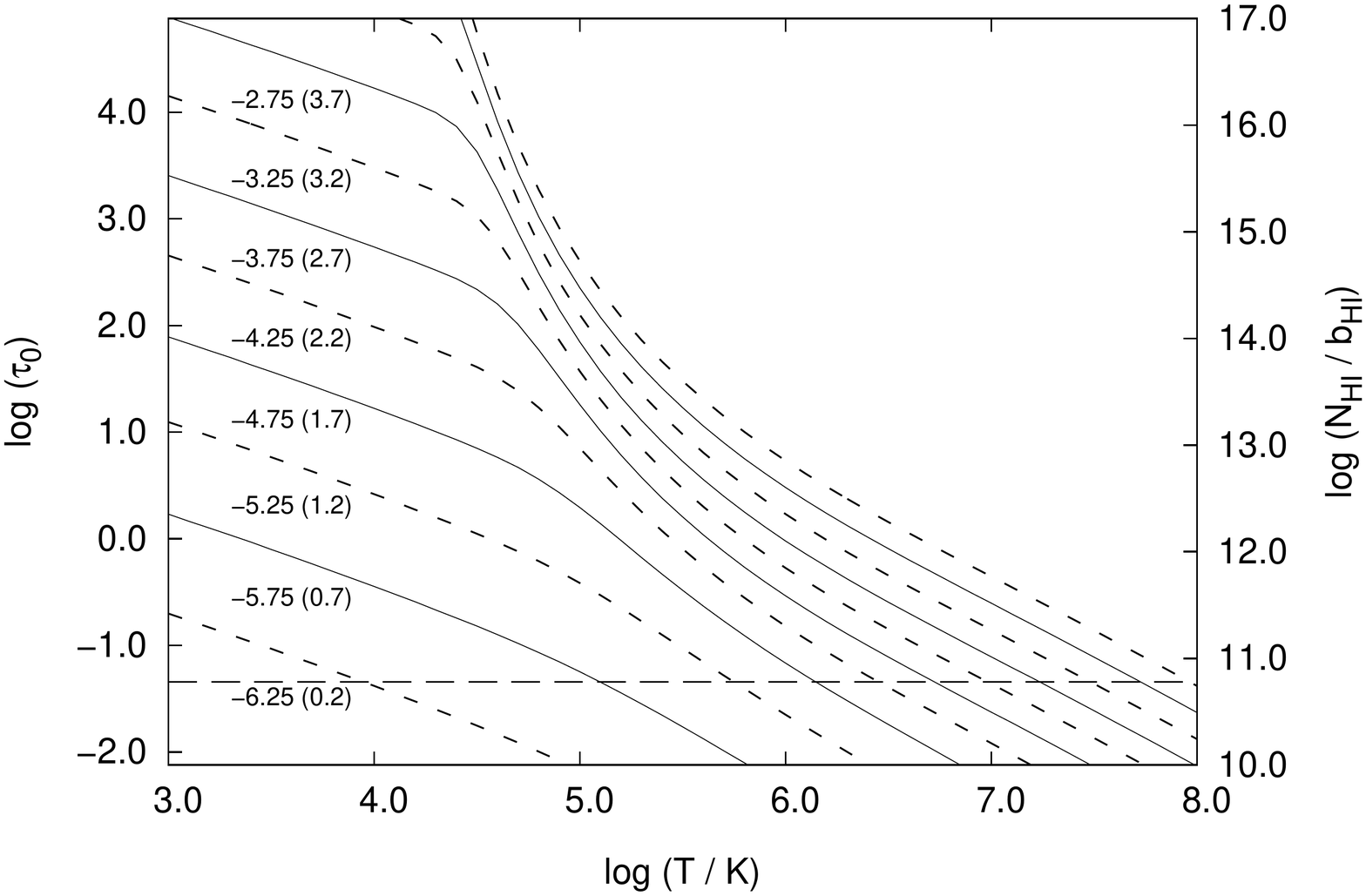}}}
\caption[]{\HI\ \lya\ central optical depth, \mbox{$\tau_0$}, as given by eq.~\eqref{eq:tau0_jeans}, required to detect gas with a given temperature for a range of densities at \mbox{$z=0.25$}. The value along each curve indicates the corresponding logarithmic hydrogen particle density, \mbox{$\log \left( \nH / \pcc \right)$}. The number in parentheses indicates the corresponding logarithmic overdensity. For reference, we include the alternative $y$-axis which shows the corresponding value of the \HI\ \lya\ absorption strength at the line centre, \mbox{$\log \left( \NHI / \bhi \right)$} in units of \mbox{$\psc / \kms$} (see equation \ref{eq:tau_0}). The dashed horizontal line indicates the adopted BLA detection threshold given by eq.~\eqref{eq:sl_tau0}.}
\label{fig:Noverb_vs_T}
\end{figure}

Fig.~\ref{fig:Noverb_vs_T} shows the \HI\ \lya\ central optical depth, $\tau_0$, as a function of gas temperature for a range of densities typical of intergalactic gas, as given by eq.~\eqref{eq:tau0_jeans}. The alternative $y$-axis shows the corresponding values for \mbox{$\log \left( \NHI / \bhi \right)$} in units of \mbox{$\psc / \kms$}. The value along each curve indicates the assumed logarithmic hydrogen particle density, \mbox{$\log \left( \nH / \pcc \right)$}, and the value in parentheses indicates the corresponding logarithmic overdensity, \mbox{$\log \Delta$}, at \mbox{$z=0.25$}. The horizontal dashed line indicates our adopted sensitivity limit as given by eq.~\eqref{eq:sl_tau0}. This figure demonstrates that the detectability of \HI\ absorbing gas at a given density drops sharply with temperature. For example, the detection of gas with \mbox{$\log ( \nH / \pcc) = -4.75 $} (which corresponds to \mbox{$\Delta \sim 50 $} at \mbox{$z=0.25$}) and \mbox{$\log \left( T / \K \right) \sim 6 $} requires a minimum sensitivity which is two orders of magnitude higher than the one required to detect gas with the same density and \mbox{$\log \left( T / \K \right) \sim 5 $}. The reason behind the strong dependence of \mbox{$\log \left( \NHI / \bhi \right)$} on temperature is that, at a given density, $\tau_0$ is completely dominated by the neutral fraction, $\fHI$. 

\section{Convergence with respect to the physical model} \label{sec:cnvg_phys}

In this section, we demonstrate that the predicted \HI\ observables are robust with respect to variations of the adopted physical model.

\begin{figure}
\resizebox{\colwidth}{!}{\includegraphics{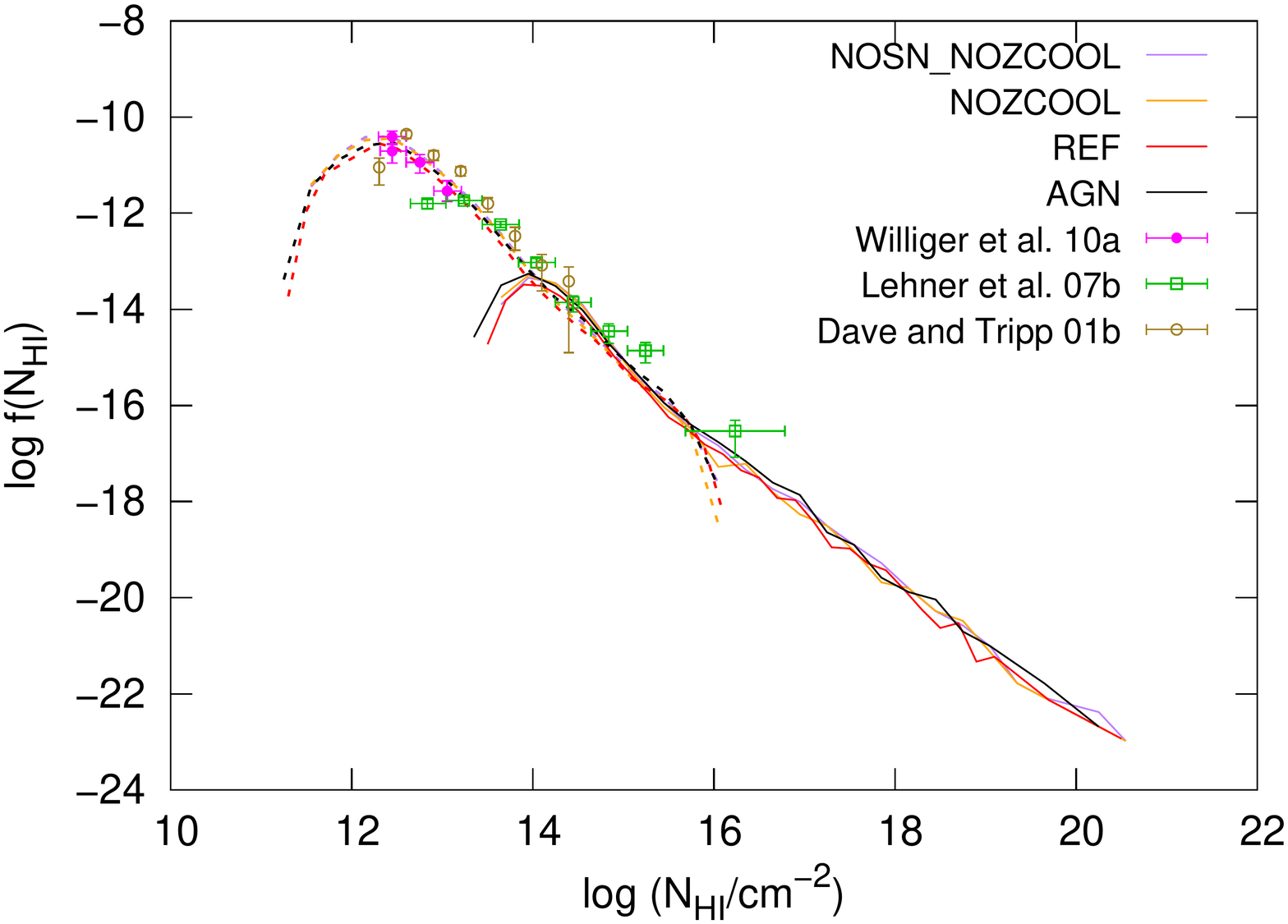}}
\caption[ ]{CDDF for different model runs in a box $L = 100 \hMpc$ per side at $z = 0.25$ using 5000 random \loss. Dashed lines show the distribution of column densities for individual absorption lines obtained from fitting the synthetic spectra adopting S/N=50. Solid lines show the distribution of {\em integrated} $\NHI$ along each \los. Note that the adopted S/N value is higher than the S/N of the data, and the latter are only included for reference. The distribution of column densities for individual absorption components extends to much lower values since a spectrum (corresponding to a single physical \los) is generally fitted with more than one component.}
\label{fig:convg_cddf}
\end{figure}

Fig.~\ref{fig:convg_cddf} shows the \HI\ column-density distribution (CDDF) for different models run in a box \mbox{$L = 100 \hMpc$} per side at \mbox{$z = 0.25$}, using 5000 random \loss. Dashed lines show the distribution of column densities for individual absorption components obtained from fitting the synthetic spectra adopting S/N=50, as described in Sec.~\ref{sec:obs}; solid lines show the distribution of the {\em integrated} \HI\ column density along each \los. The latter are important in order to remove the uncertainty introduced in the CDDF by our fitting algorithm. We include various data sets in this figure for reference, but note that our adopted S/N value is higher than the S/N of the data. The distribution of column densities for {\em individual} absorption components extends over a range of much lower values since a spectrum (corresponding to a single physical \los) is generally fitted with more than one component. Also, as mentioned in Sec.~\ref{sec:obs}, our version of \textsc{autovp} tends to fit (highly) saturated lines with more than one component, thus generally yielding lower $\NHI$ values for each individual component with respect to the $\NHI$ value obtained by integrating over the corresponding absorption feature.

\begin{figure}
\resizebox{\colwidth}{!}{\includegraphics{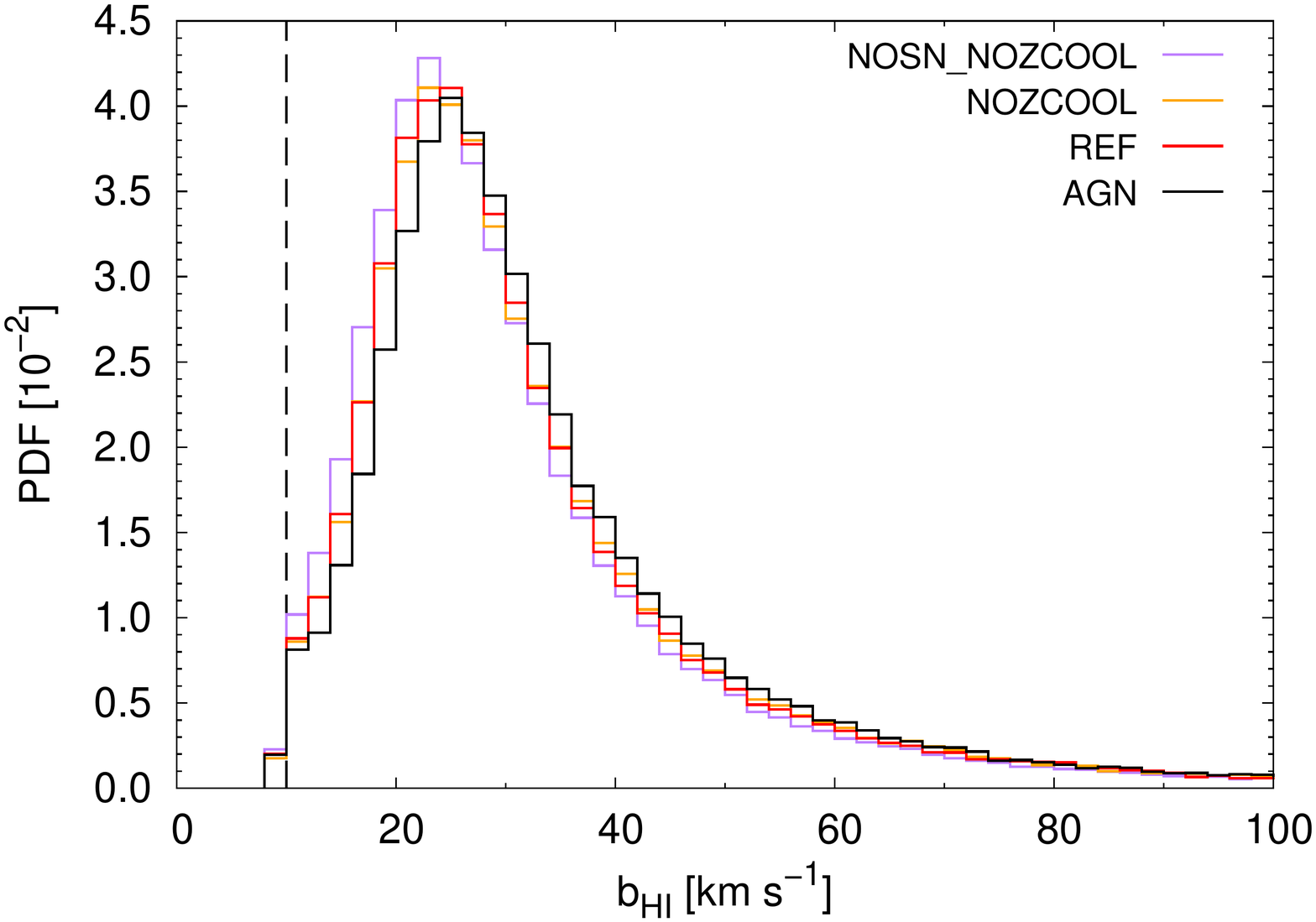}}
\caption[ ]{Distribution of Doppler parameters of individual components identified in 5000 spectra at \mbox{$z = 0.25$} with S/N=50 adopting different physical models in a \mbox{$L = 100 \hMpc$} box. The dashed vertical line indicates the minimum Doppler parameter during the fitting process (see Sec.~\ref{sec:obs}). Note that feedback (both by SNe and AGN) leads to a slightly larger number of broad lines, although the effect is small.}
\label{fig:convg_bv}
\end{figure}

Fig.~\ref{fig:convg_bv} shows the distribution of Doppler parameters of individual components identified in 5000 spectra at \mbox{$z = 0.25$} with S/N=50 adopting different physical models run in a box \mbox{$L = 100 \hMpc$} per side. Note that feedback (both by SNe and AGN) leads to a slightly larger number of broad lines, although the effect is not large.

\begin{figure}
\resizebox{\colwidth}{!}{\includegraphics{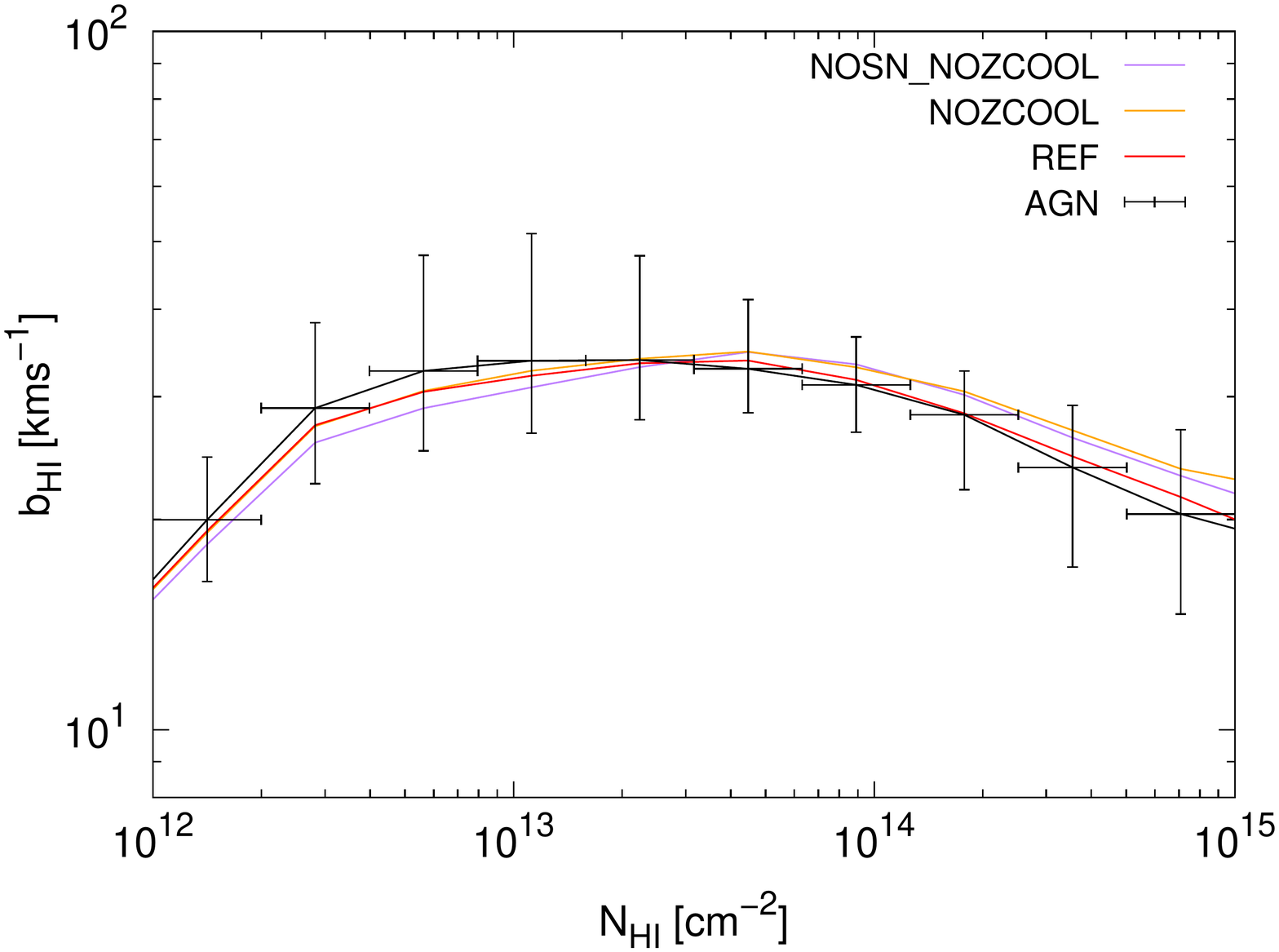}}
\caption[ ]{$\bhi - \NHI$ distribution of individual components identified in 5000 spectra at $z = 0.25$ with S/N=50  adopting different physical models in a box \mbox{$L = 100 \hMpc$} per side. For clarity, errors bars (corresponding to the bin size along the $x$-axis and to the 25-/75-percentiles along the $y$-axis) are shown only for the AGN result. We note that the scatter for the other models is similar.}
\label{fig:convg_bN}
\end{figure}

Finally, Fig.~\ref{fig:convg_bN} shows the corresponding $\bhi - \NHI$ distribution of individual components. The $\bhi - \NHI$ distribution for each model has been obtained as described in Sec.~\ref{sec:bnc}. For each model, the corresponding median values at each bin are connected by continuous lines. For simplicity, only the result for model \mbox{\em AGN} displays $x$- and $y$-error bars indicating, respectively, the bin size and 25-/75-percentiles; but note that the bin-size is identical and the scatter similar for all other models. Although the models differ slightly from each other, the various $\bhi - \NHI$ distributions are fully consistent with each other.\\

Clearly, feedback by supernovae and AGN does not significantly affect the observed properties (\HI\ column density distribution, Doppler parameter distribution,  \mbox{$\bhi - \HI$} correlation) of the gaseous structures giving rise to \HI\ absorption. This result is consistent with previous results by \citet[][]{the02a} who, however, did not consider AGN feedback or metal-line cooling.

\section{Numerical convergence} \label{sec:cnvg_num}

In this section, we address the convergence of our results with respect to varying the box size and the mass and spatial resolution. To this end, we compare the \HI\ CDDF and Doppler parameter distribution of individual components identified in 5000 spectra obtained from simulation runs with different box sizes and resolutions, which all adopt the model \mbox{\em REF} at  \mbox{$z=0.25$}. Note that the use of this particular model does not affect our results, since we have shown in Appendix \ref{sec:cnvg_phys} that both the \HI\ CDDF and the distribution of Doppler parameters are insensitive to the adopted model. In the following, the simulation runs we use are denoted by $LxxxNyyy$, where $xxx$ corresponds to the linear size of the cubic box in $\hMpc$, and $yyy$ to the number of (dark matter, baryonic) particles per side.

To investigate the convergence with the box size, we use the simulation runs $L025N128$, $L050N256$, and $L100N512$, which all have the same mass- and spatial resolution. The convergence with resolution is investigated using simulations run in a box \mbox{$L = 50 \hMpc$} per side, and varying the (dark matter, baryonic) particle number; more specifically, we use the runs $L050N128$, $L050N256$, and $L050N512$, whose mass (spatial) resolution varies in factors of 8 (2). The choice of this particular box size is arbitrary but justified, since our results are converged with respect to the box size, as we will show next.

\begin{figure}
{\resizebox{\colwidth}{!}{\includegraphics{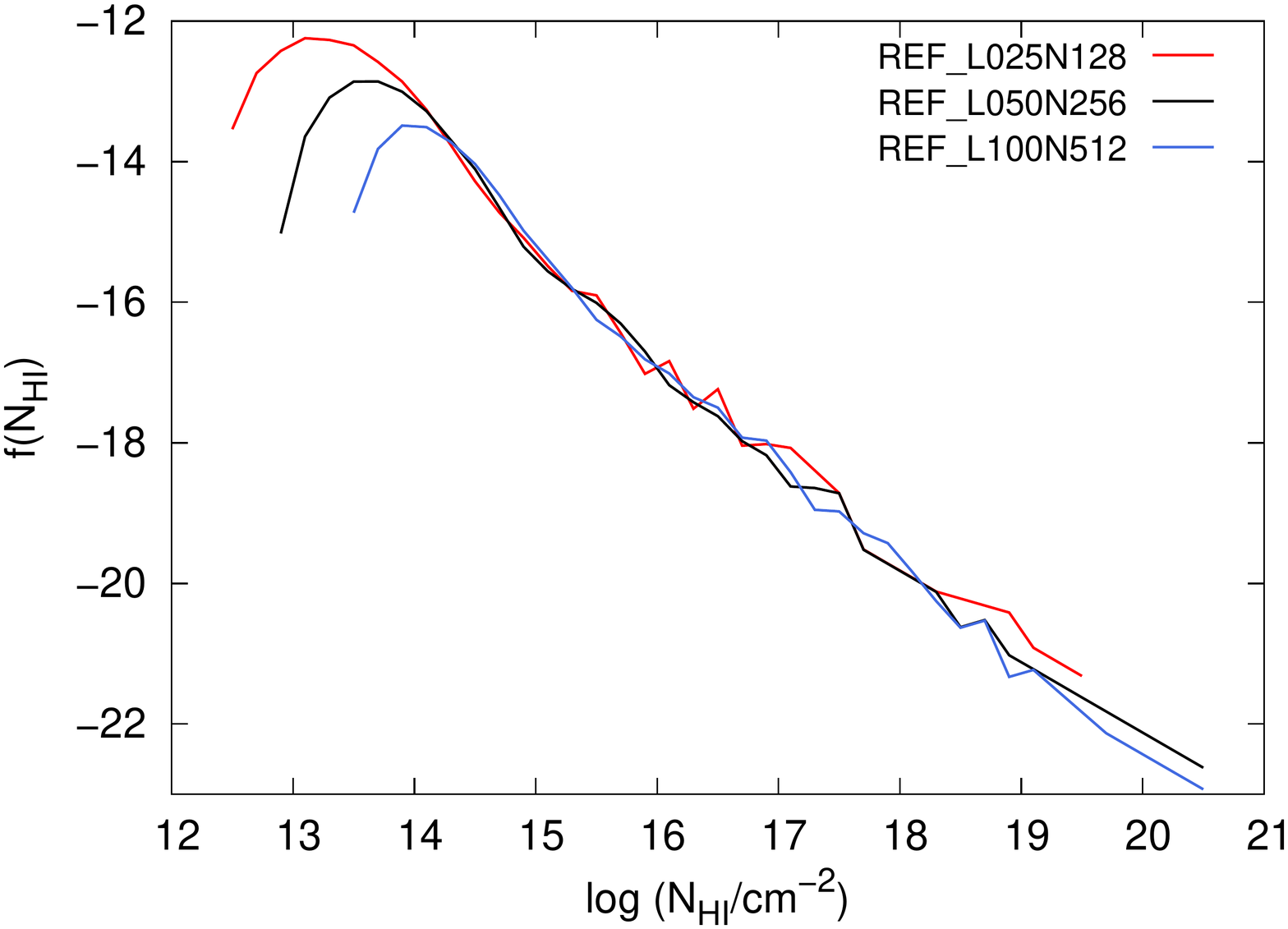}}}
{\resizebox{\colwidth}{!}{\includegraphics{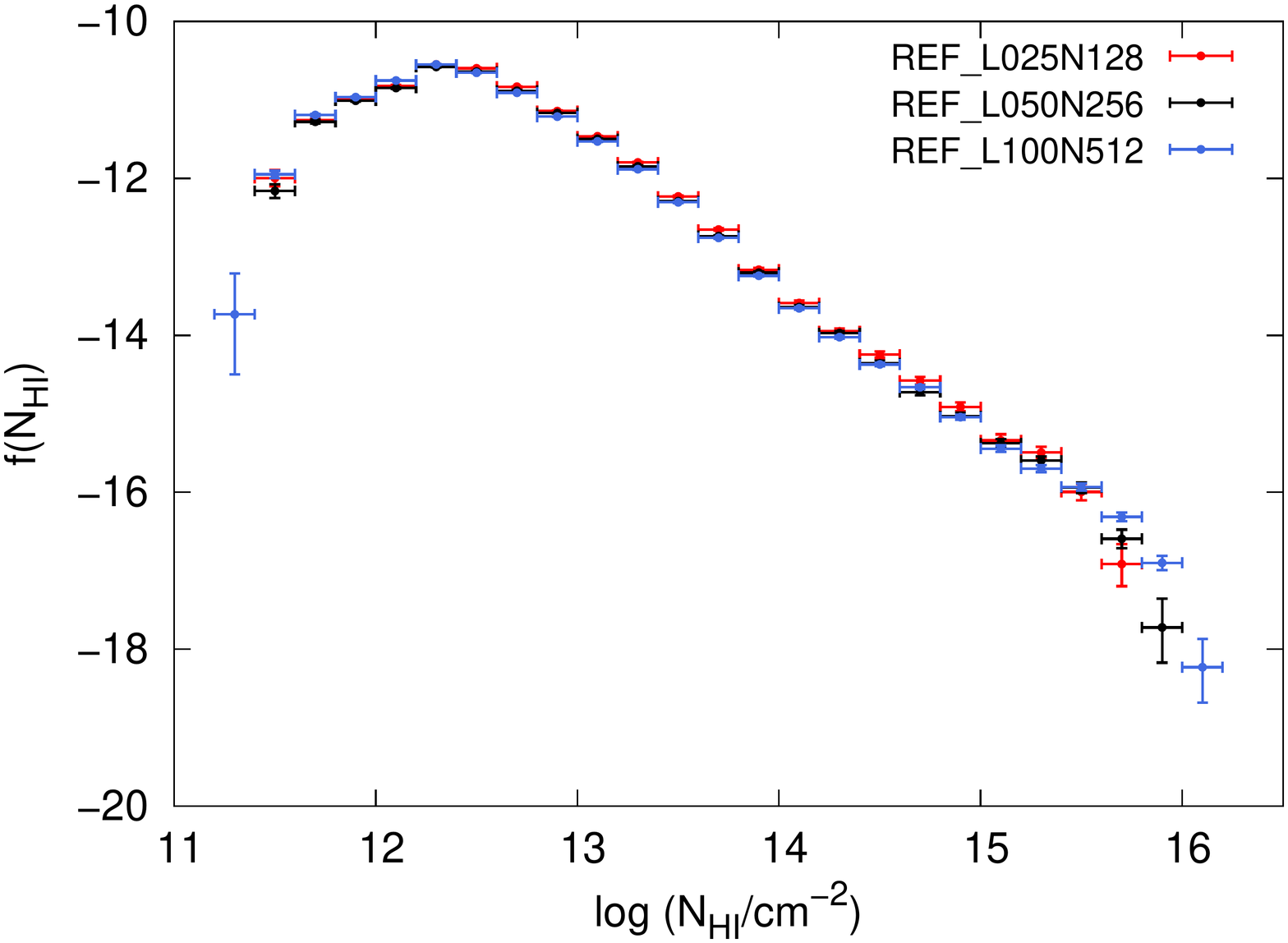}}}
{\resizebox{\colwidth}{!}{\includegraphics{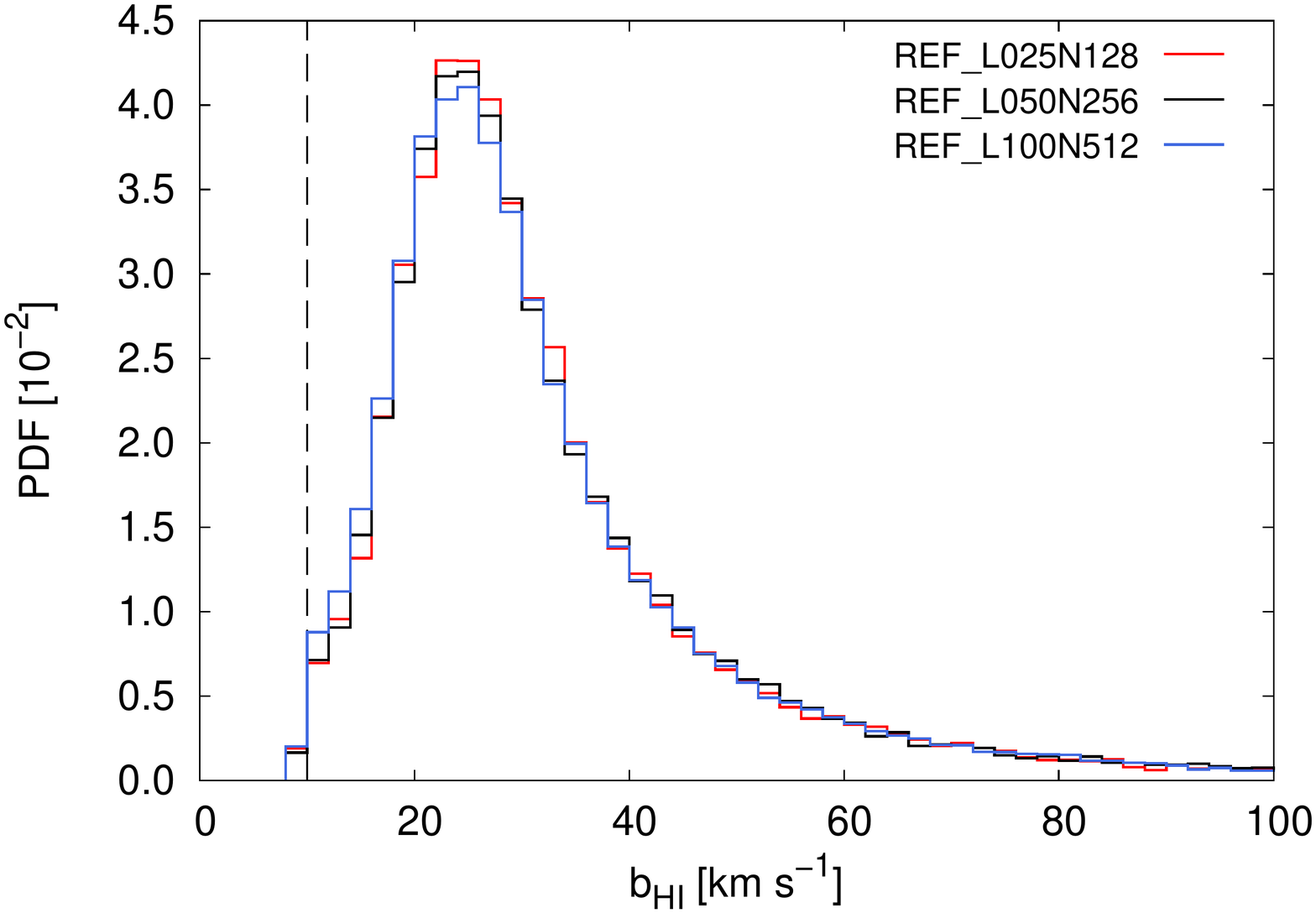}}}
\caption[]{Numerical convergence with respect to the box size at a fixed mass and spatial resolution using 5000 \loss\ and corresponding spectra obtained from a simulation which adopts the model \mbox{\em REF} in a \mbox{$L = 100 \hMpc$} box per side at \mbox{$z = 0.25$}. {\em Top:} \HI\ CDDF obtained from the \HI\  column density integrated along each \los. {\em Middle:} \HI\ CDDF of the \HI\ absorbing components identified in the corresponding spectra with S/N=50. Note the difference in the plotted range between the top and middle panels. {\em Bottom:} Distribution of Doppler parameters. The dashed vertical line indicates the minimum Doppler parameter allowed during the fitting process (see Appendix \ref{sec:fit}).}
\label{fig:num_conv_box}
\end{figure}

Fig.~\ref{fig:num_conv_box} shows the \HI\ CDDF and the distribution of Doppler parameters using 5000 \loss\ obtained from simulation runs with different box size but fixed mass- and spatial resolution. The top panel shows the distribution obtained from the \HI\ column density integrated along each individual \los; the middle panel shows the distribution obtained from the \HI\ column density of each component identified in the corresponding spectra with S/N=50. The \HI\ CDDF of integrated column densities is fully converged for column densities  \mbox{$\NHI < 10^{17} \psc$}. At higher column densities, we do not expect a good convergence, since these column densities correspond to the optically thick regime, while our calculations assume optically thin gas. The \HI\ CDDF of individual components is fully converged at column densities  \mbox{$\NHI < 10^{15} \psc$}, which is the relevant column density range for this study. The difference between the various \HI\ CDDFs of individual components in the range \mbox{$\NHI > 10^{15} \psc$} is due to the inability of our fitting algorithm to accurately determine the column density of saturated \HI\ absorption features. Finally, the bottom panel demonstrates that the Doppler parameter distribution in the range \mbox{$10 \kms \leq \bhi \leq 100 \kms$} is fully converged with respect to the box size.\\

\begin{figure}
{\resizebox{\colwidth}{!}{\includegraphics{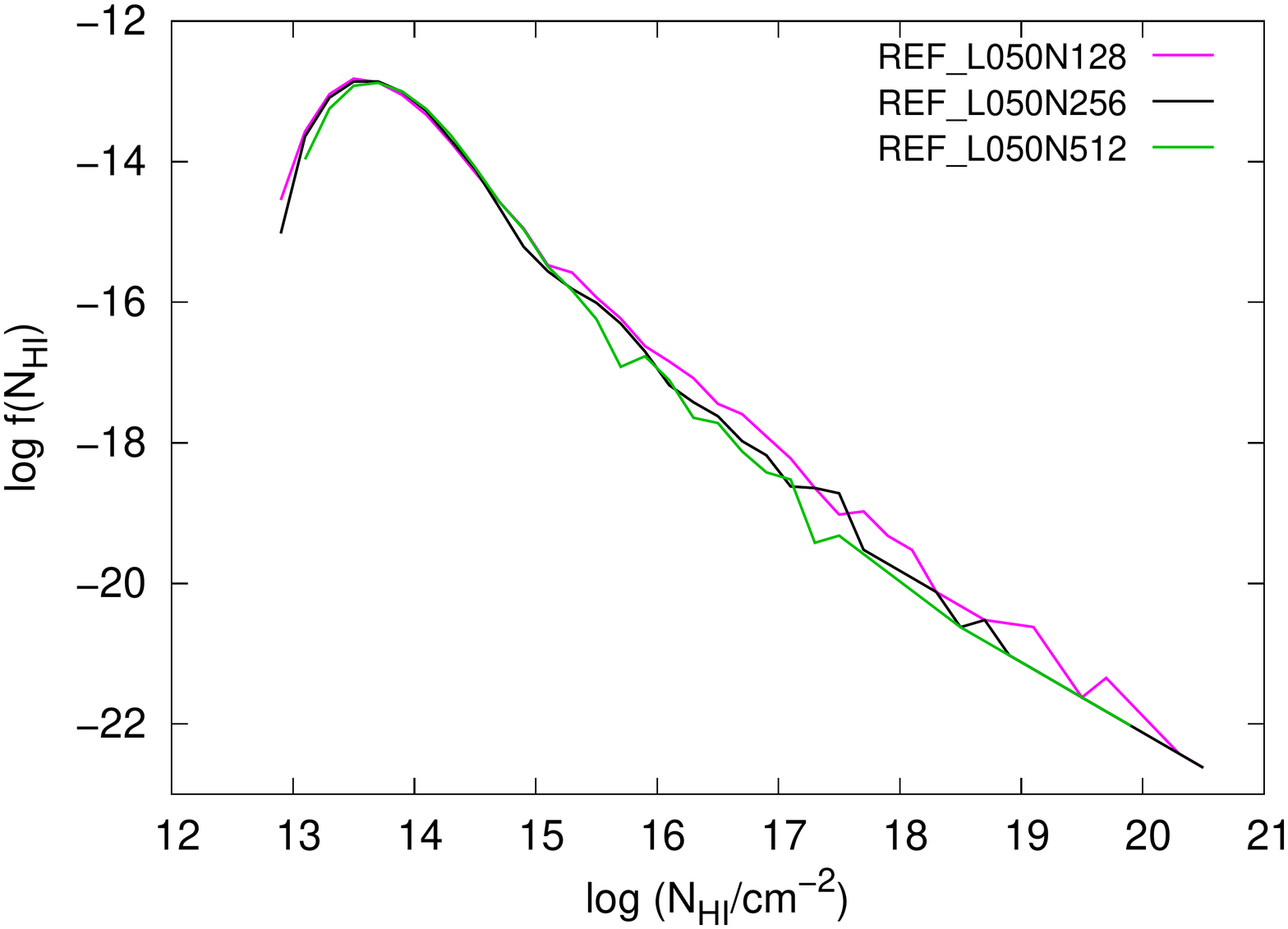}}}
{\resizebox{\colwidth}{!}{\includegraphics{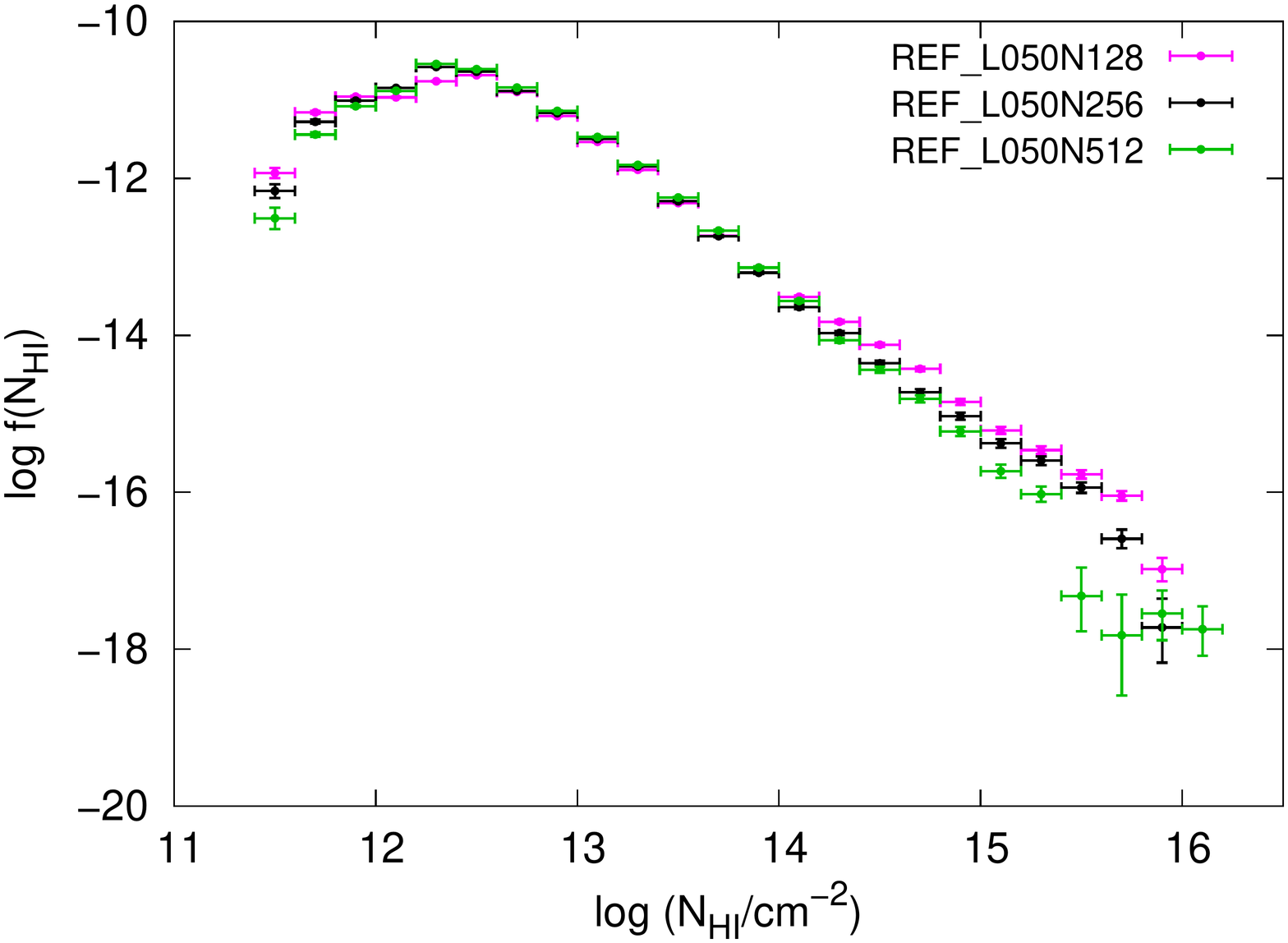}}}
{\resizebox{\colwidth}{!}{\includegraphics{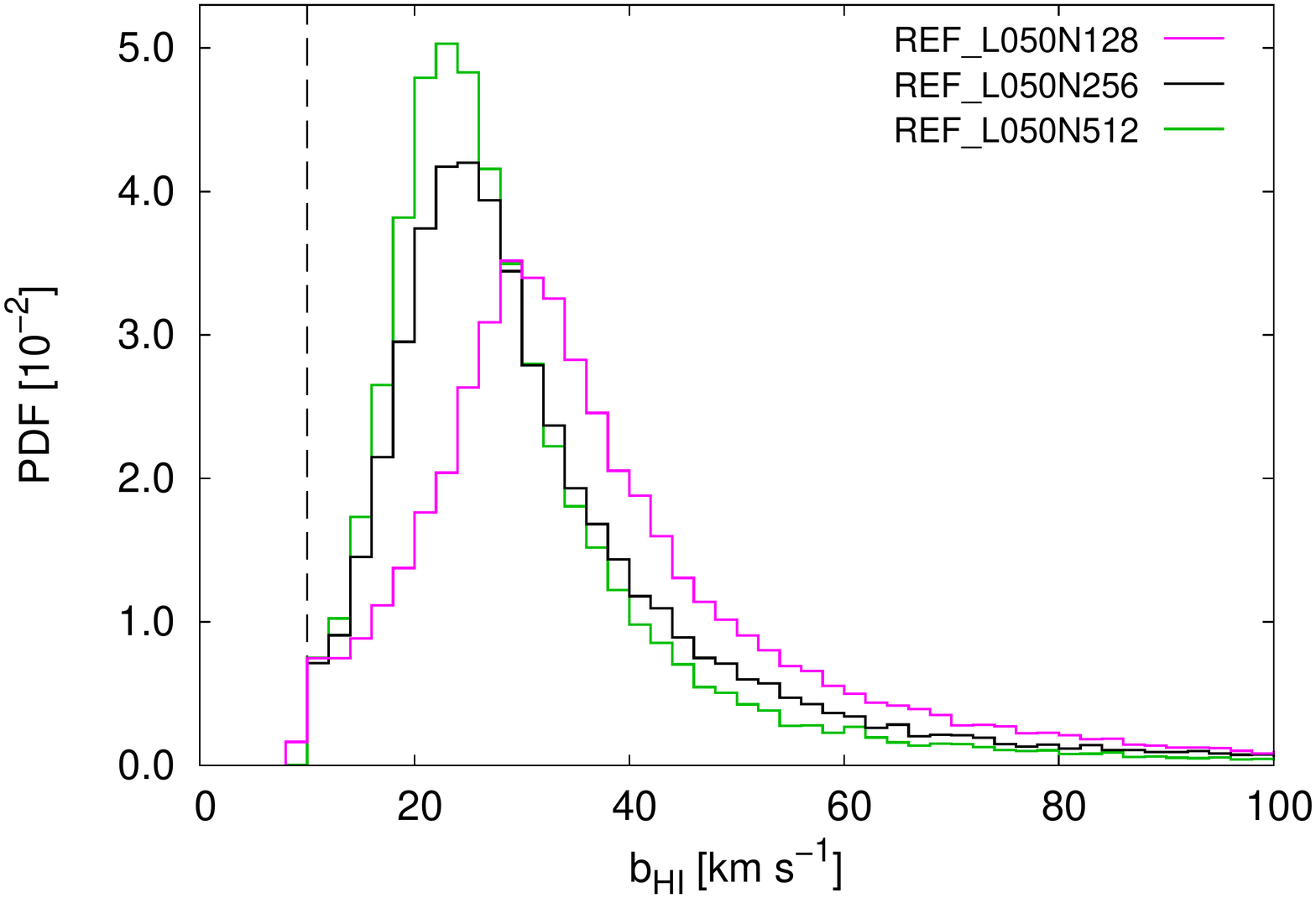}}}
\caption[]{Same as Fig.~\ref{fig:num_conv_box} for the numerical convergence with respect to the mass and spatial resolution at a fixed box size.}
\label{fig:num_conv_res}
\end{figure}

Fig.~\ref{fig:num_conv_res} shows the \HI\ CDDF and the distribution of Doppler parameters using 5000 \loss\ obtained from simulations run in a box \mbox{$L = 50 \hMpc$} per side with different mass and spatial resolution. Both the distribution of \HI\ column densities integrated along each \los\ (top panel) and the distribution of \HI\ column densities of individual components identified in the corresponding spectra (middle panel) appear to be fully converged with respect to resolution at column densities \mbox{$\NHI < 10^{15} \psc$}. At higher column densities, neither distribution is fully converged, although the difference between the two highest resolution runs, $L050N256$ and $L050N512$,  is very small for the integrated column densities. In the case of the \HI\ CDDFs for individual components at \mbox{$\NHI > 10^{15} \psc$}, this is again in part due to the difficulty in determining the true column density of saturated \HI\ absorption features.

The Doppler parameter distribution shown in the bottom panel of Fig.~\ref{fig:num_conv_res} indicates that the resolution of the $L050N128$ run is not high enough. The distribution in the $L050N256$ run, which has the same resolution as our fiducial run, does not show full convergence at Doppler parameters in the BLA regime (\mbox{$\bhi \geq 40 \kms$}) with respect the higher resolution run, $L050N512$, although the difference is small \citep[see also][]{the98b}.\\

In summary, our results are robust with respect to varying the size of the simulation box, and our adopted resolution is high enough to guarantee the convergence of our results  in the range of column densities relevant for this study. However, the distribution of Doppler parameters is slightly sensitive to the adopted resolution in the range of interest for BLAs, i.e. for Doppler parameters \mbox{$\bhi \geq 40 \kms$}. Thus, some caution is advised when interpreting results based on or making predictions for this observable.

\end{document}